\documentclass[fleqn,usenatbib]{mnras}
\pdfoutput=1

\usepackage[T1]{fontenc}
\usepackage{ae,aecompl}

\usepackage{graphicx}	
\usepackage{amsmath}	
\usepackage{amssymb}	
\usepackage{longtable}
\usepackage{enumitem}
\usepackage{xspace}

\newcommand{\Msun}{M$_{\odot}$\xspace}
\newcommand{\Mvir}{M$_{\rm vir}$\xspace}
\newcommand{\Mmax}{M$\rm_{max}$\xspace}
\newcommand{\Rvir}{R$_{\rm vir}$\xspace}
\newcommand{\Vvir}{V$_{\rm vir}$\xspace}
\newcommand{\Rvirz}{R$_{\rm vir}$(z)\xspace}
\newcommand{\kms}{km s$^{-1}$\xspace}
\newcommand{\tcross}{t$\rm_{cross}$\xspace}
\newcommand{\footnoteref}[1]{\textsuperscript{\ref{#1}}}

\title[Orbits of the LMC and M33]{Orbits of Massive Satellite Galaxies: I. A Close Look at the Large Magellanic Cloud and a New Orbital History for M33}

\author[E. Patel et al.]{
Ekta Patel,$^{1}$\thanks{E-mail: ektapatel@email.arizona.edu}
Gurtina Besla,$^{1}$
and Sangmo Tony Sohn$^{2}$
\\ 
$^{1}$Department of Astronomy, University of Arizona, 933 North Cherry Avenue, Tucson, AZ 85721, USA\\
$^{2}$Space Telescope Science Institute, 3700 San Martin Drive, Baltimore, MD 21218, USA\\}

\date{Accepted XXX. Received YYY; in original form ZZZ}

\pubyear{2016}

\begin{document}
\label{firstpage}
\pagerange{\pageref{firstpage}--\pageref{lastpage}}
\maketitle

\begin{abstract}
The Milky Way (MW) and M31 both harbor massive satellite galaxies, the Large Magellanic Cloud (LMC) and M33, which may comprise up to 10 per cent of their host's total mass. Massive satellites can change the orbital barycentre of the host-satellite system by tens of kiloparsecs and are cosmologically expected to harbor dwarf satellite galaxies of their own. Assessing the impact of these effects depends crucially on the orbital histories of the LMC and M33. Here, we revisit the dynamics of the MW-LMC system and present the first detailed analysis of the M31-M33 system utilizing high precision proper motions and statistics from the dark matter-only Illustris cosmological simulation. With the latest {\em Hubble Space Telescope} proper motion measurements of M31, we reliably constrain M33's interaction history with its host. In particular, like the LMC, M33 is either on its first passage ($\rm t_{inf}$ < 2 Gyr ago) or if M31 is massive ($\geq2 \times 10^{12}$ \Msun), it is on a long period orbit of about 6 Gyr. Cosmological analogs of the LMC and M33 identified in Illustris support this picture and provide further insight about their host masses. We conclude that, cosmologically, massive satellites like the LMC and M33 are likely completing their first orbits about their hosts. We also find that the orbital energies of such analogs prefer a MW halo mass $\sim$$1.5\times10^{12}$ \Msun and an M31 halo mass $\geq1.5\times10^{12}$ \Msun. Despite conventional wisdom, we conclude it is highly improbable that M33 made a close (< 100 kpc) approach to M31 recently ($\rm t_{peri}$ < 3 Gyr ago). Such orbits are rare (< 1 per cent) within the 4$\sigma$ error space allowed by observations. This conclusion cannot be explained by perturbative effects through four body encounters between the MW, M31, M33, and the LMC. This surprising result implies that we must search for a new explanation for M33's strongly warped gas and stellar discs.
\end{abstract}




\section{Introduction}
\label{sec:intro}
Both the Milky Way (MW) and M31 host systems of satellite dwarf galaxies that are relics of their assembly history. These satellite galaxies are typically assumed to exert minimal gravitational forces on their hosts or on each other. As such, satellites are often considered point mass tracers of their host potentials. However, this assumption breaks down if the total mass of the satellite is a significant fraction of the host mass.  

The MW and M31 both host such massive satellites, the Large Magellanic Cloud (LMC) and M33, respectively. With stellar masses of 3-5 $\times 10^9$ \Msun, both galaxies are cosmologically expected to have dark matter masses of order $10^{11}$ \Msun, roughly 10 per cent of the total mass of the MW or M31 \citep{moster13}. 

While we generally think of satellites as being heavily affected by their hosts (via tides, ram pressure, etc.), massive satellites can in turn affect the dynamics and structure of their hosts as well. They can induce warps in the host galactic disc \cite[e.g. ][]{weinberg98,weinberg06}, shift the orbital barycentre of the host-satellite system by tens of kiloparsecs \cite[e.g. ][hereafter G15]{gomez15}, and are cosmologically expected to host dwarf satellite galaxies of their own \citep{sales13, deason15}. 

The past orbital trajectory of these galaxies is critical to understanding the origin of and magnitude to which these effects play a role in the dynamical history of the Local Group. More specifically, the accretion time, number of pericentric approaches, and the host-satellite separation at those pericentric passages are key determinants for these phenomena. The survivability of satellites in the environment of their host haloes is also directly connected to the time-scale over which satellites approach and potentially interact with their hosts. 

Knowledge of warps in the gas discs of the MW and M31 date back to the mid-20$\rm^{th}$ century \citep[e.g. ][]{oort58, roberts75, newton77}. Both warps reach heights up to several kiloparsecs above the disc plane. Several authors have argued that the LMC could be responsible for inducing the warp in the MW's disc \citep[e.g. ][]{weinberg98, weinberg06, tsuchiya02} or whether the influence of other satellites, such as the Sagittarius dSph, also play a role in this phenomena \citep{laporte16, gomez13}. Similarly, M31 satellites are suspects in the formation of the warp in M31's gas disc and other prominent substructures, such as its star forming ring \citep{block06, fardal09}. M33 being the most massive of those satellites may contribute to the current structure of M31's warped disc if it once reached a similar pericentric distance as the LMC ($\sim$50 kpc). On a lower mass scale, satellites of massive satellite galaxies could have similar impacts on their dwarf hosts if their orbits allow for close passages \citep[e.g. the LMC's disc may be warped owing to interactions with the Small Magellanic Cloud;][]{besla12,besla16}.

 In 2015, a slew of ultra-faint dwarf galaxy candidates were discovered in the Southern hemisphere, many of which are located in close proximity to the LMC \citep{bechtol15, koposov15, martin15, dwagner15, kim15a, kim15b}. Several studies have suggested that the new ultra-faint dwarfs are dynamical companions of the LMC \citep[e.g. ][]{deason15, jethwa16, martin15}. This association depends directly on the orbital history of the LMC and its purported system of satellites about the MW \citep{sales11}. For example, in a first infall scenario for the LMC, there has not been enough time for the MW's tides to disrupt the infalling system and consequently remove traces of common orbital trajectories and similar kinematics. Whether M33 may harbor faint satellite companions today will similarly depend sensitively on its orbital history about M31.
 
As massive satellites approach distances within tens of kpc from their hosts, the high mass ratio of the system becomes exacerbated. For example, at a separation of 50 kpc, the total mass of the LMC may be up to 25 per cent of the MW mass enclosed within a similar radius. G15 illustrate that the MW experiences a strong gravitational influence due to the massive LMC residing nearby. As a result, the orbital barycentre of the MW-LMC sloshes back and forth over Gyr time-scales--this effect can also modify the orbital planes of other MW satellites like the Sagittarius dSph (G15). The M31-M33 system is similarly susceptible to this gravitational effect, as \citet{dierickx14} have shown for the past M31-M32 orbital history. Thus a constraint on the closest passage of M33 about M31 is crucial to understanding the current and future dynamics of M31 and its satellites. 

Furthermore, the host-satellite separation determines the morphological impact that processes such as tidal stripping will have on the satellite. The time-scales over which satellite galaxies deplete their gas reservoirs and cease forming stars (quenching) is also sensitive to the orbital histories of the satellites about their hosts \citep[e.g. ][]{wetzel14, wetzel15}. It is curious that the MW and M31 both host a massive, gas-rich satellite at distances where other satellites are gas poor. These abundant gas reservoirs suggest that the LMC and M33 have only recently been accreted by their hosts. A recent accretion scenario is consistent with proper motion measurements of the LMC \citep{k13,b07} and cosmological expectations \citep{bk11, busha11, gonzalez13}. However, to date the orbital history of M33 about M31 has not been similarly examined. 
 
 A scenario under which M33 makes a close passage to M31 is presented by \cite{mcconnachie09} and \citet{putman09}. The gas and stellar discs of M33 are substantially warped. These studies require that M33 made a close (50-100 kpc) encounter with M31 in the past 3 Gyr. We will use these models as a foundation to our assessment of M33's orbital history in analytic models and cosmological simulations.

To date, a rigorous, simultaneous study of the orbital history of M33, both numerically and cosmologically, has not been conducted. The major missing component for such an analysis has been a precise measurement of M31's proper motion. Recent {\em Hubble Space Telescope (HST)} observations have constrained the tangential velocity of M31 to $\rm v_{tan}=17\pm17$ \kms \citep{sohn12,vdm12ii}. Others have inferred the tangential velocity component of M31 by using the kinematics of M31 satellites \citep{vdmG08, salomon16}, and the latter reports a value as high as $\rm v_{tan}\!\sim\!150$ \kms. The discrepancy between these values has severe implications for the history and current state of the Local Group. Depending on its orbital history, M33 may help minimize this discrepancy or it may simultaneously impact the motions of other M31 satellites, further complicating such analyses.

M33's proper motion was measured recently by \cite{brunthaler05} using the {\em Very Long Baseline Array (VLBA)}. Together, the proper motion measurements of M33 and M31 allow us to constrain the relative motion of the two galaxies, enabling us to quantify the plausibility of a recent, close M31-M33 encounter for the first time. 

In this study we develop a self-consistent picture linking the observed morphological structure of M33 and the LMC with their numerically derived orbital histories and statistics from the {\em Illustris} cosmological simulation \citep{nelson15, vogelsberger14B, vogelsberger14A, genel14}. We will constrain the orbit of M33 using the latest astrometric data and place it in a cosmological context for the first time. We also compare the similarities and differences in the orbital properties of the two most massive satellite galaxies in the Local Group. Using orbits extracted for massive satellite analogs in the dark matter-only Illustris simulation, we will not only place their present-day kinematics in a cosmological context, but also their full orbital histories. Finally, we assess the impact of massive satellites on the structure of their hosts, on other satellites, and discuss implications of this picture regarding their own morphological evolution.

This paper is structured as follows. Section~\ref{sec:observed} highlights the observed properties of the LMC and M33 including their morphology, proper motions, and mass estimates. In Sections~\ref{sec:analyticmethods} and~\ref{sec:orbitanalysis}, we develop and analyse orbital histories for each host-satellite system based on astrometric data. Section~\ref{sec:illustris} describes the Illustris simulation and our sample selection methods for host-satellite analogs. Section~\ref{sec:orbitalprops} compares the average dynamical properties of the LMC and M33 to the cosmological sample of massive satellite analogs. Section~\ref{sec:discussion} assesses the viability of a close M31-M33 encounter in both a cosmological context and in light of the astrometric data. Finally, Section~\ref{sec:conclusions} contains our final remarks on the link between proper motions, analytic orbital models, and cosmological analogs for massive satellite galaxies and their hosts.

This is the first in a pair of affiliated papers. In Paper II (Patel et al. 2016b), the orbits of massive satellite analogs in cosmological simulations are used to constrain the halo mass of the MW and M31. While this has been previously done for the MW-LMC by \citet{bk11,busha11,gonzalez13}, we will constrain the mass of M31 for the first time in this fashion. 

\section{Observed Properties of the LMC and M33}
\label{sec:observed}

\subsection{HI Structure}
\label{subsec:HI}
The LMC and M33 are gas-rich satellites. They are both outliers with respect to the distance-morphology relation exhibited by Local Group satellites. Most satellite galaxies nearest their host galaxies typically contain the least amount of gas and little to no star formation while the farthest satellites tend to host larger gas reservoirs and have increased star formation activity \citep{vdb06}. Similar results are found for dwarf galaxies in the Local Volume \citep[ANGST survey,][]{weisz11}. 

At only 49.6 and 202.6 kpc, respectively, from their host galaxies, the LMC and M33 are amongst the satellite galaxies with the highest gas fractions given their separations. The high HI masses of these satellites suggest they may have followed similar orbital histories about their respective hosts. Both galaxies exhibit highly disturbed HI morphologies, which have been traditionally used to constrain their orbital interaction history in the absence of well-constrained 3D velocities.

Here, we discuss the detailed morphological structures of the LMC and M33, mainly focusing on the distribution of HI gas. We also provide an overview of the traditional orbital histories suggested due to these specific structural features.

\subsubsection{LMC}
The HI mass of the LMC is $\sim\!5 \times 10^{8}$ \Msun \citep{bruns04, bruns05, staveleysmith02}. In addition to hosting a large gas reservoir \citep{staveleysmith03}, the LMC and its companion, the Small Magellanic Cloud (SMC), are trailed by a larger gaseous system known as the Magellanic Stream (MS). The MS is a band of HI gas composed of filaments and clumps stretching more than $100^{\circ}$ across the sky \citep{mathewson77, putman98, putman03, bruns05, nidever08, nidever10}. Prior to the first set of proper motions for the Magellanic Clouds (MCs) reported by \cite{k06a,k06b}, the MS was the main feature used to constrain the orbits of the MCs \citep{murai80, gardiner96, connors06}.

In this paradigm, the MCs and the MS have a shared orbital history such that the orbits of the Clouds are directly related to the formation of the stream \citep{fich91}. The most common theories for MS formation mechanisms invoke multiple passages of the MCs about the MW. Careful orbital analysis with numerical and cosmological simulations using the MCs' proper motions has shown they are actually more likely to be on their first orbital passage \citep[e.g. ][hereafter B07, BK11]{b07, bk11}. The presence of the SMC lends support to this picture, as binary LMC-SMC configurations are unlikely to survive multiple passages about a massive host \citep[][BK11]{gomez13}. Tidal interactions between such tenuous binaries can instead facilitate the formation of gaseous streams like the MS, without aid from their hosts, as several authors have shown \citep[e.g.][]{diaz11, diaz12, besla12, guglielmo14}.

Galaxies like the MW are surrounded by gaseous haloes referred to as the circumgalactic medium (CGM) \citep[e.g.][]{werk14}. Recent simulations by \citet{salem15} illustrate that the gas disc of the LMC is affected by the MW's CGM. Its gas disc is truncated to a radius of 6 kpc in the direction of motion--this truncation depends sensitively on the density of the CGM and the pericentric approach of the LMC to the MW. If the LMC recently passed its first pericentre, the truncation is naturally explained as the asymmetry washes out over time \citep{salem15, besla16}. Simulations show the CGM effects are also maximized at pericentre, where the CGM density the LMC faces and its relative speed are highest. Therefore, the distribution of gas in massive satellites needs to be taken into account to develop a self-consistent picture of their orbital histories.

\subsubsection{M33}
While M33 is not trailed by a gaseous stream, it has a very extended gas disc with a total HI mass of $1.4 \times 10^9$ \Msun. The disc stretches nearly 22 kpc from its centre of mass and it shows evidence of a strong warp \citep[][hereafter P09]{putman09}. The Pan-Andromeda Archaeological Survey (PAndAS) of M31 and its environment also show that M33 contains a previously unknown warped stellar disc that extends about 30 kpc in projected distance across the sky from its centre in both the North and South \citep[][hereafter M09]{mcconnachie09}. These strongly disturbed features have been traditionally explained by a recent close approach of M33 about M31.

M09 reproduces the stellar distortions in M33 using N-body simulations where M33 reaches perigalacticon (relative to M31) at 53 kpc nearly 2.6 Gyr ago and an apogalacticon of 264 kpc just under a Gyr ago. This modeled apocentre and pericentre pair results in an eccentricity of about 0.67. It also implies M33 is currently receding from apocentre and is heading towards its second pericentric approach, consistent with the relative radial velocity of M33 with respect to M31. P09's models find that about 60 per cent of their orbits with a perigalacticon of $\lesssim$ 100 kpc in the last 3 Gyr would recover a tidal interaction between M31 and M33, where the tidal radius is $\leq$ 15 kpc. They claim this interaction could cause the observed distortions in its gas disc. Taking today's position as apogalacticon, this model implies a minimum eccentricity of about 0.34. These studies suggest that, if M33 did have a recent, close encounter with M31, its orbital trajectory is unlikely circular. Both the stellar and gas disc warps require a recent, close encounter and its lack of gas depletion suggests M33 was not accreted at early times. 

Aside from its warped morphology, the distribution of gas in M33's HI disc may hold clues to its interaction history with M31. Unlike the LMC, M33's immense gas disc does not show any significant signs of truncation. The marked lack of such truncation in the disc of M33 suggests either a much more diffuse CGM about M31 or a pericentre distance much larger than $\sim$50 kpc. The CGM of M31 is fairly similar to other $\rm L_*$ galaxies \citep[e.g.][]{lehner15}. On the other hand, a larger pericentre distance would be in contention with the models of M09 and P09. These scenarios can be disentangled with an accurate picture of M33's orbital history, further motivating this work.

\subsection{Proper Motions}
\label{subsec:propermotions}

\begin{table*}
\centering
\caption{The position and velocity components for the LMC and M33, each with respect to their host galaxy. Errors are computed in a Monte Carlo fashion (see K13) by sampling the proper motion error space of the LMC (K13), M33 (B05), and M31 (S12, vdM12), combined with uncertainties in their positions and the solar quantities. The Local Standard of Rest velocity at the solar circle $V_0 = 239\pm5$ \kms \citep{mcmillan11} is used, rather than the IAU value of $V_0=220$ \kms. The solar peculiar velocity is adopted from \protect\cite{schonrich10}, who find $\rm (U,V,W)_{\odot}=(11.1^{+0.69}_{-0.75}, 12.24^{+0.47}_{-0.47}, 7.25^{+0.37}_{-0.36})$ \kms. }
\label{table:vectors}
\begin{tabular}{p{0.8cm}p{1.6cm}p{1.7cm}p{1.7cm}p{1.5cm}p{1.3cm}p{1.3cm}p{1.3cm}p{1.3cm}p{1.3cm}p{1.3cm}}\hline\hline 
Galaxy & $x$ & $y$ & $z$ & r & $v_X$ & $v_Y$ & $v_Z$ & $v_{\rm tot}$ & $v_{\rm rad}$ & $v_{\rm tan}$ \\ 
            &  (kpc) & (kpc) & (kpc) & (kpc) & (\kms)   &  (\kms)   & (\kms)   & (\kms)              & (\kms)               & (\kms)      \\[2mm]  \hline
LMC & $-1.1\pm0.3$ & $-41.1\pm1.9$ & $-27.9\pm1.3$ & $49.6\pm2.3$ & $-57\pm13$ & $-226\pm15$ & $221\pm19$ & $321\pm24$ & $64\pm7$ & $314\pm24$ \\ \\
M33 & $-97.2\pm23.8$ & $-121.5\pm35.1$ & $-129.7\pm19.1$ & $202.6\pm46.5$ & $-24\pm34$ & $177\pm30$ & $94\pm39$ & $202\pm38$ & $-152\pm48$ & $133\pm37$ \\ \hline \\
\end{tabular}
\end{table*}

The proper motion measurements of the MCs \citep{k13} and M33 \citep[hereafter B05]{brunthaler05} provide a foundation to analytic studies of their orbital histories. Recently, space based observatories, such as HST and the upcoming {\em Gaia} satellite, have enabled the measurement of proper motions to an accuracy of microarcsecond per year, providing a precise, instantaneous picture of the 3D motions of Local Group galaxies. With constrained dynamics, we can now readily identify kinematic analogs to these in cosmological simulations, where there exists a statistically significant population of massive satellites. In the following, we review the latest proper motion measurements of the most massive members of our Local Group: the LMC, M33 and M31.

\subsubsection{LMC}
\label{subsubsec:lmcpms}
The proper motion of the LMC was most recently measured by K13. They used observations taken with HST of 22 fields in the LMC over a $\sim$7 year baseline to measure the motion of LMC stars with respect to background quasars. In this study, the proper motion of the LMC is transformed to a Galactocentric position and velocity using the methods described in \citet{vdm02}. The uncertainties on the mean values are determined by a Monte Carlo scheme that propagates all uncertainties in the position and velocity of the LMC and the Sun. This Monte Carlo technique yields a sample of 10,000 position and velocity vectors from which the mean Galactocentric velocity and errors are computed (see K13 and references therein). We will make use of these 10,000 Monte Carlo drawings in later sections. The resulting total position and velocity of the LMC from these drawings are $\rm R_{LMC} = 49.6\pm2.3$ kpc and $\rm v_{LMC} = 321\pm24$ \kms. The LMC's current radial velocity is $\rm v_{rad}=64\pm7$ \kms with respect to the MW. 3D position and velocity vectors are reported in Table~\ref{table:vectors}.

\subsubsection{M31}
\label{subsubsec:M31pms} 
Proper motions of M31 have been inferred indirectly from satellite kinematics where the line of sight velocities are used to fit for the transverse motion of M31 \citep[][hereafter vdMG08]{vdmG08}. The resulting tangential velocity component of M31 is $\rm v_{tan}=42\pm18$ \kms. Recently, \citet[][hereafter S12]{sohn12} directly measured the proper motion of M31 for the first time by tracking the motions of stars in three fields with respect to thousands of background galaxies. The observations were taken with HST over a 5-7 year baseline. 

\citet[][hereafter vdM12]{vdm12ii} corrected the S12 measurements for M31's internal kinematics and viewing perspective. In the end, they quote a weighted average for M31's proper motion using both HST direct measurements (S12) and M31's satellite kinematics (vdMG08). In this analysis, we use this weighted average where the 68.3 per cent confidence level is $\rm v_{tan}=17\pm17$ \kms. Again, the errors on the mean values for M31's position and velocity components are computed in a Monte Carlo fashion. 

It should be noted that more recent estimates of M31's tangential velocity vector in \citet[][hereafter S16]{salomon16} find different values than those reported by vdM12. S16 reports a mean tangential velocity of $\sim$150 \kms. Their method utilizes Markov Chain Monte Carlo techniques to statistically maximize the likelihood of $\rm v_{tan}$ by using only satellite kinematics. They then weight the likelihoods by $\Lambda$CDM halo velocity dispersion profiles to estimate their best-fitting parameters. We will discuss the implications of these conflicting values further in Section~\ref{sec:discussion}. 

\subsubsection{M33}
M33's proper motion was measured by B05 using the VLBA. The observations were taken over $\sim$3 years to measure the emission of water masers in two regions of M33 (IC133 and M33/19). By tracking emission features over 8 epochs, they compute a weighted average of their motions across the sky. They also propagate the errors in the velocity offset between specific maser features and the HI gas. These proper motions on the sky are transformed to Cartesian position and velocity coordinates in the MW reference frame through the same Monte Carlo methods as in K13. 

This scheme is also used to obtain 10,000 Monte Carlo drawings from the 4$\sigma$ proper motion error space of both M31 and M33. In addition to the proper motions and solar motion quantities, the Monte Carlo method incorporates distance errors into the analysis. Therefore, we adopt the M09 distance to M33 relative to the MW ($\sim$800 kpc). It should be noted that other authors \citep[e.g.][]{u09, bonanos06} have measured a significantly higher absolute distance to M33 ($\sim$960 kpc). The maximum distance probed by the Monte Carlo scheme is $\sim$880 kpc. We will discuss the impact of a larger M33 distance in Section~\ref{sec:discussion} where we explore if M33 could have reached within a close distance to M31. 

The two sets of 10,000 unique position and velocity vectors are combined to form the relative position and velocity vectors of the M31-M33 system (see Table~\ref{table:vectors}). The magnitude of the position and velocity of M33 are $\rm R_{M33} = 202.6\pm46.5$ kpc and $\rm v_{M33} = 202\pm38$ \kms. At present, M33 has a radial velocity of $-152\pm48$ \kms relative to M31. Hereafter, the subscript \emph{M33} refers to the position or velocity relative to M31 just as \emph{LMC} refers to kinematics relative to the MW.

\subsection{Mass Estimates of the LMC and M33}
\label{subsec:masslmcm33}
Orbit determination and the identification of satellite analogs in cosmological simulations require knowledge of the satellite mass. In the following, we provide an overview of known mass constraints on the LMC and M33.

\subsubsection{LMC}
 \vspace{0.75cm}
The LMC's rotation curve is well-defined, peaking at $\rm v_{circ}=91.7\pm18.8$ \kms and remains flat to about 8.7 kpc \citep{vdmnk14}. Basic dynamical mass arguments ($V^2=GM/r$) give an enclosed mass of $\rm M(8.7\; kpc) = 1.7 \times 10^{10}$ \Msun. 
 
The stellar mass of the LMC is $\rm M_{*}\!\sim\!2.7\times10^{9}$ \Msun with a neutral gas mass of $\rm M_{gas}\!\sim\!0.5 \times 10^9$ \Msun \citep{kim98}. This yields a total baryonic mass for the LMC of $\rm M_{bary}=3.2 \times 10^9$ \Msun. Several lines of evidence point to the radius of the LMC extending to at least 15 kpc \citep{majewski09, saha10, mackey16} at its outermost limits. At this radial extent, the total enclosed mass of the LMC is $\rm M(15\;kpc)\!\sim\!3\times 10^{10}$ \Msun, assuming the rotation curve of the disc remains flat to this distance. While this is the total dynamical mass measured today, the LMC may have been significantly more massive at its time of infall. G15 and \citet{besla15} propose its mass at infall is between $\rm M_{vir,inf}=6-20 \times 10^{10}$ \Msun.

Abundance matching techniques find similar values for the infall mass of the LMC. For example, \cite{guo11} finds $\rm M_{vir,LMC}= 1.6 \times 10^{11}$ \Msun. In our cosmological study, we focus on satellites with a maximal mass in the range encompassing a factor of two about the LMC dark matter halo mass inferred from abundance matching, $\rm 8\times 10^{10}$ \Msun $\rm < M_{max} < 3.2\times10^{11}$ \Msun (see also BK11). Our analytic models probe a wider range of masses, extending down to the dynamical mass estimate: $3-25 \times 10^{10}$ \Msun.

\subsubsection{M33}
The rotation curve of M33 is similarly well-defined using 21cm gas maps \citep{corbellisalucci}. Unlike the LMC, M33's rotation curve continues to rise out to its most distant data point. Using the peak of the rotation curve, $\rm v_{circ}(15 kpc)=130$ \kms, the dynamical mass of M33 is $\rm M_{dyn}(15\; kpc)\!\sim\!5\times 10^{10}$ \Msun. We adopt the combined stellar mass measured by \cite{corbelli03} and inferred by \cite{guo10}, averaging to $\rm M_{*} = 3.2 \times 10^{9}$ \Msun (vdM12). \cite{corbelli03} also measures a total gas mass of $\rm M_{gas}\!\sim\!3.2 \times 10^{9}$ \Msun. Therefore, the average baryonic mass is $\rm M_{bary}=6.4\times10^{9}$ \Msun. 

Using the dynamical mass estimate of M33, the baryon fraction is $\rm M_{bary}/M_{tot} =12.8$ per cent. This baryon fraction is a factor of a few more compared to average disc galaxies, suggesting that the total dark matter mass of M33 at infall was much larger than the dynamical mass inferred from the rotation curve. To get a more typical $\rm M_{bary}/M_{tot}=3-5$ per cent appropriate for disc galaxies, M33's mass at infall would have to be $\rm M_{vir,inf}=1.3-2.1 \times 10^{11}$ \Msun. This range is well within the 1$\sigma$ errors of abundance matching where $\rm M_{vir,M33}=1.7\pm0.55\times10^{11}$ \Msun \citep{guo11}. In our cosmological analysis, we examine satellites with maximal masses between $\rm 8\times10^{10}$ \Msun $\rm < M_{max} < 3.2\times10^{11}$ \Msun. These values encompass the full range of masses inferred from abundance matching for both the LMC and M33. In our analytic models, we adopt a similar mass range as for the LMC, except we account for M33's larger dynamical mass, giving $5-25 \times 10^{10}$ \Msun.

\section{Analytic Methods}
\label{sec:analyticmethods}
Here, we describe methods to take the observed range of LMC and M33 positions, velocities and masses listed in the previous sections and extrapolate orbital histories. These analytic models represent the orbits preferred by the astrometric data, independent of cosmological or morphological arguments. We follow the strategy outlined in G15 to track the orbital history of these satellites and the corresponding motions of their hosts. However, we implement a different scheme to account for dynamical friction (following Appendix A of \citet{vdm12iii}). We consider the MW-LMC system to be independent of the M31-M33 system. This choice is justified in Section~\ref{subsec:4body}, where we show that M33 has not closely approached the MW within the past 6 Gyr, nor the LMC to M31.

To compute past orbital histories, the equations of motion are numerically integrated backwards in time. We adopt two mass models for both the MW and M31. For the MW, we use a total virial mass{\footnote{Virial mass is the mass enclosed within the virial radius (\Rvir). \Rvir is the radius at which the average density within that radius reaches an overdensity of $\rm \Delta_{vir}$ in a spherical `top-hat' perturbation model. This $\rm \Delta_{vir}$ factor depends directly on the cosmological parameters. The Illustris cosmology yields $\rm \Delta_{vir} = 357$ (or $\rm \Delta_{vir}/\Omega_m = 97.4$). See \citet{brynorman98}.}} of $1\times 10^{12}$ \Msun and $1.5 \times 10^{12}$ \Msun. For M31, we use slightly higher mass models, which are $1.5\times 10^{12}$ \Msun and $2 \times 10^{12}$ \Msun. The MW and M31 potentials are constructed to include three components: a Navarro-Frank-White (NFW) dark matter halo \citep{nfw96}
\begin{equation}\rm \Phi_{halo} = -\frac{GM_{h}}{r[ln(1+c_{vir}) - c_{vir}/(1+c_{vir})]} ln\left(1+\frac{r}{r_s} \right), \label{eq:nfw} \end{equation}
a Miyamoto-Nagai disc \citep{mn75}
\begin{equation} \rm \Phi_{disc} = -\frac{GM_{d}}{\sqrt{r^2 +\left(R_d + \sqrt{z^2 + z_d^2}\right)^2}}, \end{equation}
and a Hernquist bulge \citep{hernquist90}
\begin{equation}\rm \Phi_{bulge}= -\frac{GM_{b}}{r+R_b}. \end{equation} For the NFW halo, $\rm M_h = M_{vir} - M_d - M_b$. 

\begin{figure*}
\centering
\includegraphics[scale=0.5]{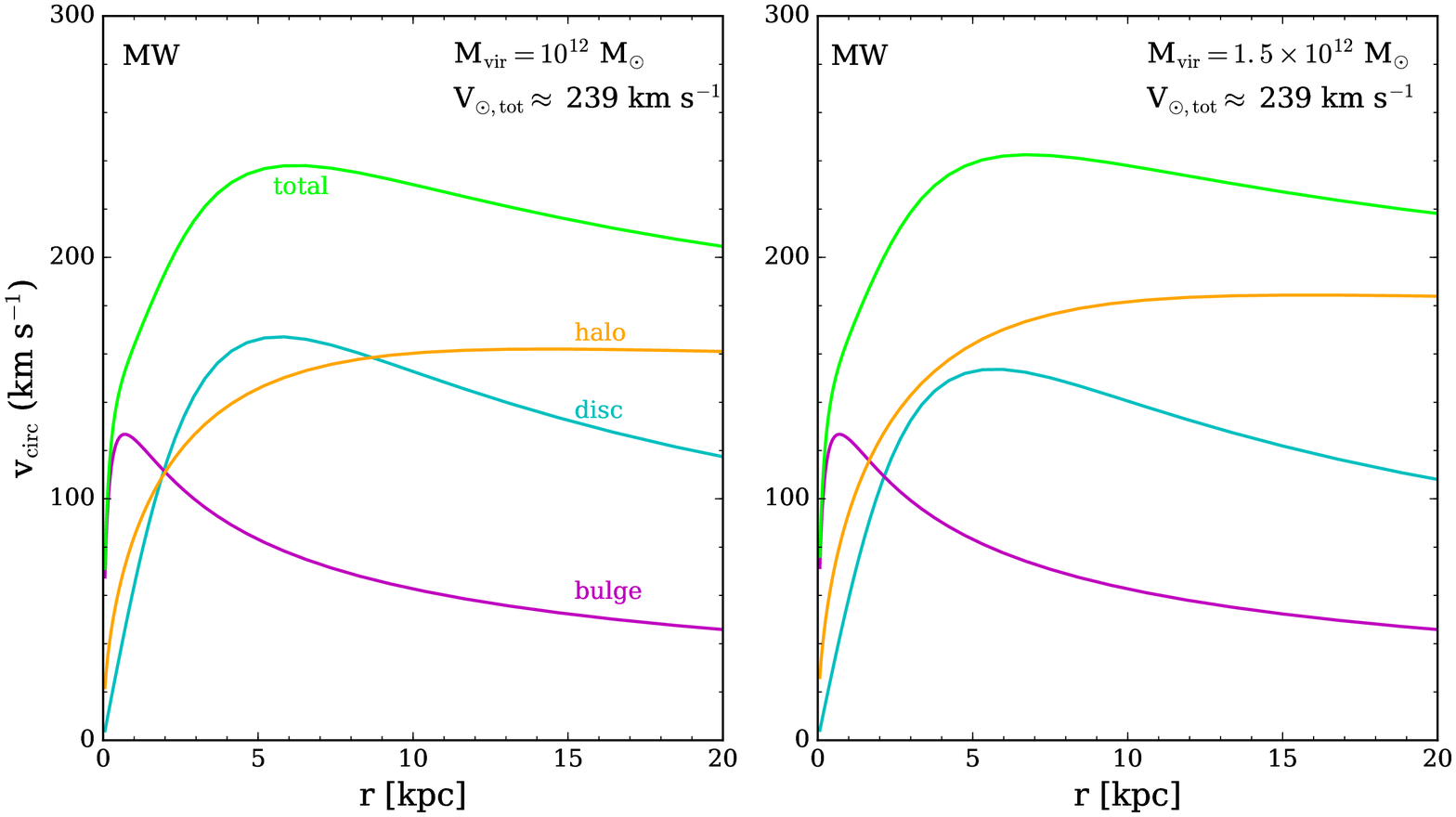}
\includegraphics[scale=0.5, trim=0mm 7mm 0mm 0mm]{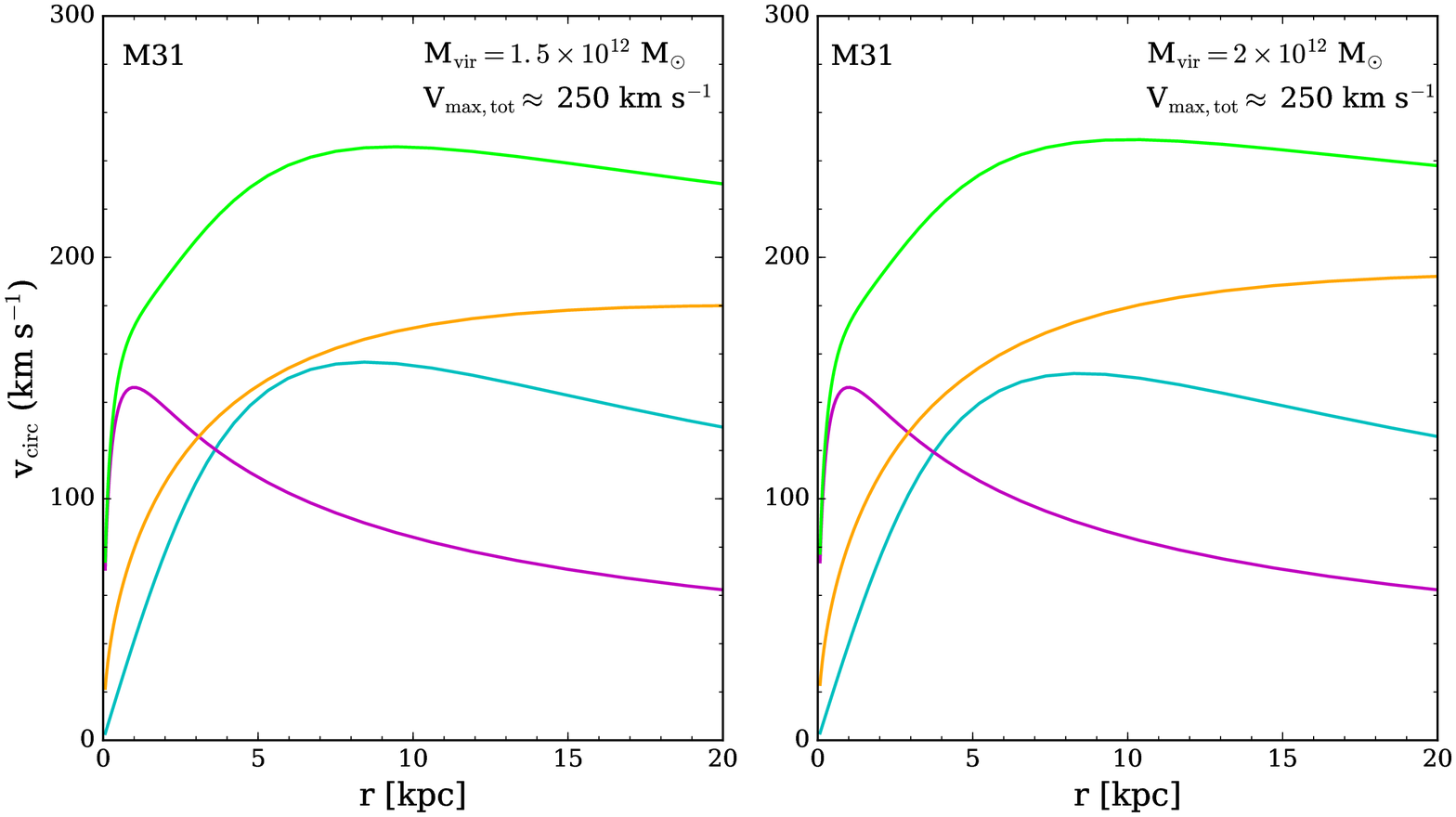}
\caption{Model rotation curves of the MW and M31 used to calculate the orbital histories in Fig.~\ref{fig:orbits}. {\em Top:} Two different virial mass models for the MW have been constructed to match the observed rotation curve: \Mvir =[$1\times10^{12}$, $1.5 \times 10^{12}$] \Msun. {\em Bottom:} Slightly higher virial mass models, \Mvir =[$1.5\times10^{12}$, $2 \times 10^{12}$] \Msun, are used for M31. In each model, the individual contributions from the disc (cyan), halo (orange), and bulge (purple) are indicated. The disc mass was chosen to approximately reproduce the observed maximum circular velocity for each galaxy: $V_c \approx 239$ \kms at the solar radius for the MW \citep{mcmillan11}, $V_c \approx 250$ \kms for M31 \citep{corbelli10}. All haloes have been adiabatically contracted due to the presence of the disc using the CONTRA code \citep{contra}.
\label{fig:RCs}}
\end{figure*}

The NFW dark matter halo of the host galaxy is adiabatically contracted due to the presence of the disc with the CONTRA code \citep{contra}. The dark matter density profile is truncated at the virial radius of the host galaxy in each model. 

The disc mass in each model is chosen to approximately reproduce the observed maximum circular velocity for each galaxy: $V_c\approx239$ \kms at the solar radius for the MW \citep{mcmillan11}, $V_c\approx250$ \kms for M31 \citep{corbelli10} at its peak. In all cases, the bulge scale length and mass remain fixed. The disc scale length and scale height are also held constant. All host galaxy parameters are listed in Table~\ref{table:orbitparams}. 

The rotation curves for each MW mass model are given in the top panels of Fig.~\ref{fig:RCs}. These show the circular velocity profile as a function of radius from the Galactic centre for each of the halo (orange), bulge (purple), and disc (cyan) components as well as the total circular velocity curve (green). All MW models have been constructed to match the velocity of the Local Standard of Rest at the solar circle. Rotation curves for M31 are constructed in the same fashion as the MW's. The M31 models peak at its maximum circular velocity. They are given in the bottom panels of Fig.~\ref{fig:RCs}. 

\begin{table}
\centering
\caption{Initial conditions for the mass distribution in host galaxies used in the orbit integrations. The disc masses have been chosen to match the observed maximum circular velocities (see also G15 and vdM12). See Fig.~\ref{fig:RCs} for more details.}
\label{table:orbitparams}
\begin{tabular}{lllll}\hline\hline 
& MWa & MWb & M31a & M31b \\ \hline
\Mvir [10$^{10}$ \Msun] & 100 & 150 & 150 & 200 \\ 
c$_{\rm vir}$ & 9.86 & 9.56 & 9.56 & 9.36 \\ 
\Rvir [kpc] & 261 & 299 & 299 & 329 \\
M$_{\rm d}$ [10$^{10}$ \Msun] & 6.5 & 5.5 & 8.5 & 8 \\ 
R$_{\rm d}$ [kpc]& 3.5 & 3.5 & 5.0 & 5.0 \\
z$_{\rm d}$ [kpc] &0.53 & 0.53 & 1.0 & 1.0\\ 
M$_{\rm b}$ [10$^{10}$ \Msun] & 1 & 1 & 1.9 & 1.9  \\ 
R$_{\rm b}$ [kpc] & 0.7 & 0.7 & 1.0 & 1.0 \\ \hline
\end{tabular}
\end{table}

We consider three different mass models for the LMC and M33 respectively. For the LMC, we consider infall masses of $(3, 10, 25)\times 10^{10}$\Msun. For M33, we use $(5, 10, 25)\times10^{10}$ \Msun (see Section~\ref{subsec:masslmcm33}). There will thus be six orbital models for each host-satellite pair.

The satellites are represented as Plummer spheres such that their gravitational potentials are:
\begin{equation} \rm
\Phi_{sat} = - \frac{GM_{sat}}{\sqrt{r^2 + k_{sat}^2}} 
\label{eq:plummer}
\end{equation}
with masses and softening lengths ($\rm k_{sat}$) as listed in Table~\ref{table:satparams}.
The Plummer softening lengths have been calculated to match the measured dynamical masses of the LMC and M33. See Section~\ref{subsec:masslmcm33}. 

Using a symplectic leapfrog integration method \citep{gadget}, we follow the gravitational interactions between the host galaxy and satellite by integrating the equations of motion backwards in time for 13 Gyr. In all models, we use the mean 3D position and velocity vectors of each satellite as listed in Table~\ref{table:vectors} to describe the motions of the LMC with respect to the MW and M33 with respect to M31 at the present time. We will search the full error space when quoting statistical measures of the orbital properties for each system. 

The orbit integrations also include the gravitational forces exerted on the host galaxies due to the massive satellites. G15 find that the acceleration of the MW due to the gravitational influence of the LMC is non-negligible. As a result, the orbital barycentre of the MW-LMC system is significantly displaced from the MW's centre of mass. We allow M31's centre of mass to move as a result of M33 and notice a comparable shift in the orbital barycentre of the M31-M33 system as well. The force exerted by the satellite on the host galaxy is therefore computed and updated at each time-step just like the force exerted on the satellite by the host. 

Our models also include the damping effects resulting from dynamical friction. If the orbits are integrated forward in time, the damping causes the orbit to decay. Since we integrate the orbits backwards in time, dynamical friction actually acts as an accelerating force. We approximate this acceleration by the Chandrasekhar dynamical friction formula \citep{chandrasekhar}:
 \begin{equation}\textbf{F}_{df}= \rm - \frac{4\pi G^2 M_{sat}^2 ln\Lambda \rho(r)}{v^2} \left[ erf(X) - \frac{2X}{\sqrt\pi} exp(-X^2)\right] \frac{\textbf{v}}{v}.\label{eq:df} \end{equation} 
Here, $X=v/\sqrt{2\sigma}$ where $\sigma$ is the one-dimensional galaxy velocity dispersion. We adopt the $\sigma$ approximation for an NFW profile derived in \citet{zentner03}. $\rho(r)$ is the density of the contracted NFW dark matter halo at a distance \textbf{r} from the centre of the host galaxy. 
For the Coulomb factor, ln$\rm\Lambda$, we implement the parametrization described in \citet[][Appendix A]{vdm12iii}:

\begin{equation}\rm ln \Lambda = max[L, ln(r/Ca_s)^{\alpha}]\label{eq:dfvdm} \end{equation} 

$L$, $C$, and $\alpha$ are constants. $a_s$ is the softening length of the satellite, or $\rm k_{sat}$, the Plummer softening length for our models (see Table~\ref{table:satparams}). \cite{hashimoto03} notes the importance of using ln$\rm\Lambda$ which varies with the distance of the secondary from the primary ($r$). \cite{vdm12iii} fits for ln$\rm\Lambda$ using N-body simulations where both the host and satellite are modeled as live, extended masses. They report the best-fitting parameters for a roughly equal mass orbit and a 10:1 host-satellite mass ratio tuned to match the future evolution of the M31-M33 system. We use the latter in our models for the MW-LMC and M31-M33 systems since both systems exhibit roughly this mass ratio at infall. The best-fitting results for unequal masses are L=0, C=1.22, $\alpha$=1.0. 

Note that this implementation of dynamical friction differs from that adopted in K13 and G15, as both studies implement the \cite{hashimoto03} ln$\rm \Lambda$ with a fixed softening length of 3 kpc for all satellite masses. If we keep the softening length fixed, we recover the same orbits for the LMC as K13 and G15. Finally, we ignore the dynamical friction effects on the MW (M31) due to the LMC (M33). With all of the relevant components, the total acceleration felt by the satellites is:
\begin{equation} \rm \ddot r_{sat} = \frac{d\Phi_{bulge}}{d\textbf{r}} + \frac{d\Phi_{disc}}{d\textbf{r}} + \frac{d\Phi_{halo}}{d\textbf{r}} + \frac{\textbf{F}_{df}}{M_{sat}}\end{equation}
and the total acceleration felt by the hosts is:
\begin{equation} \rm \ddot r_{host} =  \frac{d\Phi_{sat}}{d\textbf{r}}. \end{equation}
{\bf r} is always measured as the position vector between the host and the satellite, where both galaxies are free to move.

\begin{table}
\centering
\caption{Initial satellite parameters for the analytic orbit integrations of the LMC and M33. The satellites are modeled as Plummer spheres with the given mass and the softening lengths are calculated to match the measured dynamical masses. These are $\rm M_{LMC} (8.7~kpc) = 1.7 \times 10^{10}$ \Msun \protect\citep{vdmnk14} and $\rm M_{M33}(15~kpc) \gtrsim 5 \times 10^{10}$ \Msun \protect\citep{corbellisalucci}.}
\label{table:satparams}
\begin{tabular}{lccc}\hline\hline 
M$_{\rm LMC}$ [10$^{10}$ \Msun] & 3 & 10 & 25 \\ 
k$_{\rm LMC}$ [kpc]  & 5.9 & 13.1 & 19.5 \\ \hline 
M$_{\rm M33}$ [10$^{10}$ \Msun] & 5 & 10 & 25 \\        
k$_{\rm M33}$ [kpc] \hfill & 1 & 11.5 & 21 \\ \hline
\end{tabular}
\end{table}

\section{Analysis of Numerically Integrated Orbits}
\label{sec:orbitanalysis}

\begin{figure*}
\centering
\includegraphics[scale=0.43]{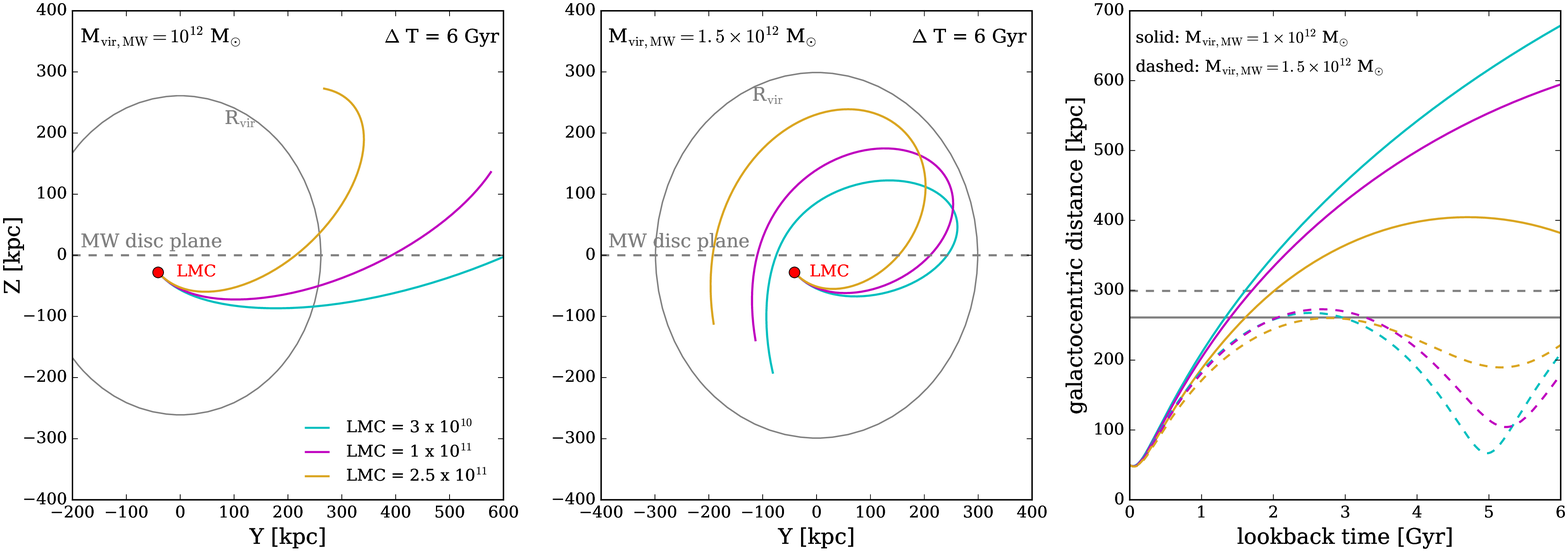}
\includegraphics[scale=0.43, trim=0mm 10mm 0mm 0mm]{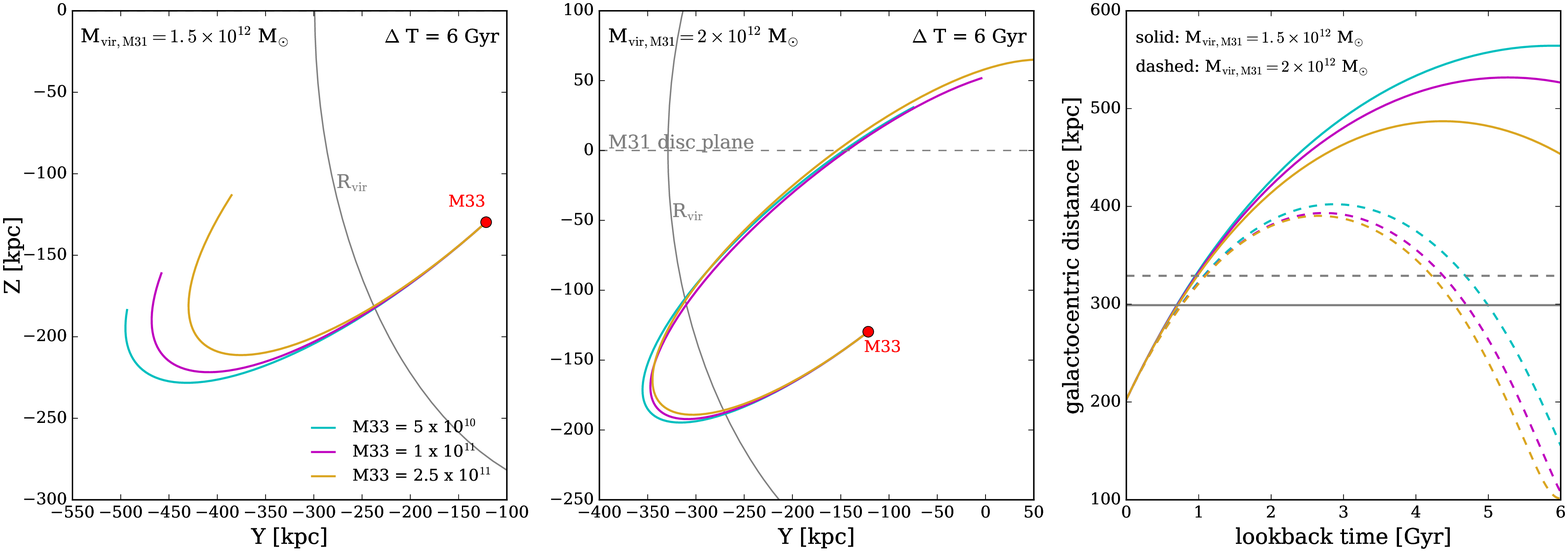}
\caption{Numerical orbit integrations for the MW-LMC and M31-M33 models presented in Fig.~\ref{fig:RCs} and Tables~\ref{table:orbitparams} and~\ref{table:satparams}. The Y-Z orbital cross sections are shown for two different virial mass models for the MW (top) and M31 (bottom), respectively. The orbits have been integrated backwards in time for 6 Gyr using the mean position and velocity vectors as observed today for the initial conditions. Three different masses are considered for each of the LMC and M33 in all models (magenta, cyan, and gold solid lines). The disc planes of the host galaxies are indicated by the gray dashed line in the first and second column. The gray solid curves indicate the extents of the virial radius in each model. The third column shows the orbits for each of the two host galaxy models in time versus galactocentric distance. The coloured solid lines indicate the lower host galaxy mass for the MW and M31, while the dashed coloured lines show the resulting orbits in the higher host galaxy mass models. The dashed and solid gray lines represent the virial radius in the high and low virial mass models of the host galaxies.
\label{fig:orbits}}
\end{figure*}

We consider the evolution of the LMC about the MW and M33 about M31 as two independent systems following the prescriptions outlined in the previous section. In the following, we analyse the resulting six orbital models for each host-satellite pair. For instance, in the case of the LMC, we determine if the analytic orbital models are in agreement with previous work. One main goal for M33 is to compare the resulting orbits to those presented by M09 and P09 to determine if a past interaction with M31 is a plausible cause of M33's warped structures.
\subsection{LMC}
In the top panel of Fig.~\ref{fig:orbits}, we present the orbital histories derived for the LMC about the MW over the past 6 Gyr, following the methods outlined in Section~\ref{sec:analyticmethods}. We plot only the last 6 Gyr of the integration period because we do not include a live halo potential or satellite mass loss prescriptions. These processes may significantly affect the resulting orbits at earlier times.

For the lower MW virial mass model, where \Mvir$=1\times10^{12}$ \Msun, all LMC masses are consistent with a first infall scenario. The LMC's orbit indicates an infall time\footnote{Infall time is defined as the first time the satellite crosses into the virial radius of its host. \label{infall}} of about 1.5-2 Gyr ago consistent with the results of B07, K13, and G15. One major difference between this work and G15 is that dynamical friction is reduced for the more massive satellites since we allow the Plummer softening length to vary with satellite mass. In G15, the softening lengths were fixed, therefore the lowest satellite mass experienced the least dynamical friction. Here, we see the opposite trend.

The higher MW mass model, when \Mvir$=1.5\times10^{12}$ \Msun, invokes at least one pericentric passage of the LMC about the MW with orbital periods of order 5 Gyr, regardless of LMC mass. For this MW mass model, the only difference between this analysis and previous work is the orbit of the highest LMC mass (gold dashed line). In G15, this mass combination does not result in a pericentric approach around 5 Gyr ago as seen here. Once again, this is a result of the modified dynamical friction term, and specifically the varying softening length. Note that a 5 Gyr orbital period is too long to explain the origin of the Magellanic Stream via MW tides, as the Stream is unlikely to have survived for greater than 2 Gyr \citep{blandhawthorn07}. 

The models of the MW-LMC orbits presented in Fig.~\ref{fig:orbits} only represent six specific orbital trajectories of the LMC as we have used the mean position and velocity vectors from Table~\ref{table:vectors}. However, we also have 10,000 unique position and velocity vectors sampling the proper motion, position, and velocity error space of the LMC (see Section~\ref{subsubsec:lmcpms}), which we use to compute statistical measures of the orbital properties allowed by the astrometric data. Holding the LMC's mass constant at $1\times10^{11}$ \Msun, we integrate the orbits backwards in time for 13 Gyr in both MW mass models. The first, second, and third quartile of important orbital properties are listed in Table~\ref{table:lmcstats}.

\begin{table}
\centering
\caption{The first, second, and third quartiles for infall time, pericentric distance, and time of pericentre for the 10,000 orbits calculated from the LMC velocity error space. All orbits are integrated backwards for 13 Gyr. The listed times are lookback times. For both MW masses, the LMC's mass remains fixed at $1\times10^{11}$ \Msun.}
\label{table:lmcstats}
\begin{tabular}{lccc}\hline\hline 
MW \Mvir &  $\rm t_{inf}$ [Gyr] & $\rm r_{peri}$ [kpc] &  $\rm t_{peri}$ [Gyr] \\ 
& $q_1$, $q_2$, $q_3$ & $q_1$, $q_2$, $q_3$ & $q_1$, $q_2$, $q_3$ \\ \hline
$1\times 10^{12}$ \Msun & 1.2, 1.4, 8.1 & 46.3, 48.0, 49.7  & 0.04, 0.05, 0.05  \\ 
$1.5\times 10^{12}$ \Msun  & 5.9, 6.9, 8.7 & 46.1, 47.8, 49.6  & 0.05, 0.05, 0.06 \\ \hline 
\end{tabular}
\end{table}

For all orbits in the LMC velocity error space, at least one pericentric approach is inevitable within a Hubble time \citep{ruzicka09}. Regardless of infall time, the orbits of the LMC are found to be consistent with recent studies--its first pericentric passage certainly occurs about $\sim$ 48 kpc from the MW in the last 50 Myr. 

In agreement with G15, in the last 2 Gyr the MW's centre of mass gains a velocity of $\sim$10-100 \kms (depending on the LMC mass) owing to the shift in the orbital barycentre of the MW-LMC system. This has been shown by G15 to cause noticeable shifts in the orbital planes of satellites like the Sagittarius dSph. We will explore similar effects for the M31-M33 system in the following section. 

\subsection{M33}
Plausible orbital histories for M33 about M31 are presented in the bottom row of Fig.~\ref{fig:orbits}, using the mean relative velocity quoted in Table~\ref{table:vectors}. In the low M31 mass model (\Mvir$=1.5\times10^{12}$ \Msun), all of the M33 orbits are on first infall. These results are consistent with the orbital solutions found in \citet{shaya13}, who suggest that M33 is currently at its closest approach to M31.  Furthermore, the star formation history in the disc of M33 shows no evidence of elevated star formation in the last 2-4 Gyr \citep{williams09}. Instead it appears to have grown steadily, contrasting the expectation if M33 was actually recovering from a recent interaction with M31.

M33's morphology may not naturally favor an orbital scenario other than a recent M31-M33 encounter. However, its extended stellar structures are still understandable without a close passage if M33 has interacted with other M31 satellites or if it has hosted satellites of its own in the past. We also note that \citet{lewis13} have pointed out that a spatial offset exists in the HI disk and stellar structures of M33. This offset could be attributed to accretion events or even mild ram pressure stripping in the halo outskirts ($\sim$100 kpc). Moreover, given that M33's total mass is $\sim10^{11}$ \Msun, it could have hosted several less massive satellite galaxies \citep[see][]{sales13}, and the accretion of these satellites could have formed the stellar halo of M33 \citep{mcmonigal16} through traditional hierarchical evolution. Furthermore, \citet{berentzen03} and \citet{besla12} have shown that off-center collisions with less massive satellites can perturb the stellar disk of host galaxies. One example of such phenomena is the multi-armed extended spiral structure detected in the outskirts of the LMC, a feature that \citet{besla16} suggests was induced by the SMC rather than MW tides.

For the high M31 mass model (\Mvir=$2\times10^{12}$ \Msun), there is a chance of a pericentric passage at about 6 Gyr ago. However, this orbital period is almost twice that suggested by the P09 and M09 models, which were designed to match the observed morphological structure of M33. Both sets of orbits in Fig.~\ref{fig:orbits} support the future M31-M33 orbits calculated in \citet{vdm12iii}, which find M33's next pericentric passage about M31 will occur in the next 1-2 Gyr. 

For the explored M33 mass range, we have also numerically integrated its orbit using the mean velocity of M31 resulting from the HST proper motion only (S12). By contrast, all previous orbits used the weighted average of the HST measurements and satellite kinematics (vdM12). We find that independent of M31 mass, all six orbits result in a first infall scenario and shows no signs of a pericentric passage < 12 Gyr ago. 

The large M31 tangential velocity inferred by S16, however, has different implications for the orbit of M33. Using the S16 M31 velocity vector, in the low mass M31 model, we find that one pericentric passage between 2-3 Gyr ago is inevitable for all M33 masses explored here. However, the distance at pericentre is still large, reaching $\sim$175 kpc at best. In the high M31 mass model, all orbits evidence one or more pericentric passages. The most recent pericentre occurs between 2-3 Gyr ago and again, the separations are no less than $\sim$140 kpc.

Fig.~\ref{fig:orbits} relied on the average M33 velocity but we can also explore the full velocity error space to quantify the most typical orbital histories. Calculating 10,000 orbits spanning the M33 and M31 velocity error space with a fixed M33 mass of $1 \times 10^{11}$ \Msun results in the median orbital properties listed in Table~\ref{table:m33quartilestats}. A first infall scenario is favored for the low M31 mass model, as proposed in Fig.~\ref{fig:orbits}. At this M31 mass, only about 48 per cent of all orbits contain a pericentric passage in an orbital period of 13 Gyr, but the distance of closest approach is more than twice that suggested by M09 and P09\footnote{We have also relaxed our assumption of halo truncation and turned off dynamical friction at the virial radius, but the resulting statistics only improve mildly for a recent, close passage scenario, independent of M31 mass.}. The median pericentric distances listed ($\sim$105 kpc) are consistent with the findings of \citet{vdm12iii}. The orbits with a pericentric passage unanimously have an infall time > 5 Gyr ago. In each case, the infall time occurs earlier than the time of pericentre. 

If M31 is massive ($2 \times 10^{12}$ \Msun), M33 follows a long period orbit of order 6 Gyr. In this scenario, it is unlikely that M33 made more than one pericentric passage about M31 in a Hubble time. Based on the values in Table~\ref{table:m33quartilestats}, this host-satellite mass combination prefers early accretion, or $\rm t_{inf}\gtrsim$5 Gyr ago. At this M31 mass, 77 per cent of orbits reflect a pericentric passage also $\gtrsim$ 5 Gyr ago. Similar to the low mass M31 model, the average pericentric distances are generally much higher than what is required to justify M33's warped structures by a close interaction with M31. 

Therefore, the low mass M31 model typically favors a first infall scenario for M33, while a higher mass M31 suggests M33 made a distant pericentric approach about M31 about 5-6 Gyr ago. Consequently, a large fraction of the recent infall scenarios do not allow for a recent close encounter (within 100 kpc of M31) as suggested by M09 and P09. These results support the lack of a truncated gas disc in M33, which might be an artefact of a large distance at pericentre (> 100 kpc), consistent with our analysis. 

The proper motions of the M31-M33 system are thus in direct conflict with the conventional orbital history of M33 adopted in the literature. We examine other host-satellite mass combinations in further detail in Section~\ref{sec:discussion} to reconcile a recent pericentric passage of M33 about M31. 

Following G15, we inspect the shift in M31's velocity due to the presence of M33. We find that M33 increases M31's velocity by $\sim$5-25 \kms for our explored mass range. Since M33's closest approach to M31 is much larger than that of the LMC to the MW, the magnitude of M31's velocity shift is lower. However, this shift is not negligible and may also result in observable signatures in the relative kinematics of M31 and its other satellites. 

\begin{table}
\centering
\caption{Column 2 lists the first, second, and third quartiles in infall time for the 10,000 orbits spanning the M31-M33 velocity error space. Columns 3 and 4 give the quartiles in pericentric distance and time of pericentre for the fraction of orbits (48 per cent and 77 per cent, respectively, for low and high M31 mass) where M33 reaches distances closer than its current separation to M31. All orbits are integrated backwards for 13 Gyr. The listed times are lookback times. For both M31 masses, M33's mass remains fixed at $1\times10^{11}$ \Msun.}
\label{table:m33quartilestats}
\begin{tabular}{lccc}\hline\hline 
M31 \Mvir &  $\rm t_{inf}$ [Gyr] & $\rm r_{peri}$ [kpc] & $\rm t_{peri}$ [Gyr] \\
& $q_1$, $q_2$, $q_3$ & $q_1$, $q_2$, $q_3$ & $q_1$, $q_2$, $q_3$ \\ \hline
$1.5\times 10^{12}$ \Msun &  0.3, 0.4, 7.9 & 75.8, 104.9, 140.3  &  5.6, 7.4, 9.6\\ 
$2\times 10^{12}$ \Msun  & 4.3, 6.3, 8.4 & 74.3, 104.4, 137.0 & 4.6, 5.9, 8.0 \\ \hline 
\end{tabular}
\end{table}

\subsection{The Four Body Orbit}
\label{subsec:4body}

\begin{figure*}
\centering
\includegraphics[scale=0.48, trim=5mm 10mm 0mm 0mm]{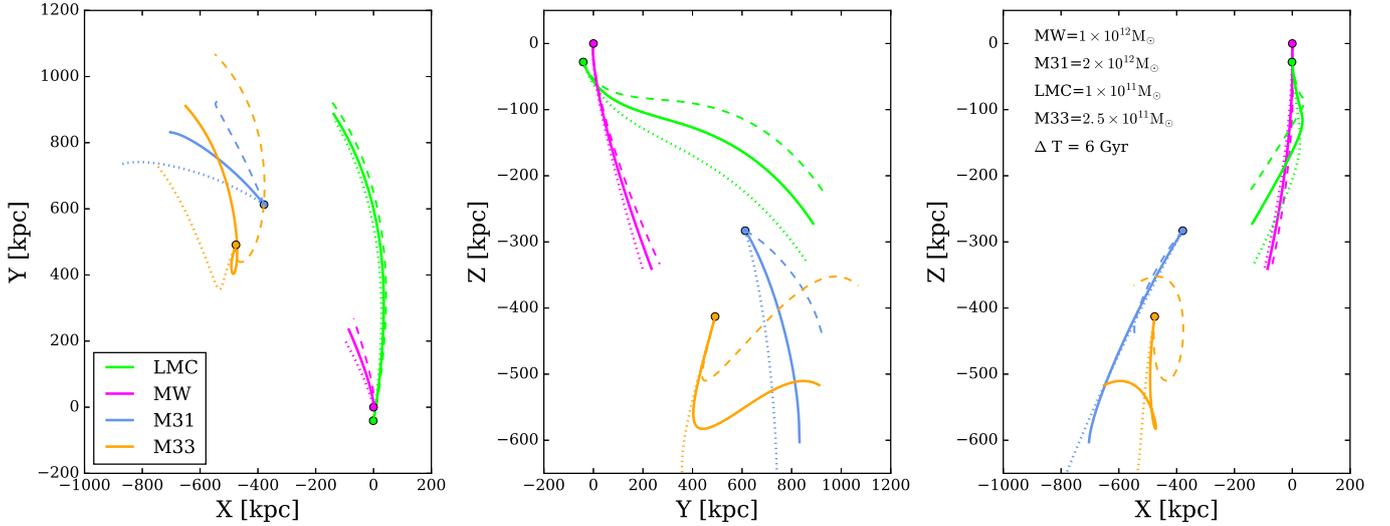}
\caption{The numerically integrated orbital trajectories of the MW, LMC, M31, and M33 for the last 6 Gyr. All positions are plotted with respect to the MW's position as a function of time. Each galaxy exerts a gravitational force on all other galaxies. Dynamical friction is implemented in two regimes--for the host-satellite interactions and the host-host interactions. The solid lines are the resulting orbits computed at the mean velocity vectors of all galaxies. The dashed lines indicate the resulting orbital trajectory for a -1$\sigma$ deviation in all four velocity vectors while the dotted lines are the +1$\sigma$ deviation. The deviations represent the errors in the observed proper motion, position, and velocity error space for each galaxy.
\label{fig:4body}}
\end{figure*}

\begin{figure}
\centering
\includegraphics[scale=0.45, trim=0mm 10mm 0mm 0mm]{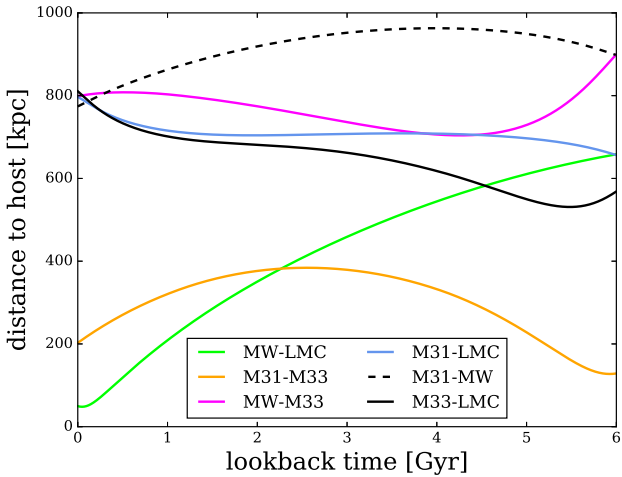}
\caption{The orbital trajectories as a function of time for each pair of galaxies calculated at their mean velocities. The positions are plotted such that the first galaxy in the pair is the {\em host} and therefore the position of the corresponding {\em satellite} is plotted relative to that galaxy. Aside from the MW-LMC and M31-M33 pairs, no galaxies move closer than about 500 kpc from each other. The LMC is negligibly affected by M31 and M33 experiences little to no perturbations as a result of the MW.
\label{fig:2body}}
\end{figure}

Throughout this work, we treat the MW-LMC system and the M31-M33 as two, separate, isolated systems. Here, we validate this choice by numerically integrating the orbits of all four galaxies backwards in time simultaneously. Doing so ensures that the LMC has not been affected by the gravitational influence of M31, nor M33 by the MW during the past 6 Gyr. As \cite{vdm12ii, vdm12iii} have shown, the MW and M31 are currently moving towards one another and will likely experience their first pericentric passage in about 4 Gyr. Therefore, in the past, the MW and M31 are moving away from one another. 

Following the methodology of Section~\ref{sec:analyticmethods}, we numerically integrate each galaxy's equations of motion backwards in time as they simultaneously experience the gravitational influence of each other. The satellites (LMC, M33) exert forces on both hosts (M31, MW) and dynamical friction is implemented for the satellites as described by Equations~\ref{eq:df} and~\ref{eq:dfvdm}. The masses of the LMC and M33 are held constant at $1\times10^{11}$ \Msun and $2.5\times10^{11}$ \Msun, respectively. 

To approximate the gravitational interactions between the MW and M31, we use the dynamical friction tuning for a 1:1 mass ratio as given in Appendix A of \citet{vdm12iii}. The softening lengths for the MW and M31 are chosen to match their observed masses within some radius compared to the total virial masses used in this model. The MW's virial mass is fixed at $1\times10^{12}$ \Msun and M31's is $2\times 10^{12}$ \Msun. For both galaxies, a Hernquist profile \citep{hernquist90} is used to determine the softening length. The MW's mass enclosed within 50 kpc is about $3.8\times10^{11}$ \Msun (B07 and references therein). M31's observationally constrained mass within 300 kpc is about $1.4\times10^{12}$ \Msun \citep{watkins10}. Therefore, the resulting softening lengths are approximately $\rm k_{MW}=31.11$ kpc and $\rm k_{M31}=58.57$ kpc. 

In Fig.~\ref{fig:4body}, we present the orbital planes of the four-body interaction between the MW, LMC, M31, and M33 during the last 6 Gyr. All positions are plotted relative to the MW. The circular markers denote the position of each galaxy today. The dashed lines indicate the orbital trajectories resulting from a -$1\sigma$ deviation in the velocity vector of each galaxy and the dotted lines are due to a +$1\sigma$ deviation in each velocity vector. 

Fig.~\ref{fig:2body} shows the orbital trajectory as a function of lookback time for each pair of galaxies amongst the four considered here. Notice that aside from the MW-LMC and M31-M33 pairs, the closest separation between any two galaxies is no more than $\sim$500 kpc. In particular, the LMC does not reach less than $\sim$650 kpc from M31 and M33 reaches no closer than $\sim$700 kpc from the MW. While the LMC's orbit may be perturbed slightly by M31 > 6 Gyr ago, there is no strong evidence that this perturbation exists at more recent times. These results are consistent with \citet{k09} who find that M31 does not significantly affect the LMC's orbit within the proper motion error space, unless M31 is sufficiently massive (i.e. > $3.5\times 10^{12}$). Thus, treating the MW-LMC and M31-M33 as isolated systems in the past is a reasonable simplification.

\section{Analogs in the Illustris Simulation}
\label{sec:illustris}
Using the observed and inferred infall masses of the LMC and M33, we identify analogous host-satellite pairs in the {\em Illustris Project} \citep{nelson15, vogelsberger14B,vogelsberger14A, genel14}, an N-body and hydrodynamic simulation spanning a cosmological volume of (106.5 Mpc)$^3$, carried out with the moving-mesh code \texttt{AREPO} \citep{springel10}. In this analysis, we use the highest resolution dark matter-only run, {\em Illustris-1-Dark} (hereafter Illustris-Dark), which follows the evolution of $1820^3$ dark matter particles from $z=127$ to $z=0$, stored in a series of 136 snapshots. Illustris-Dark adopts the following cosmological parameters, which are consistent with \emph{WMAP-9} measurements \citep{hinshaw13}: $\Omega_m =0.2726$, $\Omega_{\Lambda} = 0.7274$, $\Omega_b =0.0456$, $\sigma_8 = 0.809$, $n_s = 0.963$ and $h = 0.704$. 

Dark matter haloes and their bound substructures (i.e. subhaloes) are identified in each of the 136 snapshots of the simulation using the \texttt{SUBFIND} halo-finding routine \citep{springel01, dolag09}, which proceeds in the following way. First, dark matter haloes are identified with the friends-of-friends (FoF) algorithm \citep{davis85}, which links together any two particles separated by less than 0.2 times the mean interparticle separation. Then, for each of these haloes (also known as FoF groups), subhaloes are identified as overdense regions which are also determined to be gravitationally bound. Usually, each halo hosts a massive, central subhalo which contains most of the loosely bound material in the halo. 

In order to follow such haloes and subhaloes across time, we use the merger trees\footnote{The Illustris merger trees, halo catalogs, and group catalogs are all publicly available at www.illustris-project.org} created with the recently developed SUBLINK code \citep{rg15}. The merger trees allow us to trace the histories of massive satellite analogs and therefore identify their properties at various epochs in cosmic history. 
 
The large volume and high resolution in the Illustris-Dark simulation provide an ideal data set for studying the dynamics and properties of massive satellite galaxies in MW/M31-mass systems. The simulation achieves a dark matter particle mass resolution of $\rm m_{DM} = 7.5 \times 10^{6}~M_{\odot}$. Therefore, a MW/M31-mass halo in Illustris-Dark is composed of up to a few times $10^5$ dark matter particles and an LMC/M33-mass analog consists of up to $\sim4\times 10^4$ dark matter particles. 

In the following, we identify a sample of MW/M31 mass analogs at $z=0$. Within this set of haloes, we search for those that host a massive subhalo analogous to the LMC or M33. With this population of massive satellite analogs and their hosts, we will examine the orbital dynamics of the analogs compared to the observed present-day dynamics of the real LMC and M33.

\subsection{Sample Selection: Milky Way/M31 Analogs}
\label{subsec:hostanalogs}
Estimates for the virial mass of the MW's halo range from $\approx$ [0.7-2.5] $\times 10^{12}$ \Msun \citep[e.g. ][]{belokurov14, gibbons14, bk13, sakamoto03, battaglia05, dehnen06, liwhite08, gnedin10, watkins10}. M31's halo mass has been estimated via methods like abundance matching, satellite orbital dynamics, the timing argument, and cosmological simulations. Its halo mass is inferred to be as massive as the MW's or larger, especially by timing argument studies. A plausible virial mass range for M31 from the literature is $\approx$ [1-3] $\times 10 ^{12}$ \Msun \citep[e.g. ][]{evans00, watkins10, tollerud12, guo10, fardal13}. The M31 rotation curve peaks at a velocity greater than that of the MW and both its bulge and disc mass are also greater than the MW's \citep[e.g. ][and references therein]{guo10, wang06, loeb05, peebles96, k09, vdm12ii}. Therefore, we allow for a wide dark matter halo mass range to encompass a broad distribution from the literature. 

\begin{figure}
\begin{center}
\includegraphics[scale=0.6,  trim= 0mm 2mm 0mm 5mm, clip=true]{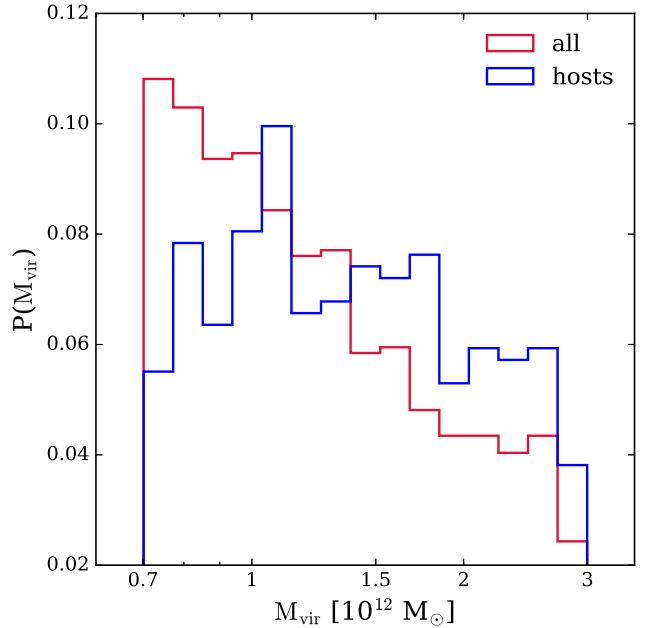} 
\caption{The distribution of virial mass ($\sim$ halo mass) for all MW/M31 mass analogs in the Illustris-Dark simulation at $z=0$ (red). The distribution of virial mass for the MW/M31 mass analogs which also host a massive satellite analog at $z=0$ (host haloes, blue). This subset comprises 24.4 per cent of the full MW/M31 analogs sample. The bins are normalized to the size of each data set such that $\rm \sum_{i=1}^{N_{bins}} M_{vir, i} =1$.}
\label{fig:mvir}
\end{center}
\end{figure}

MW/M31 analogs in Illustris-Dark are chosen as all central subhaloes, as determined by \texttt{SUBFIND}, at $z=0$ whose halo (or FoF group) virial masses are between $7 \times 10^{11}$ \Msun $<$ $\rm M_{vir}$ $<$ $3\times10^{12}$ \Msun. We have checked that the virial mass of each MW/M31 analog's halo is comparable to the central subhalo mass as given by \texttt{SUBFIND}. Since these quantities have nearly a one to one ratio, we take the virial mass of the halo (FoF group) as the halo mass of all hosts throughout this analysis. Our sample contains 1933 dark matter haloes that satisfy these criteria. The distribution of virial mass for these MW/M31 analogs is indicated by the red histogram in Fig.~\ref{fig:mvir}. There are many more low mass haloes, as one expects, due to the hierarchical evolution of cold dark matter haloes. 

\subsection{Sample Selection: Hosts of Massive Satellites}
\label{subsec:hosts}
The subset of MW/M31 analogs which host a massive satellite analog will be referred to as the host halo sample. About 24.4 per cent of the MW/M31 mass analogs host a massive subhalo. Throughout this analysis, {\em host halo} will exclusively refer to the dark matter halo of a MW/M31 mass analog which hosts a massive subhalo analogous to the LMC or M33. The blue histogram in Fig.~\ref{fig:mvir} shows the probability of finding a host halo with a given mass from the full sample of MW/M31 analogs. The peak of the distribution lies at $\sim$ $10^{12}$ \Msun and only a few percent of host haloes reach a mass close to the lower and upper limits: $0.7 \times 10^{12}$ \Msun, $3 \times 10^{12}$ \Msun. Thus, we allow for a broad range in the host halo mass. In practice, very few haloes at the extrema of this range host massive satellite analogs.

\Rvir and \Vvir, the virial circular velocity, can be computed from \Mvir as follows

\begin{equation} \rm R_{vir} = 260 \left(\frac{M_{vir}}{10^{12} ~M_{\odot}}\right)^{1/3}~kpc,\end{equation} 
\begin{equation} \rm V_{vir} = 128.6 \left(\frac{M_{vir}}{10^{12} ~M_{\odot}}\right)^{1/3}~ km~ s^{-1}. \end{equation}

These virial quantities will be used to calculate satellite orbital dynamics (i.e. orbital energy and angular momentum) throughout this work.

\subsection{Sample Selection: Massive Satellites}
\label{subsec:satanalogs}
We follow the basic methods of BK11 to identify LMC/M33 analogs. BK11 used the Millennium-II \citep{bk09} $\rm \Lambda$CDM cosmological simulation to study the dynamics of LMC and SMC analogs. However, in that study, the dynamics of the MCs were computed using the \citet{k06a,k06b} proper motions, which have since been revised in K13. Here we extend their analysis to M33 and also update results for the LMC using the K13 proper motions. 

For each MW/M31 analog identified in Section~\ref{subsec:hostanalogs}, we search through all subsequent satellite subhaloes in the same FoF group to identify the subset of massive satellite analogs and their corresponding host haloes. Below, we define the two samples used to accomplish this.

\begin{itemize}[leftmargin=*]
\item[](i.) Preliminary Massive Satellite Analogs:
\newline the most massive subhalo, identified by the maximal mass (\Mmax) ever attained, residing within \Rvir of a MW/M31 analog's centre at $z=0$.
\item[] 
\item[](ii.) Massive Satellite Analogs:
\newline the subset of preliminary analogs with $\rm 8\times10^{10}$ \Msun $\rm < M_{max} < 3.2\times10^{11}$ \Msun. The corresponding MW/M31 analog is then classified as a host halo (Section~\ref{subsec:hosts}).
\end{itemize}

By this construction, each host halo is limited to one massive satellite analog. We use \Mmax to relate dark matter subhaloes in simulations like Illustris-Dark to the observed galaxy properties because abundance matching techniques typically correlate stellar mass to the maximal mass (in the form of $\rm M_{200crit}$\footnote{The mass contained within $\rm R_{200}$, the radius at which the average overdensity of the universe is 200 times the critical density} or the FoF group mass) of haloes. If we use the $z=0$ satellite subhalo mass to identify analogs, the mass loss due to tidal stripping after accretion would have to be accounted for, requiring the implementation of mass loss prescriptions for interacting galaxies. Choosing satellite subhaloes based on \Mmax does not necessarily mean that they are the most massive satellite subhaloes in their FoF group at $z=0$ but they have at least achieved the mass of a {\em massive satellite} (i.e. 10 per cent of the host halo mass) at some point in their history. 

We impose a mass floor of $10^{10}$ \Msun ($\sim$1300 dark matter particles) at $z=0$ for a satellite subhalo to be considered an analog of the LMC and M33. This value comes from the dynamical mass estimates of the LMC and M33 (see Section~\ref{subsec:masslmcm33}). While identifying preliminary LMC/M33 analogs, we also correct relative positions for the box edges to make sure that no subhaloes are dismissed due to the finite box volume and simulation boundary conditions.

Requiring MW/M31 analogs to be the central subhalo in a FoF group and identifying LMC/M33 analogs in this fashion ensures that there are no massive companion galaxies in each group (i.e. no Local Group analogs). This choice is justified by the study of \citet{gonzalez13} which concludes that the environment of Magellanic Cloud analogs, whether they are hosted by a MW mass analog or within an analog of the Local Group, does not strongly affect estimates of the MW's halo mass. However, the frequency of the latter is much lower cosmologically. This choice thus allows us to increase our orbital statistics.

Our host+massive satellite analogs sample consists of 472 systems. We therefore find that 24.4 per cent of MW/M31 mass haloes harbor a massive satellite analog. BK11 finds about 35 per cent of their MW sample hosts an LMC analog in the Millennium-II simulation, however their host halo mass range has a lower limit of $10^{12}$ \Msun. If we apply this lower halo mass limit to our MW/M31 host sample, about 33 per cent of them host an LMC/M33 analog, in good agreement with BK11. Observational studies of the {\em Sloan Digital Sky Survey} also show that about 40 per cent of L$_{*}$ galaxies host a bright satellite within 250 projected kpc \citep{tollerud11}. 

\begin{figure}
\begin{center}
\includegraphics[scale=0.6, trim=0mm 10mm 0mm 0mm]{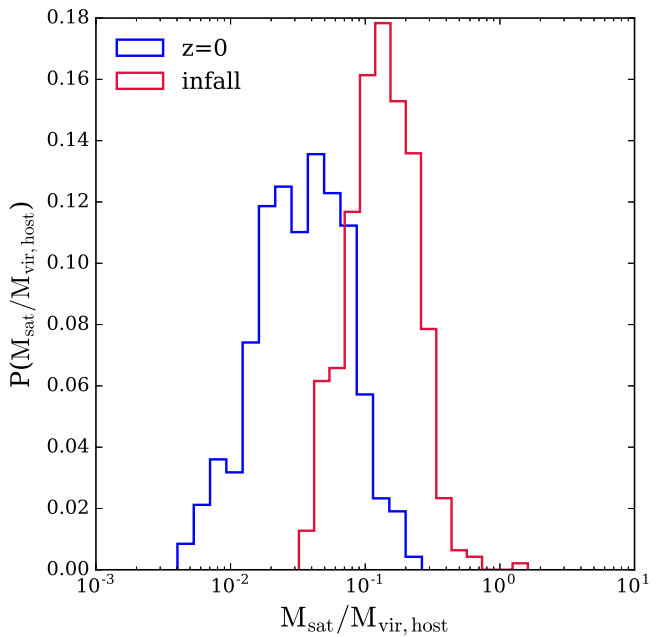}
\caption{The distribution of satellite to host dark matter mass ratios at infall (red) and at $z=0$ (blue) in the Illustris-Dark simulation. At infall\footnoteref{infall}, the mass ratio peaks at $\sim10^{-1}$. At $z=0$, the host-satellite mass ratios are no lower than $\sim10^{-2}$.}
\label{fig:lmcmassratio}
\end{center}
\end{figure}

Fig.~\ref{fig:lmcmassratio} indicates the distribution of host to satellite mass ratio at $z=0$ (blue) and at the time where the satellites reach maximal mass (red). The peak of the distribution is about $10^{-1}$ at the time of maximal mass, indicating that analogs of the LMC or M33 are a significant fraction of their host's mass at that epoch. This is consistent with the work of \cite{stewart08} who suggest MW mass dark matter haloes are built up by 1:10 mergers. Even at $z=0$, the massive satellite analogs are no less than $10^{-2}$ of their host halo mass. The offset of the two distributions suggests that the sample of host+massive satellite analogs evolve quite noticeably from one epoch to the next, a property that cannot be captured in the analytic orbital models.

\section{Orbital Analysis of LMC and M33 Analogs in Illustris}
\label{sec:orbitalprops}

With a sample of several hundred massive satellite analogs and their respective hosts, we identify the average trends in their orbital histories. The mean positions and velocities of the LMC and M33 obtained from their proper motion measurements are used to infer the present-day dynamics of the satellites as a point of reference. By comparing these dynamical properties against the properties of the cosmological sample, we place the orbits of the LMC and M33 in a cosmological context.

\subsection{Crossing Time}
\label{subsec:crossing}
The first crossing time (\tcross) is synonymous with the time at which a subhalo infalls into its host halo. From our analytic orbit analysis in Section~\ref{sec:orbitanalysis}, we find the lower mass models for the MW and M31 both suggest recent, first infall scenarios for the LMC and M33, respectively. The higher host mass models show some evidence that longer-lived orbits, and therefore earlier crossing times are possible. Here, we identify the first crossing time for all massive satellite analogs in Illustris-Dark to statistically determine the most likely infall time for the LMC and M33, respectively, in a cosmological setting. To date, M33's infall time has not been constrained by a statistically significant cosmological sample.

BK11 defines the first crossing time (or BK \tcross) as `the lookback time at which the LMC first crossed the physical $z=0$ virial radius of the MW, moving inward'. It is important to note here, however, that as the host halo mass evolves over time, \Rvir of that halo will also evolve. This halo evolution is especially important for subhaloes that survive to $z=0$ but were accreted early, or $>$ 4 Gyr ago (by the BK11 definition). The virial radius of a host halo will have changed by a factor of a few from the crossing redshift to present-day such that the radius would typically increase with time. For these types of systems, some subhaloes would falsely be identified as massive satellite analogs because these subhaloes might only reside in the extended outskirts of a halo for a majority of their lives. Such subhaloes would never achieve orbital dynamics that mimic those of the LMC or M33, and would therefore contaminate the massive satellite analogs sample.

\begin{figure}
\begin{center}
\includegraphics[scale=0.6, trim= 0mm 10mm 0mm 5mm]{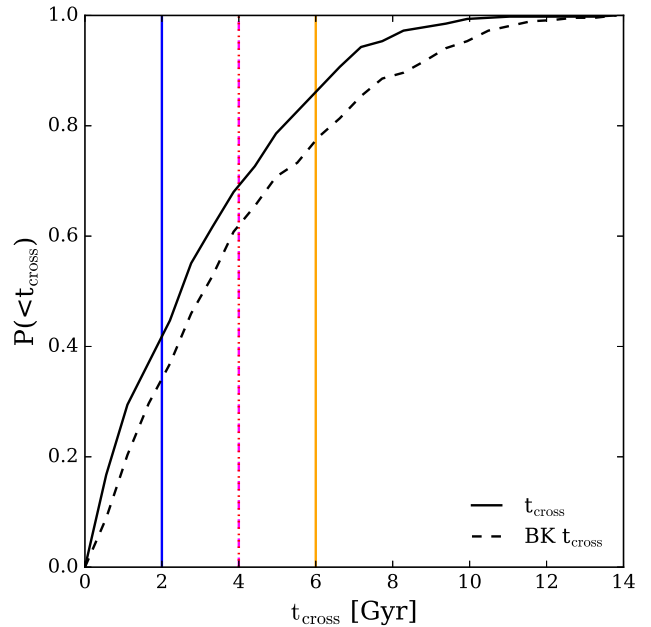} 
\caption{A cumulative distribution of the lookback time at which the satellites first crossed into their host haloes. The solid black line illustrates results when the crossing time is defined using a time evolving virial radius. The dashed black line is the method used in BK11 where crossing time is defined as the lookback time at which a satellite first crosses the $z=0$ physical radius of its host halo. \tcross yields a more recently accreted sample overall. 40 per cent of analogs have a crossing time < 2 Gyr ago, while about 70 per cent have a crossing time < 4 Gyr ago.
\label{fig:timecum}}
\end{center}
\end{figure}

To account for this discrepancy and avoid false identification of analogs, we use a modified definition for the first crossing time throughout this work. This definition, \tcross, uses the lookback time at which the subhalo crosses the time-evolving quantity \Rvirz, instead of \Rvir at $z=0$. Thus, the physically evolving virial radius (and consequently \Mvir) is accounted for and the misidentification of subhaloes with early crossing times is diminished. \Rvirz is reported in the Illustris halo catalogs for each halo at every snapshot, therefore no approximation is necessary to implement our modified definition. Note that crossing time here refers to the first simulation snapshot when the subhalo's position relative to its host is within \Rvirz.

Figure~\ref{fig:timecum} illustrates the distribution of BK \tcross compared to \tcross used throughout this analysis. Notice that only $\lesssim$ 5 per cent of our analog sample has \tcross $>$ 8 Gyr. This motivates dividing our sample by infall times in increments of 2, 4, and 6 Gyr, whereas BK11 defines the lower bound of the earliest accreted population at \tcross $>$ 8 Gyr. The divisions are indicated by the coloured lines in Fig.~\ref{fig:timecum}. The terms crossing time, infall time, and accretion time will all be used interchangeably throughout this work, but all refer to the \tcross definition. 

About 70 per cent of our massive satellite analogs have \tcross $\leq$ 4 Gyr ago. This result is consistent with previous studies of the LMC \citep[e.g. ][BK11]{busha11}. Here, we highlight that Fig.~\ref{fig:timecum} similarly implies that cosmologically, M33 is also favored to be on its first approach towards M31.

While the choice of \Rvir(z) versus \Rvir at $z=0$ does affect the most likely infall time for the analog population, we recognize the choice of virial radius as the outer spatial boundary for a $\Lambda$CDM halo is arbitrary. There are other, more physically motivated criteria which could be used as a measure for infalling satellites instead. One example is the splashback radius \citep{more15}, or the radius at which accreted mass reaches its first apocentre after the turnaround point in its orbit. Utilizing the splashback radius in the definition of crossing time as opposed to the virial radius would only shift the sample's average infall time to an earlier epoch. Since the splashback radius is considerably larger than the virial radius for a given halo at a specific redshift, the satellites would cross the splashback radius before crossing the virial radius. Therefore, the overall population would exhibit a tendency towards early infall times by definition, whereas our current method results in massive satellites crossing a smaller radius at more recent times. 

Because the splashback radius is larger, it will lead to a higher percentage of early accretion scenarios. However, in this paper, we compare to recent orbital histories of massive satellites galaxies whose orbits are well constrained in the past 5 Gyr. We specifically focus on the interaction timescales that are relevant for the LMC--a few Gyr based on previous work--as this time-scale has shown that in its current orbital configuration, the LMC is dynamically affected by the MW and vice versa. Since the virial radius is the smaller of the two definitions for the host haloes of interest in this work, its more restrictive nature within the Illustris-Dark parameter space is more suitable. A larger (i.e. splashback) radius would allow for too many orbital solutions that do not correspond to significant gravitational interactions between our host-satellite analogs and could therefore be physically misleading orbits in the context of our massive satellite analogs.

\subsection{Specific Orbital Energy}

\begin{figure}
\begin{center}
\includegraphics[scale=0.6, trim=0mm 10mm 0mm 0mm]{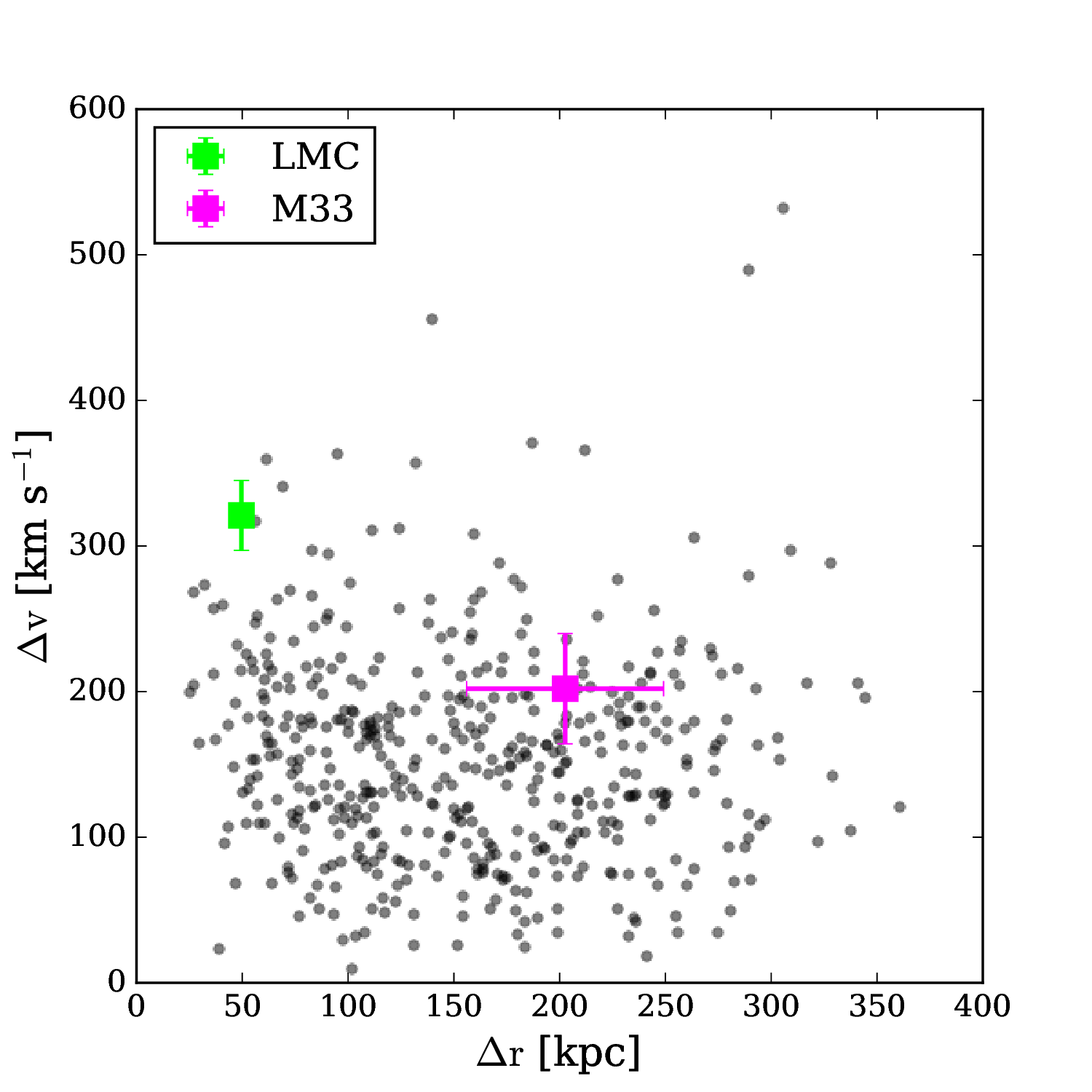}
\caption{The relative positions and velocities for massive satellites in relation to their host haloes. The observed properties of the LMC and M33 are shown in the coloured points with 1$\sigma$ error bars in each direction. The LMC does have an error in its relative position with respect to the MW ($\sim$2 kpc) but it is significantly small compared to the other errors. The LMC's phase space properties are rare in the massive satellite population (as expected since it is approximately at pericentric approach), whereas M33 is common in this sample.
\label{fig:satposvel}}
\end{center}
\end{figure}
Specific orbital energy encodes the relative position and velocity of the satellite as well its host halo mass, so it is a suitable property for determining the most favored crossing time and host halo mass of the LMC or M33. The specific orbital energy of the LMC and M33 today can be calculated from their mean position and velocity (as listed in Table~\ref{table:vectors}) in a range of host halo masses. These quantities will be useful reference points to compare against the cosmological sample.

To calculate the orbital energy of massive satellites, we approximate the gravitational potential of the host halo in each host-satellite pair with an NFW profile. It is normalized by the energy of a circular orbit at \Rvir of the host ($\tilde{E} = \rm E_{sat}/ E_{circ} (R_{vir}$)) to remove bias against the host halo mass. 

\begin{equation} \label{eq:energy}  \rm E_{sat} = \frac{1}{2}v^2 + \Phi_{NFW}(M_{vir}, c_{vir}, r) \end{equation} 

In Equation~\ref{eq:energy}, the virial concentration, $\rm c_{vir}$, is approximated with the fitting formula of the Bolshoi simulation at $z=0$ \citep{klypin11}:
\begin{equation}  c_{\rm vir}(M_{\rm vir})=9.60\left(\frac{M_{\rm vir}}{10^{12} h^{-1}M_{\odot}}\right). \end{equation}
 
$h$ will vary with the choice of cosmological parameters used in the simulation (i.e. {\em WMAP-9. Planck}, etc.) and \Mvir varies for each host-satellite pair. Evidently, orbital energy is very sensitive to the combination of host halo mass, position, and velocity. The position and velocity ranges of the satellite analog sample are plotted in Fig.~\ref{fig:satposvel}. The coloured markers with error bars indicate the observed properties of the LMC and M33 today. While the LMC is rare amongst the statistical sample, it is not surprising since it is approximately at pericentre today. Many more of the massive satellite analogs from Illustris-Dark populate the position-velocity space surrounding M33, which is reasonable since it might be somewhere between its apocentre and pericentre today. 

Lowering the $10^{10}$ \Msun mass floor at $z=0$ to $3\times 10^{9}$ \Msun such that each satellite consists of $\geq$400 particles populates the phase below 75 kpc more densely, as expected, since lower mass satellites are more likely to reside closer to their hosts. Halo-finding routines are often unable to identify subhaloes when they come within close proximity of their host haloes, so it is also possible that some massive satellite analogs which could be $\lesssim$ 50 kpc from their host may be unaccounted for at $z=0$. In these scenarios, \texttt{SUBFIND} would skip the snapshot at which a subhalo was unidentifiable and re-identify it at the next snapshot by matching particle membership. For our analysis, these effects are negligible since we aim to quantify the properties of massive satellites analogous to the LMC and M33 between their time of maximal mass and today, thus the $10^{10}$ \Msun mass floor is sufficient. 

\begin{figure}
\begin{center}
\includegraphics[scale=0.6, trim=0mm 10mm 0mm 0mm]{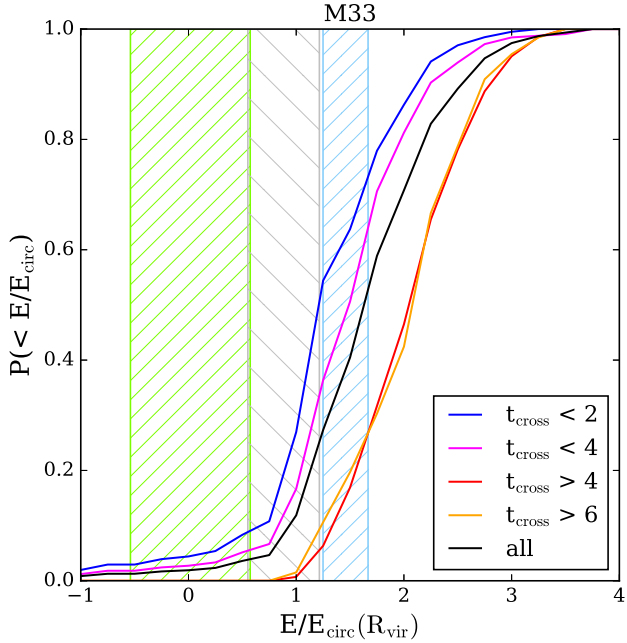}
\caption{The cumulative distribution of orbital energy scaled to the energy of a circular orbit at the virial radius of each host for the massive satellite analogs. The satellite sample is split by \tcross (blue, magenta, red, orange solid lines). The black solid line represents the orbital energy for the entire analog sample. Overplotted in the green hatched region is the mean and 1$\sigma$ errors of the M33 in a low mass host halo centred at $0.7 \times 10^{12}$ \Msun. The gray hatched region indicates the energetics of M33 in an intermediate $1.5 \times 10^{12}$ \Msun halo. The blue hatched region indicates the energetics in a high mass host halo of $3 \times 10^{12}$ \Msun. The width of the hatched regions is calculated using the Monte Carlo samples drawn to compute the mean position and velocity vectors of M33 relative to M31. M33's crossing time appears to be $\leq$ 4 Gyr ago, suggesting that it could not have arrived at its current position until recently.
\label{fig:m33energy}}
\end{center}
\end{figure}

\begin{figure}
\begin{center}
\includegraphics[scale=0.6, trim=0mm 10mm 0mm 0mm]{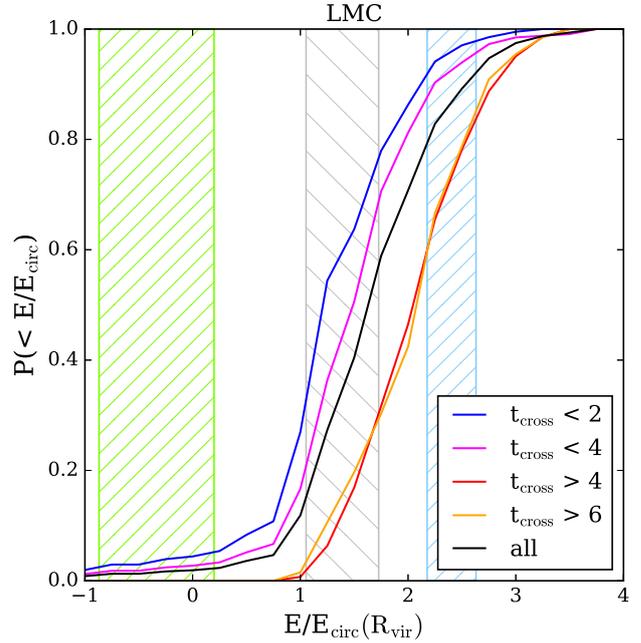} 
\caption{All solid lines are identical to those in Fig.~\ref{fig:m33energy}. The hatched regions are the corresponding quantities for the MW-LMC systems and were calculated consistently. Comparing the simulation sample to the observed properties, the crossing time of the LMC is likely $\lesssim$ 4 Gyr ago and a MW halo mass of \Mvir $\sim1.5\times 10^{12}$ \Msun is statistically favored.}
\label{fig:lmcenergy}
\end{center}
\end{figure}

Figures~\ref{fig:m33energy} and~\ref{fig:lmcenergy} show the normalized, cumulative distribution of orbital energy for the massive satellite analogs, separated by crossing time. The hatched regions in Fig.~\ref{fig:m33energy} indicate the energetics of M33 in a variety of host halo masses. From left to right, the regions are the standard deviation about the mean energy for (0.7, 1.5, 3) $\times10^{12}$ \Msun haloes. The errors have been calculated using the 10,000 Monte Carlo samples from the allowed proper motion error space. The hatched regions in Fig.~\ref{fig:lmcenergy} represent the comparable quantities for the LMC. The solid cumulative distribution lines are identical in both figures.

Overall, the population of massive satellite analogs are bound to their host haloes, i.e. $\tilde{E} > 0$. The early (\tcross $>$ 4 Gyr) and late accreted (\tcross $<$ 4 Gyr) populations exhibit distinctly different energetics. The early accreted subhaloes are statistically more bound to their host haloes as they have experienced more orbital decay, while the late-accreted subhaloes are less bound. Only a small percentage of systems are energetically unbound, likely because these systems are in a short-lived configuration at $z=0$ (i.e. a flyby satellite or a three-body encounter) or they have fallen into their host haloes on highly eccentric orbits.

The mean values of orbital energy for M33 based on its position and velocity today residing within M31's halo with masses \Mvir $=$ (0.7, 1.5, 3) $\times 10^{12}$ \Msun are $\tilde{E} = (0.02, 0.88, 1.46)$. These are the mean energies in each of the hatched regions in Fig.~\ref{fig:m33energy}. Comparing these values with the massive satellite analogs, only half of all analogs span the range of energies for M33 if M31's halo mass is 0.7-3$\times 10^{12}$ \Msun (black solid line). Those analogs which exhibit acceptable M33 energies today are dominantly satellites with crossing times $\leq$ 4 Gyr ago, suggesting that M33 could not have arrived at its current position until recently. These satellites prefer a high M31 halo mass $\geq 1.5 \times 10^{12}$ \Msun. Early infall (\tcross > 4 Gyr ago) is allowed at the 20 per cent level, but if M31's halo mass is less than $\sim$3$\times 10^{12}$ \Msun, first infall is certainly preferred. 

These conclusions are somewhat at odds with the results of the numerical orbit analysis (Section~\ref{sec:orbitanalysis}) at the high mass end ($\rm M_{vir,M31}=2\times10^{12}$ \Msun). The analytic models suggested a long period orbit for M33, whereas the cosmological analogs suggest a first infall scenario is more likely. This evidences that analytic models are not always suitable for inferring orbital histories over long ($\sim$ 6 Gyr) time-scales since they lack the appropriate physics to accurately capture the evolution of both massive satellites and their hosts \citep[see also][]{lux10}. 

For LMC-type satellites residing in a MW halo with masses \Mvir $=$ (0.7, 1.5, 3) $\times 10^{12}$ \Msun, the mean values for orbital energy are $\tilde{E} = (-0.34, 1.39, 2.40)$. These are the mean energy values for the hatched regions in Fig.~\ref{fig:lmcenergy}. Comparing to the sample of massive satellite analogs indicates the LMC's orbital energy is rather common. The black solid line representing the entire sample generally spans the full range of allowable LMC orbital energies as indicated by the hatched regions. 

If the LMC is cosmologically typical based on its orbital energy, only about 15 per cent of the massive satellite analogs exhibit preference towards a MW halo mass $\lesssim 1.5 \times 10^{12}$ \Msun. Similarly, about 85 per cent favor a MW halo mass $\gtrsim 1.5 \times 10^{12}$. Thus, independent of crossing time, a MW halo mass of $\sim1.5 \times 10^{12}$ \Msun is most favored. 

In this halo mass range, indicated by the gray hatched region, a first infall is preferred, which again differs from the orbital integration results in Section~\ref{sec:orbitanalysis}. The numerical orbit shows evidence for a pericentric passage around 5 Gyr ago. Early infall in this halo mass range is allowed at the 25 per cent level, which again likely indicates that backward integration schemes are not accurate tracers of cosmological orbits over such time-scales, especially for long period orbits.

Therefore, the distribution of orbital energy for the massive satellite analogs confirms M33 likely has an infall time within the last 4 Gyr and it prefers an M31 halo mass $\geq 1.5 \times 10^{12}$ \Msun. It also suggests that the preferred MW halo mass is $\sim\!1.5 \times 10^{12}$ \Msun based on the LMC's current position and velocity. If the LMC really is on its first infall with \tcross $\leq$ 4 Gyr ago, the likelihood of this MW halo mass increases.

BK11 used the old proper motion values of the LMC \citep{k06a} and the Millenium-II simulations to conclude the most typical MW halo mass is $> 2 \times 10^{12}$ \Msun from orbital energy studies. However, the mean total velocity of the LMC has decreased by a significant 57 \kms. This illustrates that precise proper motion measurements are required to reliably compare the properties of Local Group satellites to statistics of cosmological analogs. 

Examination of the specific orbital angular momentum of the massive satellite analogs sample results in the same conclusions as orbital energy. The angular momentum of the LMC and M33 today are common amongst the general sample of massive satellite analogs. Again, the cosmological sample prefers a recent infall time, within the last 4 Gyr, for M33 and the LMC based on their present-day angular momentum. In Paper II, we estimate the most typical halo mass for the MW and M31 in a Bayesian scheme based on the LMC and M33's angular momentum today.

\begin{figure*}
\begin{center}
\includegraphics[scale=0.55]{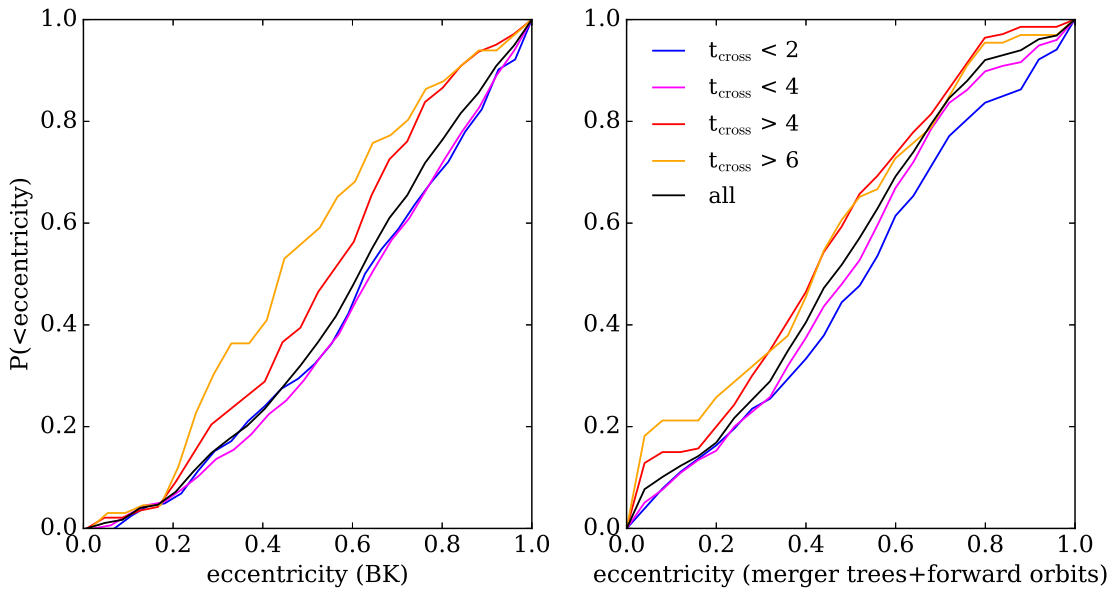}
\caption{{\em Left:} The distribution of instantaneous orbital eccentricity at $z=0$ for the massive satellite population in Illustris-Dark. Eccentricity is calculated using the position, velocity, and host halo mass at $z=0$ by approximating the host halo as a spherical NFW halo potential and the subhalo as a point mass. {\em Right:} The distribution of orbital eccentricity from the combination of merger tree data and forward orbit integrations for the massive satellite analogs. See Appendix~\ref{sec:A}.
\label{fig:ecc}}
\end{center}
\end{figure*}

\subsection{Eccentricity}

Orbital eccentricity is the final property we use to quantify the orbits of massive satellite analogs in Illustris-Dark. As discussed in \citet{hashimoto03}, the orbits of satellite galaxies tend to circularize over time, or become less eccentric \citep[e.g. ][]{murai80, ibata98}. This circularization is closely tied to the pericentric approach and mass of the satellite galaxy. Since dynamical friction is directly proportional to $\rm M_{sat}^2$ (see Equation~\ref{eq:df}), it is a determining factor in the orbital evolution of satellite galaxies, especially for massive satellites. The amount of circularization determines the ability of satellites to survive before merging with their hosts. As such, circularization of the LMC or M33 orbits may shed light on their expected infall times.

We introduce two different methods for computing the eccentricity of the massive satellite analog orbits. The first is an instantaneous eccentricity method computed with the kinematics of massive satellite analogs at $z=0$. The second uses the extracted orbital histories of the analogs to compute an eccentricity with real orbital data. In both methods, the following definition of eccentricity stemming from a combination of orbital apocentre ($r_a$) and pericentre ($r_p$) is implemented. 
\begin{equation} e\equiv \frac{r_a - r_p}{r_a+r_p}  \label{eq:ecc} \end{equation}

\subsubsection{Instantaneous Eccentricity from the Equation of Motion} 
The first eccentricity method is an instantaneous approximation given the $z=0$ properties of each massive satellite analog, identical to BK11. In this approximation, apocentre and pericentre are defined as the roots of the following equation of motion from \cite{binneytremaine}, Chapter 3. 
\begin{equation} u^2 + \frac{2[\Phi(1/u) - E]}{L^2} = 0 \label{eq:eom}\end{equation} 
$\Phi$ is the gravitational potential of the fixed massive object and $E$ is the Hamiltonian for the system. $u=1/r$, where $r$ is the distance of the satellite from its host. $L$ is the magnitude of the angular momentum vector per unit mass for the orbiting body. 

This method is similar to the reduced mass approach of \citet{wetzel11}, except we approximate the host as an extended potential. We assume the satellites are orbiting within a spherical dark matter halo that is well approximated by an NFW profile and calculate the apocentre and pericentre instantaneously with their $z=0$ properties. Unlike our numerical orbital models, the satellites are modeled as point masses. All unbound orbits (E < 0) are assigned a default eccentricity of 1, so $e=0$ describes a perfectly circular orbit. 

Table~\ref{table:ecc} lists the eccentricity values for the LMC and M33 using their current positions and velocities from Table~\ref{table:vectors} in host haloes with \Mvir $=$ (0.7, 1, 1.5, 3) $\times 10^{12}$ \Msun. The left panel of Fig.~\ref{fig:ecc} shows the cumulative probability distribution of orbital eccentricity for the massive satellite analogs split by crossing time. The black solid line indicates the distribution of orbital eccentricity for the entire analog sample. The mean value is $\sim$0.6, or fairly eccentric. Early accreted satellites (\tcross > 4 Gyr) tend to be on more circularized orbits, which suggests they have experienced the most mass loss and have become circularized by dynamical friction in the presence of their host haloes for many billions of years. 

BK11 implemented this eccentricity method and found a similar distribution of eccentricity for the population of LMC analogs in their study. Using the old LMC proper motion values, they conclude the LMC is on an unbound orbit in a $10^{12}$ \Msun MW halo when it is represented as an NFW halo that extends to infinity. However, the significant change in the updated proper motion values of the LMC allows us to re-evaluate this claim. With the new proper motion values, the LMC is indeed bound in a $10^{12}$ \Msun NFW halo with $e=0.904$. Therefore, it is possible for the MW's halo mass to be as low as $1\times10^{12}$ \Msun with the LMC's current orbital conditions even though it would be an outlier since only $\sim$10 per cent of our overall sample has $e\geq0.9$, independent of infall time.

\begin{table}
\centering
\caption{The instantaneous eccentricity for the observed position and velocity of the LMC and M33 in host haloes of varying virial mass. The eccentricity is computed using Equations~\ref{eq:ecc} and~\ref{eq:eom}. The LMC is on an unbound orbit in a host halo with \Mvir$=0.7\times 10^{12}$ \Msun, corresponding to an eccentricity > 1 (i.e. parabolic or hyperbolic orbit). }
\label{table:ecc}
\begin{tabular}{p{2.4cm}p{1.1cm}p{.9cm}p{1.1cm}p{.9cm}}\hline \hline
Host halo mass [\Msun]& $0.7 \times 10^{12}$ & $1\times10^{12}$ & $1.5\times10^{12}$ & $3\times10^{12}$ \\ \hline
LMC& unbound & 0.904 & 0.714 & 0.431 \\
M33& 0.952 & 0.808 & 0.694 &  0.623 \\ \hline
\end{tabular}
\end{table}

\subsubsection{Eccentricity from Merger Trees and Forward Orbits}
\label{subsubsec:forward}
While the previous method utilizes the $z=0$ properties to compute instantaneous eccentricities, the Illustris-Dark cosmological simulation and associated merger trees \citep{rg15} allow us to trace the orbital histories of each massive satellite analog throughout cosmic time and directly identify their most recent apocentre and pericentre distance to their host, if they exist. Apocentre and pericentre are defined here as true critical points in the satellite's distance relative to its host as a function of time. If a satellite has both an apocentre and pericentre as defined here, the eccentricity is calculated by Equation~\ref{eq:ecc}. Only 46 per cent of the massive satellite analogs sample contains an apocentre and pericentre in their past orbital trajectories since a majority of the sample is on first infall. The average value of eccentricity for that subset of analogs is about 0.4. There is no clear correlation between merger tree eccentricity and infall time for this sample.

For those satellites where only a pericentre or neither critical points are recovered in the past orbital trajectories (i.e very recent \tcross scenarios), the orbits are numerically integrated forward in time for 6 Gyr using the $z=0$ position and velocity vectors relative to their host haloes, following the methodology of Section~\ref{sec:analyticmethods}. The trajectories are then analysed to find the first pericentre and/or apocentre. In the case where the merger tree data contains a pericentre and no apocentre, only the apocentre is taken from the forward orbit. More details of this forward orbit integration are discussed in Appendix~\ref{sec:A}. 

Using both the merger tree data and the forward orbit integrations, eccentricities for 96 per cent of the massive satellite analogs sample are recovered. The right panel of Fig.~\ref{fig:ecc} shows the cumulative distribution of eccentricity separated by crossing time for this method. The remaining 4 per cent of analogs are likely fly-by satellites, so they are omitted in Fig.~\ref{fig:ecc}. Note that for the recently accreted satellites, it is not necessarily true that the pericentre and apocentre have occurred in the time between infall and today, but rather between infall and 6 Gyr in the future. The average eccentricity for the sample increases to about 0.45. 

Similar to the instantaneous eccentricity method, the real eccentricities extracted from orbital trajectories indicate some correlation with infall time. The early accreted massive satellite analogs (\tcross > 4 Gyr) are on more circular orbits than those accreted more recently. However, the correlation between eccentricity and infall time is much weaker and therefore cannot be used to discriminate between satellites with early versus recent infall times. 

We attribute this weak correlation to the rapid circularization of massive satellite analogs after infall owing to their high masses. As a result, their eccentricities are inherently more circular overall. The early accreted satellites enter their host haloes at larger separations (> 150 kpc) and with higher relative velocities (> 200 \kms) as compared to the recently accreted analogs. Consequently they are able to survive until $z=0$ because dynamical friction is less efficient at decaying their orbits quickly. 

\begin{figure}
\begin{center}
\includegraphics[scale=0.6, trim = 8mm 15mm 0mm 0mm]{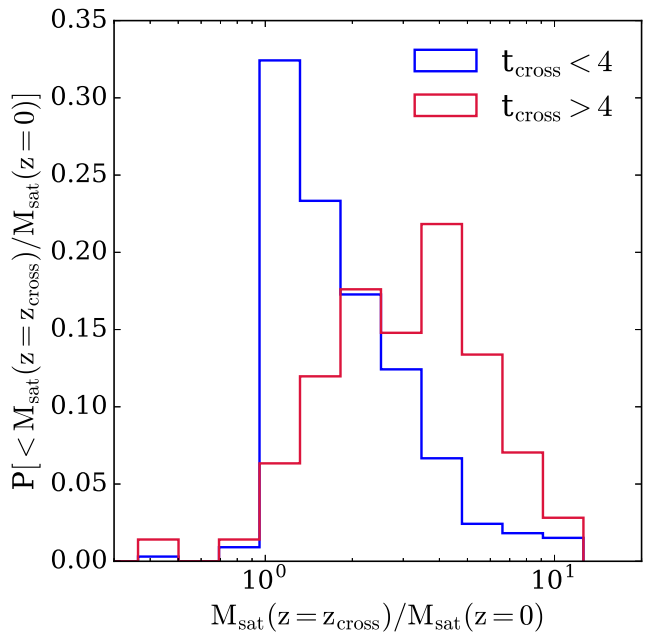}
\caption{The distribution of the ratio of mass at crossing and at $z=0$ for Illustris-Dark massive satellite analogs with recent crossing times (blue, \tcross < 4 Gyr ago) and early crossing times (red, \tcross > 4 Gyr ago). The distributions are normalized individually such that the y-axis shows the probability of a given mass ratio. 
\label{fig:massloss}}
\end{center}
\end{figure}

Figure~\ref{fig:massloss} illustrates that recently accreted (\tcross < 4 Gyr ago) massive satellites do not lose a significant fraction of their infall masses. Plotted is the ratio between total dark matter mass when the satellite first crosses the virial radius of its host relative to its bound mass at $z=0$, split by early and late crossing times. The most recently accreted satellites (blue) manage to sustain their masses since infall, while the early accreted satellites (red) experience more mass loss, decreasing in mass by a factor of 10 at most. Neither infall time nor mass loss correlate strongly with eccentricity, so massive satellite galaxies should be treated with care (i.e. treated as extended bodies with significant mass) when quantifying their survivability time-scales. 

\section{Discussion}
\label{sec:discussion}
We have numerically constrained the orbital histories of the LMC about the MW and M33 about M31 using the allowed proper motion error space of each system and a wide range of mass models. We found that both satellites favor a recent infall scenario unless the total mass of the MW or M31 is in excess of $1.5 \times 10^{12}$ \Msun or $2 \times 10^{12}$ \Msun, respectively, in which case the orbital periods of these satellites are of order 5-6 Gyr. 

We have also characterized the preferred infall times for a population of massive satellite analogs in the Illustris-Dark simulation and found that massive satellites exhibiting orbital properties similar to the LMC and M33 also prefer a recent infall scenario. While the numerical models and the simulation results are consistent with one another with respect to infall time, Illustris-Dark favors recent infall unless the host halo masses are even more massive than those used in the numerical orbit integrations: MW $\gtrsim$ $2 \times 10^{12}$ \Msun, M31 $\gtrsim3 \times 10^{12}$ \Msun. 
 
While our cosmological studies revealed tension with orbital histories over long time-scales, the analysis appears robust over time-scales < 5 Gyr. In Section~\ref{sec:orbitanalysis}, we illustrated that orbits allowed by M33's velocity error space do not exhibit a recent, close passage about M31, in tension with conventional models based on the morphology of M33. A recent, close passage scenario is also relevant for the LMC, since it is just past its pericentre about the MW. In the following we examine this controversial result in detail. Finally, we present the past orbital history of the MW, LMC, M31, and M33 as a four-body interacting system of galaxies.

\subsection{Did M33 Have a Close Encounter with M31?}
\label{subsec:M33closeencounter}
\begin{table*}
\caption{The fraction of M33 orbits for three different M31-M33 mass combinations which satisfy specific orbital criteria determined by numerical integration. In all models, 10,000 orbits are calculated in the M31-M33 error space and the mass of M31 is fixed at $2\times 10^{12}$ \Msun. The final three columns indicate the three M33 masses tested.}
\label{table:m33stats}
\centering
\begin{tabular}{c|c|c|c|c|c|c|c} \hline \hline
Identifier & N$\rm_{peri}$ & $\rm t_{peri}$ & $\rm t_{inf}$ & $\rm r_{peri}$ & $5\times 10^{10}$ \Msun & $1\times 10^{11}$ \Msun & $2.5\times 10^{11}$ \Msun  \\ \hline 
ARP & $\geq$1 &  $\leq 6$ Gyr ago  & -- & -- & 35.90\% & 38.43\% & 34.44\% \\ 
TI6 & $\geq$1 &  $\leq 6$ Gyr ago  &  $\leq$ 6 Gyr ago & -- & 18.16\% & 19.25\% & 21.03\% \\ 
RP100 & $\geq$1 &  $\leq 6$ Gyr ago  &  $\leq$ 6 Gyr ago & $\rm r_{peri} <$ 100 kpc & 6.25\% & 6.32\% & 3.59\% \\ 
RP100T & $\geq$1 &  $\leq 3$ Gyr ago  &  $\leq$ 6 Gyr ago & $\rm r_{peri} <$ 100 kpc  & 0.27\% & 0.04\% & 0.14\% \\ 
RP55 & $\geq$1 &  $\leq 6$ Gyr ago  &  $\leq$ 6 Gyr ago & $\rm r_{peri} <$ 55 kpc & 1.00\% & 1.06\% & 0.84\% \\ 
RP55T & $\geq$1 &  $\leq 3$ Gyr ago  &  $\leq$ 6 Gyr ago & $\rm r_{peri} <$ 55 kpc & 0\% & 0\% & 0\%\\ \hline
\end{tabular}
\end{table*}

\begin{figure*}
\centering
\includegraphics[scale=0.45, trim =10mm 10mm 0mm 0mm]{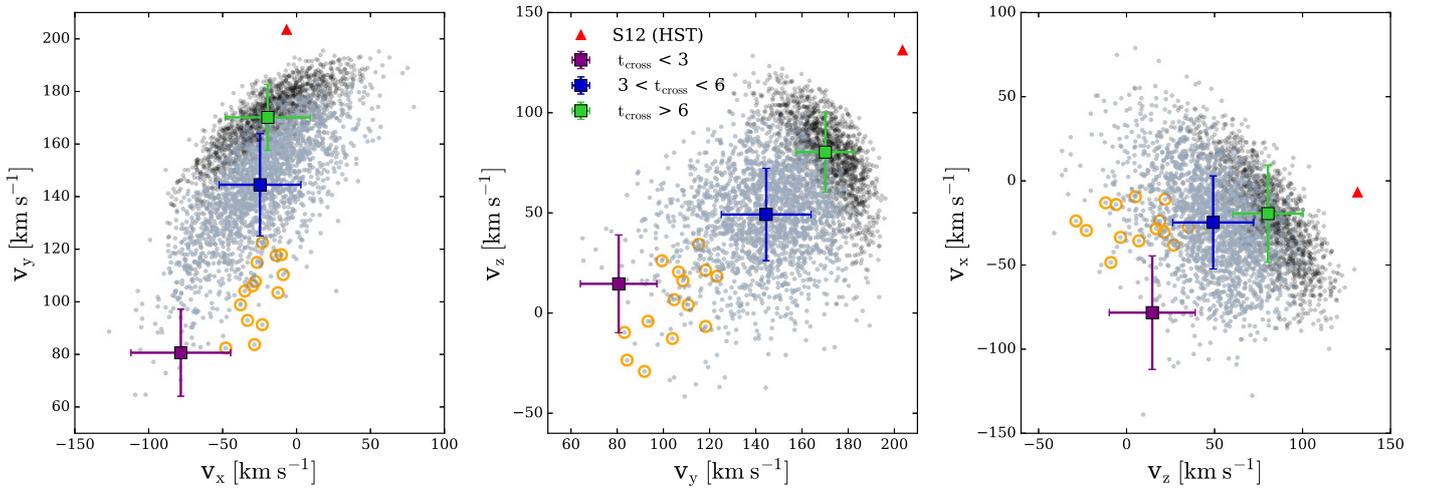}
\caption{For the ARP sample, the x, y, and z velocity components are plotted for M33's current velocity vector with respect to M31 in the highest M31 and M33 mass model (black points). Lower host mass models would only weaken the statistics. Overplotted in blue-gray points are those vectors within the ARP sample where the orbits evidence an infall time $\leq$ 6 Gyr ago (TI6 sample). The orange circles highlight only those vectors that belong to the RP100T subset. This subset represents the orbits that are most reflective of the criteria outlined in P09 and M09, which are designed to reproduce the warps in M33's gaseous and stellar discs. See Table~\ref{table:m33stats} for more details. The square markers with error bars indicate the average velocity components for all samples, binned by satellite infall time. The error bars are the standard deviation within each infall time bin for each velocity component. The red triangles denote the velocity components from the HST only proper motion of M31 (S12). The S16 velocity vector is {\bf v}=(135.30, 0.33, 117.60) \kms, but these velocity components do not lie in the same direction as the locus of RP100T orange circles.
\label{fig:analyticplanes}}
\end{figure*}

The scenarios put forth by M09 and P09 to recover the warped structure of the HI gas disc and the stellar disc of M33 require the satellite to have made a pericentric passage within 50-100 kpc of M31 in the last 3 Gyr. The minimum eccentricities from P09 and M09 described in Section~\ref{subsec:HI} are about 0.34 and 0.67, respectively. The former is computed using the current position of M33 as the apogalacticon, therefore it yields only a minimum value for eccentricity. 

From Fig.~\ref{fig:ecc}, about 70 per cent of the massive satellite analogs sample has an eccentricity $\gtrsim 0.34$, while approximately 20 per cent has an eccentricity $\gtrsim 0.67$. To first order, it is not rare to find orbits of massive satellite analogs in Illustris-Dark that resemble those theorized by M09 and P09. In the following, we examine the question of M33's pericentric approach to M31 first by exploring the allowed proper motion error space and then by using cosmological simulations. 

Two data samples are defined to carry out this analysis. The first is the analytic recent pericentre (ARP) sample, which describes the orbital histories recovered by searching 10,000 Monte Carlo drawings from the 4$\sigma$ proper motion error space as described in Section~\ref{subsec:propermotions}. Second, we define the Illustris-Dark recent pericentre (IRP) sample. This set includes all massive satellite analogs identified in Section~\ref{sec:illustris} containing a recent pericentric passage in their orbital history. 

\subsubsection{The Analytic Recent Pericentre Sample}
M33 and M31 have been modeled as a system where M33 has a recent, close (50-100 kpc) encounter with M31 (P09, M09). This close encounter may be strong enough to induce the formation of warps in the gas and stellar discs of M33. Given its gas content and cosmological expectations, it is most likely to have been accreted within the past 4 Gyr or so. Here, we seek to reconcile these two requirements given the observationally constrained parameter space. 

We follow the methodology outlined in Section~\ref{sec:analyticmethods}, assuming M33's mass is fixed at $2.5 \times 10^{11}$ \Msun and M31's virial mass is $2\times10^{12}$ \Msun. A lower M31 mass would only weaken the statistics for a recent, close passage scenario since it would be less effective at decaying the orbit of M33 via dynamical friction. We use the highest M33 mass from Fig.~\ref{fig:orbits} because its orbital trajectory exhibits the lowest eccentricity. Lower M33 mass models are tested later in this section. 

10,000 backwards orbits are computed for 6 Gyr, spanning the M31-M33 velocity error space. We first identify the orbits that allow M33 to have made a pericentric passage about M31 in the last 6 Gyr, regardless of infall time. Pericentre is defined such that it is a true critical point in the orbital trajectory and the relative position of M33 at pericentre has a magnitude less than its separation today. This sample will be referred to as the analytic recent pericentre sample (ARP). 34.44 per cent of the allowed orbits belong to the ARP sample. The average orbital properties of the ARP sample are: $\rm t_{inf} = 5.5\pm0.9$ Gyr ago, $\rm r_{peri}=130.7\pm45.6$ kpc, and $\rm t_{peri}=4.5\pm1.1$ Gyr ago.

\begin{figure*}
\centering
\includegraphics[scale=0.4, trim=0mm 15mm 30mm 0mm]{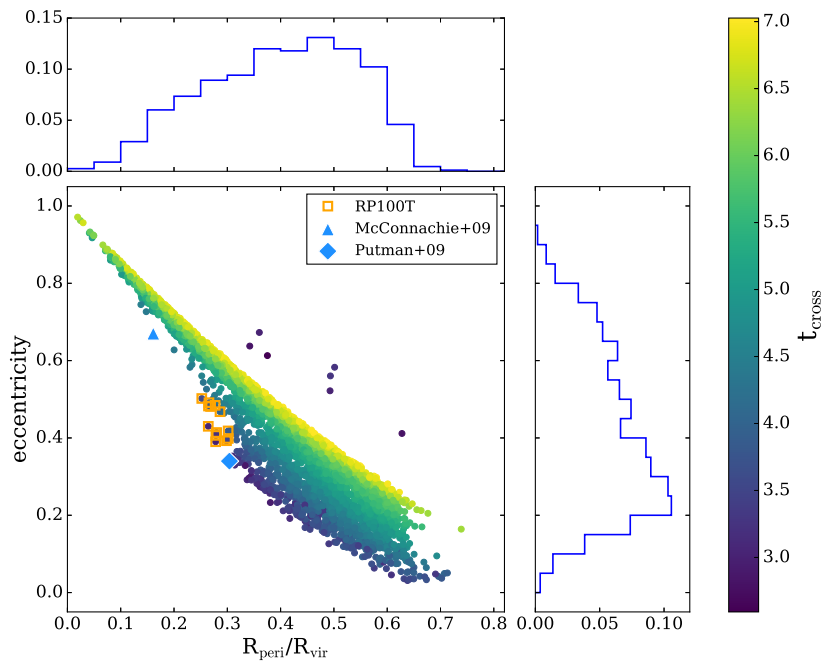}
\caption{The distribution of eccentricities and pericentres for the same ARP sample plotted in Fig.~\ref{fig:analyticplanes}. The x-axis gives the pericentric distance with respect to the virial radius of M31. M31 is $2\times10^{12}$ \Msun and its \Rvir\!=329 kpc for this model. The points are coloured by their crossing time, the time at which the satellite infalls into the host's virial radius. The orange squares highlight those which have a crossing time in the last 6 Gyr and a pericentric approach within 100 kpc of M31 in the last 3 Gyr (RP100T, 0.14 per cent). No orbits reach within 55 kpc during the last 3 Gyr as suggested by M09 (RP55T). The blue triangle shows the eccentricity and pericentre distance they conclude would result in M33's stellar warp. The blue diamond represents the eccentricity and most probable pericentre put forth by P09. The five outliers above the swath of points are position and velocity vectors that result in a positive radial velocity, suggesting M33 is moving away from M31 instead of towards it. These vectors are likely artefacts of the edges of the allowed proper motion error space. 
\label{fig:analytic2dhist}}
\end{figure*}

Fig.~\ref{fig:analyticplanes} shows the distribution of M33's velocity vector components with respect to M31 for the ARP sample (all points). Overplotted in blue-gray points are the components of M33's velocity vectors whose orbits also indicate an infall time during the last 6 Gyr. The coloured squares with error bars denote the average velocity components for the orbits within the given infall time range. 

The orange circles in Fig.~\ref{fig:analyticplanes} indicate the velocity components of the allowed proper motions where the orbits have recent infall times (\tcross $\leq$ 6 Gyr) and reach a pericentric distance within 100 kpc of M31 in the last 3 Gyr. {\em Only 0.14 per cent of the full orbital sample exhibit these criteria} (denoted by RP100T in Table \ref{table:m33stats}). These criteria most closely resemble those estimated by P09 to reproduce M33's gaseous warps. The velocity components of the RP100T sample are clearly outliers in the $y$ and $z$ directions. The mean $v_Y$ and $v_Z$ components of this sample both lie about 2.5$\sigma$ from M33's mean velocity components listed in Table~\ref{table:vectors}. 

Further restricting the allowed orbits such that the pericentric distance must be within 55 kpc of M31 in the last 6 Gyr whittles the fraction down to about 1 per cent (RP55), but there are no orbits that recover $\rm r_{peri} < 55$ kpc in the last 3 Gyr (RP55T). Therefore, M09's proposed M33 orbit cannot be recovered within the proper motion error space of M31-M33. The final column of Table~\ref{table:m33stats} indicates the percentage of orbits that achieve each of the aforementioned criteria for this M31-M33 mass combination.

To ensure that these statistics for the highest M33 and M31 masses are not sensitive to our dynamical friction prescription, we recomputed all allowed orbits without any dynamical friction term. This model would be most likely to reproduce the M09 and P09 orbits. We find that the RP100T sample increases to 7.57 per cent and the RP55T sample contains 2.09 per cent of all allowed orbits as opposed to zero. Regardless, these results do not change our conclusion that a close passage between M33 and M31 is unlikely within the error space. 

We also test two other M33 mass models to constrain whether a lower mass satellite is more statistically effective at recovering a recent, close encounter with M31 in the allowed error space. However, the statistics only improve minimally. Table~\ref{table:m33stats} provides a summary of the same constraints placed on these orbital samples. We have also considered the gravitational influence of the MW on M33, but our computations show that the MW never reached $\leq$ 770 kpc relative to M33 in the last 6 Gyr for all three M31-M33 mass combinations. 

Fig.~\ref{fig:analytic2dhist} further demonstrates the properties of the ARP sample. The distribution of pericentre to virial radius and eccentricity are shown relative to their crossing times, assuming a host mass of $2\times10^{12}$ \Msun and a virial radius of 329 kpc. The orange squares highlight the RP100T orbits, corresponding to the orange circles in Fig.~\ref{fig:analyticplanes}. They most closely resemble the orbit suggested by P09, denoted by the light blue diamond. No orbits in the M31-M33 proper motion error space resemble the orbit in M09's work. This is likely because the orbits which do recover recent pericentres generally have a $\rm r_{peri}$ that is too high ($\gtrsim$ 100 kpc), and therefore they are inconsistent with a recent, {\em close} interaction with M31. 

In Section~\ref{subsec:propermotions}, we note that several recent works have quoted larger M33 distances than that of M09. \citet{u09} and \citet{bonanos06} both quote values of approximately 960 kpc between M33 and the MW, as opposed to $\sim$800 kpc which has typically been used in previous works and which we use in this analysis. Our set of 10,000 Monte Carlo drawings considers M33 distances in the range $\sim$715-880 kpc. If M33's true distance is 880 kpc, we find that the resulting orbit only ratifies our results from the ARP sample--a recent, close pericentric passage of M33 about M31 is rare. A larger M33 distance further indicates that a first infall scenario is more favorable. Even larger distances to M33 (> 880 kpc) are expected to continue this orbital trend. 

From the ARP sample, we conclude it is not rare to find M33 in a recent infall scenario within the observationally constrained phase space of the M31-M33 system. A recent ($\sim$ 4-6 Gyr ago) pericentric passage of M33 about M31 is also allowed. Both scenarios are plausible at the 20-30 per cent level (see Table~\ref{table:m33stats}). However, it is very rare to find close pericentric passages, within 50-100 kpc from M31. M09 and P09 require M33 to achieve a separation of 53 kpc and 100 kpc, respectively. At most, we find only $\leq$ 0.27 per cent of orbits reach within 100 kpc of M31 in the last 3 Gyr and no orbits get as close as 55 kpc to M31 during that time. Therefore, the long period orbit mentioned in Section~\ref{sec:orbitanalysis} is still preferred by the proper motions of M31 and M33 when M31 is massive ($2\times10^{12}$ \Msun). At higher M31 halo masses, the statistics improve somewhat as M33's orbit will turn over at more recent times, but at a virial mass of $3\times10^{12}$ \Msun, the total mass of the Local Group would also have to increase.

While the desired trajectory is infrequent in our numerically integrated orbits, we use the Illustris-Dark massive satellite sample to infer its likelihood in a cosmological setting.

\subsubsection{The Illustris Recent Pericentre Sample}
\begin{figure*}
\centering
\includegraphics[scale=0.4, trim=0mm 15mm 30mm 0mm]{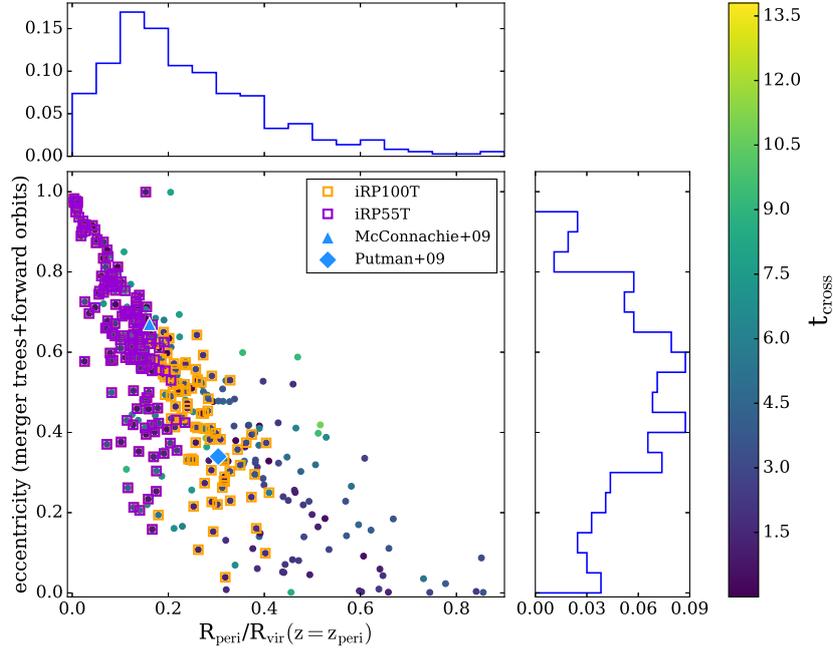}
\caption{Like Fig.~\ref{fig:analytic2dhist}, the distribution of pericentre and eccentricity is shown for the sample of massive satellite analogs in Illustris-Dark belonging to the IRP sample. The orbital data are taken from the merger trees directly or they are combined with a forward orbit integration using the $z=0$ host-satellite properties. The points are coloured by their crossing times. The x-axis indicates pericentre relative to the virial radius at the time of pericentre. For those where the pericentre was found in the forward orbit, the $z=0$ \Rvir was used. The orange squares highlight those massive satellite analogs where the pericentre reaches < 100 kpc of the host halo (49.36 per cent) within 3 Gyr of today (iRP100T). The purple squares are those where the pericentre is within 55 kpc of the host halo (32.42 per cent) in $\pm$3 Gyr of $z=0$ (iRP55T). The blue triangle references the M09 orbit and the blue diamond shows the P09 results.
\label{fig:illustris2dhist}}
\end{figure*}

We have shown that the allowed proper motion error space of M31-M33 does not favor a recent, close pericentric passage of M33 about M31. Here, we will examine the orbital trajectories of the massive satellite analogs from Section~\ref{subsec:satanalogs} to quantify the frequency of this scenario in a cosmological setting. 

In Section~\ref{subsubsec:forward}, we used the Illustris-Dark merger trees to extract the orbital histories of all massive satellite analogs. Given that a majority of the analogs were accreted recently, 71.8 per cent of orbits contain a pericentric passage but only 46 per cent of all orbits contain both an apocentre and pericentre. The average orbital properties for all that contain at least a pericentric passage are: $\rm t_{inf}=3.9\pm2.1$ Gyr ago, $\rm r_{peri}=89.8\pm60.2$ kpc, $\rm t_{peri}=1.6\pm1.2$ Gyr ago. 

The average infall time for all analogs without a pericentre in their merger tree data is $\rm t_{inf}=0.6\pm1.2$ Gyr ago. Unsurprisingly, these massive satellite analogs were accreted recently. For these analogs, we integrate their orbits forward in time from $z=0$ for only 3 Gyr (instead of 6 Gyr as in Section~\ref{subsubsec:forward}) since we are looking for recent accretion scenarios. Typically, the time between infall and 3 Gyr in the future totals to $\sim$4-6 Gyr, approximately equivalent to an average orbital period. The average orbital properties for all forward integrations that contain a pericentric passage in the future are: $\rm r_{peri}=53.3\pm61.0$ kpc and $\rm t_{peri}=1.4\pm0.8$ Gyr beyond today. 

The combined merger tree data and forward orbits increase the fraction of massive satellite analogs with both an apocentre and pericentre to 87.92 per cent. For every analog that shows evidence for a true pericentric passage in the past trajectory, we use this value and corresponding time. The remaining 12.08 per cent of satellites are either on long-period orbits or they are short-lived fly by encounters at $z=0$. 

Using this combined set of orbital histories, we define the Illustris-Dark recent pericentre (IRP) sample as the subset of 472 massive satellite analogs which have made one pericentric passage between their time of infall and $z=0$ in the merger tree data or infall and 3 Gyr in the future determined by forward orbits. The IRP sample encompasses 77.54 per cent of the massive satellite analogs population. The pericentres have been confirmed as true minima and occur at a separation less than their $z=0$ positions to be consistent with the ARP sample. The average infall time of the IRP sample is $\rm t_{inf}=3.0\pm2.5$ Gyr ago. 

Fig.~\ref{fig:illustris2dhist} shows the distribution of eccentricities and pericentre as a fraction of the host halo virial radius for the IRP sample. The points are once again coloured by crossing time. The virial radius of the host halo at the time of pericentre is adopted for all massive satellite analogs with a pericentre in their merger tree data. For all pericentres taken from the forward orbit integrations, the virial radius of the host halo at $z=0$ is used as an approximation. 

The orange squares highlight the fraction of the IRP sample where the satellites have an infall time $\leq$ 6 Gyr and a pericentric passage within 100 kpc of their host halo in the last 3 Gyr, or the iRP100T sample. This sample represents 49.36 per cent of the total population of massive satellite analogs. Further restricting the IRP sample to those where the pericentric distance is < 55 kpc from the host halo in the last 3 Gyr, we find 32.42 per cent of analogs satisfy these criteria (denoted by iRP55T). The iRP55T sample is overplotted in purple squares. Table \ref{table:illustrisstats} summarizes the fraction of all orbits that satisfy each criteria.

The IRP sample demonstrates that cosmologically, the orbits required by both P09 and M09 to reproduce M33's warped structures are not rare. A recent pericentric passage reaching within 100 kpc of the host is true for about half of all massive satellite analogs but only about one-third of analogs reach a distance < 55 kpc from their hosts in that time. The existence of a reasonable cosmological population of subhalo orbits satisfying this strict orbital criteria supports the possibility that, in general, massive satellite galaxies can be responsible for warps in the baryonic discs of their hosts. However, larger pericentric approaches are more common.

The results of the IRP sample are also generally applicable to the B07 orbital model for the LMC--its close approach of 50 kpc is not typical, but also not cosmologically rare. 

Upon further inspecting the orbits identified as the iRP100T orbits, we find that it is uncommon for those massive satellite analogs to have a virial host halo mass $\geq 1.5\times10^{12}$ \Msun. Only 15.46 per cent of all analogs belong to the iRP100T sample and have a host halo that massive, while only 1.5 per cent of all analogs belong to the iRP100T sample and have a host halo mass $\geq2.5\times10^{12}$ \Msun. This may provide an upper limit on the halo mass of the MW and also for M31 if M33 truly had a recent, close encounter. 

\begin{table*}
\caption{The fraction of 472 massive satellite analogs in Illustris-Dark satisfying the following orbital criteria from the combination of their merger tree data and forward orbits. * For the analogs whose orbits have been integrated forward in time, we search for those where the pericentre occurs between the satellite's time of infall (so long as it is < 6 Gyr ago) and 3 Gyr in the future. For the  iRP100T and iRP55T samples, the time of pericentre must occur within $\pm$3 Gyr of $z=0$.}
\label{table:illustrisstats}
\centering
\begin{tabular}{c|c|c|c|c|c|} \hline \hline
Identifier & N$\rm_{peri}$ & $\rm t_{peri}*$ & $\rm t_{inf}$ & $\rm r_{peri}$ &   \\ \hline 
IRP & $\geq$1 &  $\leq 6$ Gyr ago  & -- & -- & 77.54\%  \\ 
iTI6 & $\geq$1 &  $\leq 6$ Gyr ago &  $\leq$ 6 Gyr ago & -- & 67.58\%  \\ 
iRP100 & $\geq$1 &  $\leq 6$ Gyr ago &  $\leq$ 6 Gyr ago & $\rm r_{peri} <$ 100 kpc & 51.90\%  \\ 
iRP100T & $\geq$1 &  $\leq 3$ Gyr ago  &  $\leq$ 6 Gyr ago & $\rm r_{peri} <$ 100 kpc  & 49.36\%  \\ 
iRP55 & $\geq$1 &  $\leq 6$ Gyr ago  &  $\leq$ 6 Gyr ago & $\rm r_{peri} <$ 55 kpc & 32.42\% \\ 
iRP55T & $\geq$1 &  $\leq 3$ Gyr ago  &  $\leq$ 6 Gyr ago & $\rm r_{peri} <$ 55 kpc & 32.42\% \\ \hline
\end{tabular}
\end{table*}

\subsubsection{Ramifications for the Lack of a Close Encounter}
From our discussion of the morphologically motivated orbit for M33, we conclude that a recent, close encounter between M33 and M31 is rare within our analytic models and only as likely as a large pericentric approach cosmologically. In this case we must search for alternative scenarios to explain the origin of M33's warped morphology. 

For instance, M33 could host its own system of less massive satellites as predicted by galaxy formation models. Recent work has suggested the same could be true for the LMC (see Section~\ref{sec:intro}). If these satellites of satellites exist and have had close interactions with M33, one or more could have contributed to some degree of the warped structures observed in its disc. Of the known satellites within the M31 system, And XXII has been suggested as a potential companion of M33 since it has a similar systemic velocity.  While it may or may not be bound to M33 at present, several authors suggest that mutual interactions between M33 and And XXII could have distorted M33's discs if they were once associated in the past \citep{tollerud12, chapman13, shaya13, martin09}.

Ram pressure stripping could also play a role in warping M33's gas disc. However, the magnitude of ram pressure stripping depends on orbital eccentricity and the inclination of its disc relative to its orbital plane. Each of the above scenarios requires careful modeling and should be studied in further detail in attempt to fully understand the morphological and dynamical history of M33. These goals are beyond the scope of this paper.

The orbital history of M33, whether it really is on first infall or if it made a passage about M31 $\sim$5-6 Gyr ago, is also relevant for the proposed plane of satellites surrounding M31 \citep[e.g. ][]{ibata13} wherein, 13 satellites are suggested to be co-rotating about M31 in a plane about 13 kpc in width. While this plane does not include all of the known M31 satellites or M33, it could be affected by the massive nature of M33. For example, if M33 has been on a long-period orbit or if it is moving radially towards M31 for the first time, the presence of M33 would likely have some gravitational influence on the plane of satellites, especially for the Southern half of the plane. Therefore, M33's history is not only crucial to understanding the evolution of its own galactic features, but it may also influence the larger M31 system of satellites and their dynamical history. 

\subsection{Implications for the Proper Motion Measurements of M31}
\label{subsec:orbitimplications}

\begin{figure}
\centering
\includegraphics[scale=0.55, trim=0mm 10mm 0mm 0mm]{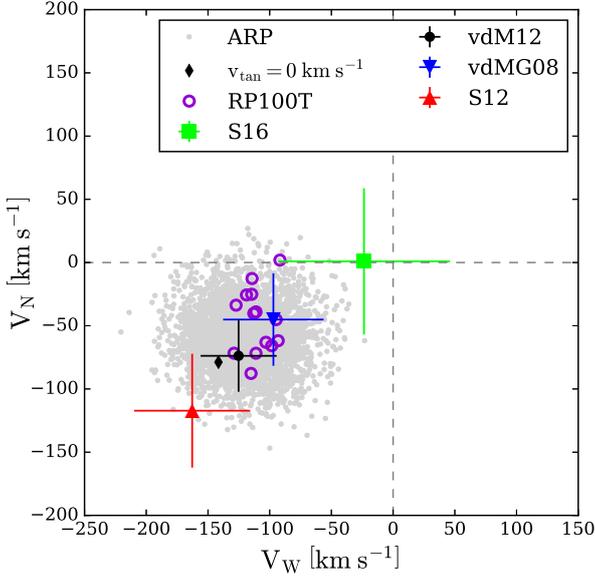}
\caption{Proper motion measurements of M31 in units of North and West tangential velocity components. The gray points are the ARP sample for the high mass M31 and $2.5\times 10^{11}$ \Msun M33. The red triangle is the HST proper motion result from S12. The blue triangle is measured by satellite kinematics in vdMG08. The black circle is from vdM12 and is the weighted average of vdMG08 and S12. The green marker indicates the proper motion results of S16 using the satellite kinematics of the M31 system estimated from $\Lambda$CDM simulations. Finally, the black diamond is the proper motion resulting from a zero tangential velocity of M31. It is offset from the origin due to the Sun's motion. The analytic orbits that have a pericentre within 100 kpc of M31 in the last 3 Gyr are indicated by purple circles (RP100T sample). 
\label{fig:pms}}
\end{figure}

In recent years, the tangential velocity component of M31 has been measured in various ways. Several results, measured directly and inferred indirectly, are plotted in Fig.~\ref{fig:pms} and summarized here. S16 used the kinematics of 40 M31 satellites to estimate the motion of M31 via $\Lambda$CDM simulations and statistical fitting methods (green square). vdMG08 performs three statistical techniques and reports a weighted average using line of sight velocities for 17 satellites, the proper motions of two satellites, and line of sight velocities for five Local Group galaxies (blue triangle). S12 recently used HST to take direct measurements of M31's proper motion (red triangle). The black circle indicates the weighted average of S12 and vdMG08, which are the values used in this analysis (vdM12). Finally, the diamond indicates the resulting velocity components for zero tangential velocity. These values are shifted away from the origin due to the Sun's motion. 

The large disparity between the S16 and S12 values is immediately evident. Consequently, the tangential velocity components reported by each team has serious implications for our understanding of the Local Group. The results of S16 imply the Local Group is not a bound system, complicating our understanding of its history. On the other hand, the S12 values imply M31's baryonic centre of mass is offset in velocity from its outer dark matter halo. The latter has been proposed by \cite{gao07} in their analysis of central galaxies in $\Lambda$CDM simulations. Their conclusion is further supported by G15, who claim the massive nature of the LMC causes a dynamical impact on the MW and therefore a velocity shift. From our numerical orbit analysis, we know that the presence of M33 does indeed cause a shift in M31's velocity up to tens of \kms as well.

Aiming to reconstruct M33's morphology, \cite{loeb05} estimated the tangental velocity for M31 by designing a numerical model where M33's stellar disc remains unperturbed by tidal disruptions over the last 10 Gyr and they recover a value of $\rm v_{tan}=100\pm20$ \kms. Similarly, we constrain M31's tangential velocity using only the theorized dynamical history of M33 which supports the formation of its warp through a past interaction with M31 (P09, M09). 

In Fig.~\ref{fig:pms}, we show the North and West tangential velocity components corresponding to the 3D velocity vectors of M31 for the entire ARP sample (gray points)\footnote{Proper motion is converted to tangential velocity components by $\mu_i = v_i/(4.74*\rm d_{M31}$).}. The points cluster in between the S12 and S16 values, with no particular preference towards either. The purple circles highlight just the RP100T orbits where a close ($\leq$ 100 kpc), recent ($\rm t_{peri} < 3$ Gyr ago) pericentre exists. These orbits trace a specific region of the West velocity component, between about -150 and -100 \kms. About half of these orbits fall within the vdM12 $1\sigma$ error space, which is the average of the S12 and vdMG08 results. 

We find that the close passage for M33 is inconsistent with the HST PM measurement of M31 by S12. However, a more precise PM for M31 will (i) better constrain the tangential motion of M31 with direct measurements and (ii) rule out whether a recent, close passage scenario for M33 has occurred given its current motion. Doing so would suggest that a past interaction between M31 and M33 is not the one and only source of significant warps in the discs of M33, motivating a re-evaluation of the dynamical history of the M31 system.

\section{Conclusions}
\label{sec:conclusions}
The orbital evolution of massive satellite galaxies, and specifically those of the LMC and M33, have been explored in three contexts in this work: by numerically integrating their orbits backwards in time using astrometric data, by studying a large sample of massive satellite analogs in the Illustris-Dark simulation \citep{nelson15, vogelsberger14A, vogelsberger14B, genel14}, and by determining their consistency with orbital expectations informed by morphology. 

We have explored plausible orbital histories for the LMC and M33 about the MW and M31, respectively, using their observationally constrained velocity error space and backward integration schemes (e.g. B07, G15). The recently determined proper motion of M31 (S12) has enabled the study of M33's orbital history for the first time in this fashion. We find consistency with previous studies of LMC's orbital history. If the MW's total mass is < $1.5 \times 10^{12}$ \Msun, the LMC is on its first approach to the MW and only recently completed its first pericentric passage. Surprisingly, we find that M33 is either also completing its first orbit about M31, or if M31 is massive (>$2 \times 10^{12}$ \Msun) then it is on a long period orbit. Note that in this study we have adopted a new dynamical friction approximation \citep{vdm12iii}, which reduces the orbital decay of satellite trajectories as their gravitational softening lengths and masses increase. 

From our sample of massive satellite analogs in Illustris-Dark, we find that the orbital energetics and eccentricities of the LMC and M33 are generally consistent with a recent infall scenario (\tcross < 4 Gyr). Comparing the kinematics from recently updated LMC proper motion measurements to the orbital energies of the massive satellite analogs in Illustris-Dark, we find that a MW halo mass of $\sim$1.5$\times 10^{12}$ \Msun is preferred for recently accreted satellites. Early accretion for an LMC analog is cosmologically likely only if the MW's halo mass is >$3\times 10^{12}$ \Msun. Applying the same analysis using M33's kinematics favors an M31 halo mass $\geq1.5\times10^{12}$ \Msun if M33 is accreted recently. Early accretion of M33 is only plausible at the 20 per cent level if M31's halo mass is $\sim\!3\times10^{12}$ \Msun. Therefore, first infall is certainly favored from energetics alone. These results are generally consistent with the results of the numerical orbit integrations. 

We conclude that both the LMC and M33 are most likely completing their first orbits about their hosts. The MW's halo mass is likely $\sim1.5 \times 10^{12}$ \Msun. M31's halo mass is likely $\geq1.5\times10^{12}$ \Msun. Paper II will focus on estimating the most typical halo masses for the MW and M31 based on the LMC and M33's present-day orbital angular momentum. We will apply Bayesian inference methods to analogs in the Illustris-Dark simulation to compute the posterior distribution of halo mass from satellite properties via importance sampling and kernel density estimation.

The orbital eccentricities of LMC and M33 cosmological analogs were extracted directly from the merger trees and also computed using the instantaneous position and velocity of the satellite, treating the satellite as a point mass. We find markedly different results in the correlation between eccentricity and infall time. In particular, the weak correlation between infall time and eccentricity computed from orbital trajectories implies that eccentricity should not be used to characterize satellites by early and late infall times.

Our orbital analysis further reveals that M33 is unlikely to have reached closer than 100 kpc to M31, regardless of its orbital history. This is at odds with conventional models, where M33 is expected to have approached within 50-100 kpc of M31 in order to reproduce its observed warped morphology (P09, M09). We find that orbits recovering this scenario are $\sim$2.5$\sigma$ outliers from the mean $v_Y$ and $v_Z$ components of M33's velocity vector relative to M31 (representing only 0.14 per cent of orbits recovered by sampling the full error space 10,000 times in a Monte Carlo fashion). Upon testing these conclusions when M31 is modeled with a total mass near its predicted upper limit of $3\times10^{12}$ \Msun, we still find that the orbits suggested by M09 and P09 are rare within the proper motion error space explored in this paper. Furthermore, high mass host haloes (>$2.5\times 10^{12}$ \Msun) are cosmologically rare in this orbital configuration.

Cosmologically, recently accreted massive satellite are about equally likely to have made recent (< 3 Gyr), close (< 100 kpc) encounters as recent wide encounters (> 100 kpc). There is no cosmological preference for either case. 32.42 per cent of such recently accreted massive satellite analogs have encounters < 55 kpc--i.e. the LMC's orbit, which brings it within 50 kpc of the MW, is not cosmologically rare. From the combined numerical integration and cosmological analysis of M33's orbit, we propose that other sources of its warped disc should be investigated (i.e. other M31 satellites, ram pressure stripping, etc.).

While the proper motions generally do not support a recent, close interaction between M33 and M31, the few numerically integrated orbits that do support this scenario are not consistent with the M31 tangential velocity components measured directly with HST (S12) or inferred by satellite kinematics (S16). More precise M31 proper motion measurements are necessary to disentangle M33's true orbital history. 

The orbital histories of the four most massive members of the Local Group are computed simultaneously to demonstrate that the LMC's trajectory has not been significantly perturbed by M31, nor M33 by the MW during the last 6 Gyr. Allowing the MW and M31 to move freely in these integrations also demonstrates that the LMC and M33 change the velocity of their hosts by tens and sometimes up to a hundred kilometers per second in just the last 2 Gyr, which may have important implications for the inferred orbital histories of their other satellites (e.g. G15), such as those used in S16 to infer properties of M31.

We conclude that the third and fourth most massive members of the Local Group, M33 and the LMC, respectively, are recent interlopers in the environment of their hosts. Such recent infall scenarios suggest they should both still contain a majority of their cosmological infall masses ($\sim$10 per cent of their host's mass) today. Therefore, we must account for their dynamical influence on all other MW and M31 satellites.

\section*{Acknowledgments}
E.P. is supported by the National Science Foundation through the Graduate Research Fellowship Program funded by Grant Award No. DGE-1143953. This research was also funded through a grant for HST program AR-12632. Support for AR-12632 was provided by NASA through a grant from the Space Telescope Science Institute, which is operated by the Association of Universities for Research in Astronomy, Inc., under NASA contract NAS 5-26555. Orbital model calculations were performed with the El Gato cluster at the University of Arizona, which is funded by the National Science Foundation through Grant No. 1228509. The Illustris simulations were run on the Odyssey cluster supported by the FAS Science Division Research Computing Group at Harvard University. Many thanks to Vicente Rodriguez-Gomez for useful discussions related to Illustris and the associated merger trees. The authors would also like to recognize Oleg Gnedin and Mary Putman whose private correspondences have further contributed to our understanding of previous orbital analyses. Finally, we thank Mike Boylan-Kolchin, Roeland van der Marel, Nicolas Garavito, Paul Torrey, Dennis Zaritsky, Nitya Kallivayalil, Daniel Stark, Benjamin Weiner, Erik Tollerud, and Beth Willman for stimulating discussions that have contributed to this paper.



\bibliographystyle{mnras}
\bibliography{myrefs} 

\begin{thebibliography}{}
\makeatletter
\relax
\def\mn@urlcharsother{\let\do\@makeother \do\$\do\&\do\#\do\^\do\_\do\%\do\~}
\def\mn@doi{\begingroup\mn@urlcharsother \@ifnextchar [ {\mn@doi@}
  {\mn@doi@[]}}
\def\mn@doi@[#1]#2{\def\@tempa{#1}\ifx\@tempa\@empty \href
  {http://dx.doi.org/#2} {doi:#2}\else \href {http://dx.doi.org/#2} {#1}\fi
  \endgroup}
\def\mn@eprint#1#2{\mn@eprint@#1:#2::\@nil}
\def\mn@eprint@arXiv#1{\href {http://arxiv.org/abs/#1} {{\tt arXiv:#1}}}
\def\mn@eprint@dblp#1{\href {http://dblp.uni-trier.de/rec/bibtex/#1.xml}
  {dblp:#1}}
\def\mn@eprint@#1:#2:#3:#4\@nil{\def\@tempa {#1}\def\@tempb {#2}\def\@tempc
  {#3}\ifx \@tempc \@empty \let \@tempc \@tempb \let \@tempb \@tempa \fi \ifx
  \@tempb \@empty \def\@tempb {arXiv}\fi \@ifundefined
  {mn@eprint@\@tempb}{\@tempb:\@tempc}{\expandafter \expandafter \csname
  mn@eprint@\@tempb\endcsname \expandafter{\@tempc}}}

\bibitem[\protect\citeauthoryear{{Battaglia} et~al.,}{{Battaglia}
  et~al.}{2005}]{battaglia05}
{Battaglia} G.,  et~al., 2005, \mn@doi [\mnras]
  {10.1111/j.1365-2966.2005.09367.x}, \href
  {http://adsabs.harvard.edu/abs/2005MNRAS.364..433B} {364, 433}

\bibitem[\protect\citeauthoryear{{Bechtol} et~al.,}{{Bechtol}
  et~al.}{2015}]{bechtol15}
{Bechtol} K.,  et~al., 2015, \mn@doi [\apj] {10.1088/0004-637X/807/1/50}, \href
  {http://adsabs.harvard.edu/abs/2015ApJ...807...50B} {807, 50}

\bibitem[\protect\citeauthoryear{{Belokurov} et~al.,}{{Belokurov}
  et~al.}{2014}]{belokurov14}
{Belokurov} V.,  et~al., 2014, \mn@doi [\mnras] {10.1093/mnras/stt1862}, \href
  {http://adsabs.harvard.edu/abs/2014MNRAS.437..116B} {437, 116}

\bibitem[\protect\citeauthoryear{{Berentzen}, {Athanassoula}, {Heller}  \&
  {Fricke}}{{Berentzen} et~al.}{2003}]{berentzen03}
{Berentzen} I.,  {Athanassoula} E.,  {Heller} C.~H.,   {Fricke} K.~J.,  2003,
  \mn@doi [\mnras] {10.1046/j.1365-8711.2003.06417.x}, \href
  {http://adsabs.harvard.edu/abs/2003MNRAS.341..343B} {341, 343}

\bibitem[\protect\citeauthoryear{{Besla}}{{Besla}}{2015}]{besla15}
{Besla} G.,  2015, preprint, \href
  {http://adsabs.harvard.edu/abs/2015arXiv151103346B} {} (\mn@eprint {arXiv}
  {1511.03346})

\bibitem[\protect\citeauthoryear{{Besla}, {Kallivayalil}, {Hernquist},
  {Robertson}, {Cox}, {van der Marel}  \& {Alcock}}{{Besla} et~al.}{2007}]{b07}
{Besla} G.,  {Kallivayalil} N.,  {Hernquist} L.,  {Robertson} B.,  {Cox} T.~J.,
   {van der Marel} R.~P.,   {Alcock} C.,  2007, \mn@doi [\apj]
  {10.1086/521385}, \href {http://adsabs.harvard.edu/abs/2007ApJ...668..949B}
  {668, 949}

\bibitem[\protect\citeauthoryear{{Besla}, {Kallivayalil}, {Hernquist}, {van der
  Marel}, {Cox}  \& {Kere{\v s}}}{{Besla} et~al.}{2012}]{besla12}
{Besla} G.,  {Kallivayalil} N.,  {Hernquist} L.,  {van der Marel} R.~P.,  {Cox}
  T.~J.,   {Kere{\v s}} D.,  2012, \mn@doi [\mnras]
  {10.1111/j.1365-2966.2012.20466.x}, \href
  {http://adsabs.harvard.edu/abs/2012MNRAS.421.2109B} {421, 2109}

\bibitem[\protect\citeauthoryear{{Besla}, {Mart{\'{\i}}nez-Delgado}, {van der
  Marel}, {Beletsky}, {Seibert}, {Schlafly}, {Grebel}  \& {Neyer}}{{Besla}
  et~al.}{2016}]{besla16}
{Besla} G.,  {Mart{\'{\i}}nez-Delgado} D.,  {van der Marel} R.~P.,  {Beletsky}
  Y.,  {Seibert} M.,  {Schlafly} E.~F.,  {Grebel} E.~K.,   {Neyer} F.,  2016,
  \mn@doi [\apj] {10.3847/0004-637X/825/1/20}, \href
  {http://adsabs.harvard.edu/abs/2016ApJ...825...20B} {825, 20}

\bibitem[\protect\citeauthoryear{{Binney} \& {Tremaine}}{{Binney} \&
  {Tremaine}}{2008}]{binneytremaine}
{Binney} J.,  {Tremaine} S.,  2008, {Galactic Dynamics: Second Edition}.
Princeton University Press

\bibitem[\protect\citeauthoryear{{Bland-Hawthorn}, {Sutherland}, {Agertz}  \&
  {Moore}}{{Bland-Hawthorn} et~al.}{2007}]{blandhawthorn07}
{Bland-Hawthorn} J.,  {Sutherland} R.,  {Agertz} O.,   {Moore} B.,  2007,
  \mn@doi [\apjl] {10.1086/524657}, \href
  {http://adsabs.harvard.edu/abs/2007ApJ...670L.109B} {670, L109}

\bibitem[\protect\citeauthoryear{{Block} et~al.,}{{Block}
  et~al.}{2006}]{block06}
{Block} D.~L.,  et~al., 2006, \mn@doi [\nat] {10.1038/nature05184}, \href
  {http://adsabs.harvard.edu/abs/2006Natur.443..832B} {443, 832}

\bibitem[\protect\citeauthoryear{{Bonanos} et~al.,}{{Bonanos}
  et~al.}{2006}]{bonanos06}
{Bonanos} A.~Z.,  et~al., 2006, \mn@doi [\apj] {10.1086/508140}, \href
  {http://adsabs.harvard.edu/abs/2006ApJ...652..313B} {652, 313}

\bibitem[\protect\citeauthoryear{{Boylan-Kolchin}, {Springel}, {White},
  {Jenkins}  \& {Lemson}}{{Boylan-Kolchin} et~al.}{2009}]{bk09}
{Boylan-Kolchin} M.,  {Springel} V.,  {White} S.~D.~M.,  {Jenkins} A.,
  {Lemson} G.,  2009, \mn@doi [\mnras] {10.1111/j.1365-2966.2009.15191.x},
  \href {http://adsabs.harvard.edu/abs/2009MNRAS.398.1150B} {398, 1150}

\bibitem[\protect\citeauthoryear{{Boylan-Kolchin}, {Besla}  \&
  {Hernquist}}{{Boylan-Kolchin} et~al.}{2011}]{bk11}
{Boylan-Kolchin} M.,  {Besla} G.,   {Hernquist} L.,  2011, \mn@doi [\mnras]
  {10.1111/j.1365-2966.2011.18495.x}, \href
  {http://adsabs.harvard.edu/abs/2011MNRAS.414.1560B} {414, 1560}

\bibitem[\protect\citeauthoryear{{Boylan-Kolchin}, {Bullock}, {Sohn}, {Besla}
  \& {van der Marel}}{{Boylan-Kolchin} et~al.}{2013}]{bk13}
{Boylan-Kolchin} M.,  {Bullock} J.~S.,  {Sohn} S.~T.,  {Besla} G.,   {van der
  Marel} R.~P.,  2013, \mn@doi [\apj] {10.1088/0004-637X/768/2/140}, \href
  {http://adsabs.harvard.edu/abs/2013ApJ...768..140B} {768, 140}

\bibitem[\protect\citeauthoryear{{Br{\"u}ns} \& {Kerp}}{{Br{\"u}ns} \&
  {Kerp}}{2004}]{bruns04}
{Br{\"u}ns} C.,  {Kerp} J.,  2004, Astronomische Nachrichten Supplement, \href
  {http://adsabs.harvard.edu/abs/2004ANS...325...61B} {325, 61}

\bibitem[\protect\citeauthoryear{{Br{\"u}ns} et~al.,}{{Br{\"u}ns}
  et~al.}{2005}]{bruns05}
{Br{\"u}ns} C.,  et~al., 2005, \mn@doi [\aap] {10.1051/0004-6361:20040321},
  \href {http://adsabs.harvard.edu/abs/2005A%26A...432...45B} {432, 45}

\bibitem[\protect\citeauthoryear{{Brunthaler}, {Reid}, {Falcke}, {Greenhill}
  \& {Henkel}}{{Brunthaler} et~al.}{2005}]{brunthaler05}
{Brunthaler} A.,  {Reid} M.~J.,  {Falcke} H.,  {Greenhill} L.~J.,   {Henkel}
  C.,  2005, \mn@doi [Science] {10.1126/science.1108342}, \href
  {http://adsabs.harvard.edu/abs/2005Sci...307.1440B} {307, 1440}

\bibitem[\protect\citeauthoryear{{Bryan} \& {Norman}}{{Bryan} \&
  {Norman}}{1998}]{brynorman98}
{Bryan} G.~L.,  {Norman} M.~L.,  1998, \mn@doi [\apj] {10.1086/305262}, \href
  {http://adsabs.harvard.edu/abs/1998ApJ...495...80B} {495, 80}

\bibitem[\protect\citeauthoryear{{Busha}, {Marshall}, {Wechsler}, {Klypin}  \&
  {Primack}}{{Busha} et~al.}{2011}]{busha11}
{Busha} M.~T.,  {Marshall} P.~J.,  {Wechsler} R.~H.,  {Klypin} A.,   {Primack}
  J.,  2011, \mn@doi [\apj] {10.1088/0004-637X/743/1/40}, \href
  {http://adsabs.harvard.edu/abs/2011ApJ...743...40B} {743, 40}

\bibitem[\protect\citeauthoryear{{Chandrasekhar}}{{Chandrasekhar}}{1943}]{chandrasekhar}
{Chandrasekhar} S.,  1943, \mn@doi [\apj] {10.1086/144517}, \href
  {http://adsabs.harvard.edu/abs/1943ApJ....97..255C} {97, 255}

\bibitem[\protect\citeauthoryear{{Chapman} et~al.,}{{Chapman}
  et~al.}{2013}]{chapman13}
{Chapman} S.~C.,  et~al., 2013, \mn@doi [\mnras] {10.1093/mnras/sts392}, \href
  {http://adsabs.harvard.edu/abs/2013MNRAS.430...37C} {430, 37}

\bibitem[\protect\citeauthoryear{{Connors}, {Kawata}  \& {Gibson}}{{Connors}
  et~al.}{2006}]{connors06}
{Connors} T.~W.,  {Kawata} D.,   {Gibson} B.~K.,  2006, \mn@doi [\mnras]
  {10.1111/j.1365-2966.2006.10659.x}, \href
  {http://adsabs.harvard.edu/abs/2006MNRAS.371..108C} {371, 108}

\bibitem[\protect\citeauthoryear{{Corbelli}}{{Corbelli}}{2003}]{corbelli03}
{Corbelli} E.,  2003, \mn@doi [\mnras] {10.1046/j.1365-8711.2003.06531.x},
  \href {http://adsabs.harvard.edu/abs/2003MNRAS.342..199C} {342, 199}

\bibitem[\protect\citeauthoryear{{Corbelli} \& {Salucci}}{{Corbelli} \&
  {Salucci}}{2000}]{corbellisalucci}
{Corbelli} E.,  {Salucci} P.,  2000, \mn@doi [\mnras]
  {10.1046/j.1365-8711.2000.03075.x}, \href
  {http://adsabs.harvard.edu/abs/2000MNRAS.311..441C} {311, 441}

\bibitem[\protect\citeauthoryear{{Corbelli}, {Lorenzoni}, {Walterbos}, {Braun}
  \& {Thilker}}{{Corbelli} et~al.}{2010}]{corbelli10}
{Corbelli} E.,  {Lorenzoni} S.,  {Walterbos} R.,  {Braun} R.,   {Thilker} D.,
  2010, \mn@doi [\aap] {10.1051/0004-6361/200913297}, \href
  {http://adsabs.harvard.edu/abs/2010A%26A...511A..89C} {511, A89}

\bibitem[\protect\citeauthoryear{{Davis}, {Efstathiou}, {Frenk}  \&
  {White}}{{Davis} et~al.}{1985}]{davis85}
{Davis} M.,  {Efstathiou} G.,  {Frenk} C.~S.,   {White} S.~D.~M.,  1985,
  \mn@doi [\apj] {10.1086/163168}, \href
  {http://adsabs.harvard.edu/abs/1985ApJ...292..371D} {292, 371}

\bibitem[\protect\citeauthoryear{{Deason}, {Wetzel}, {Garrison-Kimmel}  \&
  {Belokurov}}{{Deason} et~al.}{2015}]{deason15}
{Deason} A.~J.,  {Wetzel} A.~R.,  {Garrison-Kimmel} S.,   {Belokurov} V.,
  2015, \mn@doi [\mnras] {10.1093/mnras/stv1939}, \href
  {http://adsabs.harvard.edu/abs/2015MNRAS.453.3568D} {453, 3568}

\bibitem[\protect\citeauthoryear{{Dehnen}, {McLaughlin}  \&
  {Sachania}}{{Dehnen} et~al.}{2006}]{dehnen06}
{Dehnen} W.,  {McLaughlin} D.~E.,   {Sachania} J.,  2006, \mn@doi [\mnras]
  {10.1111/j.1365-2966.2006.10404.x}, \href
  {http://adsabs.harvard.edu/abs/2006MNRAS.369.1688D} {369, 1688}

\bibitem[\protect\citeauthoryear{{Diaz} \& {Bekki}}{{Diaz} \&
  {Bekki}}{2011}]{diaz11}
{Diaz} J.,  {Bekki} K.,  2011, \mn@doi [\mnras]
  {10.1111/j.1365-2966.2011.18289.x}, \href
  {http://adsabs.harvard.edu/abs/2011MNRAS.413.2015D} {413, 2015}

\bibitem[\protect\citeauthoryear{{Diaz} \& {Bekki}}{{Diaz} \&
  {Bekki}}{2012}]{diaz12}
{Diaz} J.~D.,  {Bekki} K.,  2012, \mn@doi [\apj] {10.1088/0004-637X/750/1/36},
  \href {http://adsabs.harvard.edu/abs/2012ApJ...750...36D} {750, 36}

\bibitem[\protect\citeauthoryear{{Dierickx}, {Blecha}  \& {Loeb}}{{Dierickx}
  et~al.}{2014}]{dierickx14}
{Dierickx} M.,  {Blecha} L.,   {Loeb} A.,  2014, \mn@doi [\apjl]
  {10.1088/2041-8205/788/2/L38}, \href
  {http://adsabs.harvard.edu/abs/2014ApJ...788L..38D} {788, L38}

\bibitem[\protect\citeauthoryear{{Dolag}, {Borgani}, {Murante}  \&
  {Springel}}{{Dolag} et~al.}{2009}]{dolag09}
{Dolag} K.,  {Borgani} S.,  {Murante} G.,   {Springel} V.,  2009, \mn@doi
  [\mnras] {10.1111/j.1365-2966.2009.15034.x}, \href
  {http://adsabs.harvard.edu/abs/2009MNRAS.399..497D} {399, 497}

\bibitem[\protect\citeauthoryear{{Drlica-Wagner} et~al.,}{{Drlica-Wagner}
  et~al.}{2015}]{dwagner15}
{Drlica-Wagner} A.,  et~al., 2015, \mn@doi [\apj]
  {10.1088/0004-637X/813/2/109}, \href
  {http://adsabs.harvard.edu/abs/2015ApJ...813..109D} {813, 109}

\bibitem[\protect\citeauthoryear{{Evans}, {Wilkinson}, {Guhathakurta}, {Grebel}
   \& {Vogt}}{{Evans} et~al.}{2000}]{evans00}
{Evans} N.~W.,  {Wilkinson} M.~I.,  {Guhathakurta} P.,  {Grebel} E.~K.,
  {Vogt} S.~S.,  2000, \mn@doi [\apjl] {10.1086/312861}, \href
  {http://adsabs.harvard.edu/abs/2000ApJ...540L...9E} {540, L9}

\bibitem[\protect\citeauthoryear{{Fardal}, {Guhathakurta}, {Gilbert}, {Babul},
  {Dodge}, {Weinberg}  \& {Lu}}{{Fardal} et~al.}{2009}]{fardal09}
{Fardal} M.,  {Guhathakurta} P.,  {Gilbert} K.,  {Babul} A.,  {Dodge} C.,
  {Weinberg} M.~D.,   {Lu} Y.,  2009, in {Jogee} S.,  {Marinova} I.,  {Hao} L.,
    {Blanc} G.~A.,  eds,  Astronomical Society of the Pacific Conference Series
  Vol. 419, Galaxy Evolution: Emerging Insights and Future Challenges. p.~118

\bibitem[\protect\citeauthoryear{{Fardal} et~al.,}{{Fardal}
  et~al.}{2013}]{fardal13}
{Fardal} M.~A.,  et~al., 2013, \mn@doi [\mnras] {10.1093/mnras/stt1121}, \href
  {http://adsabs.harvard.edu/abs/2013MNRAS.434.2779F} {434, 2779}

\bibitem[\protect\citeauthoryear{{Fich} \& {Tremaine}}{{Fich} \&
  {Tremaine}}{1991}]{fich91}
{Fich} M.,  {Tremaine} S.,  1991, \mn@doi [\araa]
  {10.1146/annurev.aa.29.090191.002205}, \href
  {http://adsabs.harvard.edu/abs/1991ARA%26A..29..409F} {29, 409}

\bibitem[\protect\citeauthoryear{{Gao} \& {White}}{{Gao} \&
  {White}}{2007}]{gao07}
{Gao} L.,  {White} S.~D.~M.,  2007, \mn@doi [\mnras]
  {10.1111/j.1745-3933.2007.00292.x}, \href
  {http://adsabs.harvard.edu/abs/2007MNRAS.377L...5G} {377, L5}

\bibitem[\protect\citeauthoryear{{Gardiner} \& {Noguchi}}{{Gardiner} \&
  {Noguchi}}{1996}]{gardiner96}
{Gardiner} L.~T.,  {Noguchi} M.,  1996, \mn@doi [\mnras]
  {10.1093/mnras/278.1.191}, \href
  {http://adsabs.harvard.edu/abs/1996MNRAS.278..191G} {278, 191}

\bibitem[\protect\citeauthoryear{{Genel} et~al.,}{{Genel}
  et~al.}{2014}]{genel14}
{Genel} S.,  et~al., 2014, \mn@doi [\mnras] {10.1093/mnras/stu1654}, \href
  {http://adsabs.harvard.edu/abs/2014MNRAS.445..175G} {445, 175}

\bibitem[\protect\citeauthoryear{{Gibbons}, {Belokurov}  \& {Evans}}{{Gibbons}
  et~al.}{2014}]{gibbons14}
{Gibbons} S.~L.~J.,  {Belokurov} V.,   {Evans} N.~W.,  2014, \mn@doi [\mnras]
  {10.1093/mnras/stu1986}, \href
  {http://adsabs.harvard.edu/abs/2014MNRAS.445.3788G} {445, 3788}

\bibitem[\protect\citeauthoryear{{Gnedin}, {Kravtsov}, {Klypin}  \&
  {Nagai}}{{Gnedin} et~al.}{2004}]{contra}
{Gnedin} O.~Y.,  {Kravtsov} A.~V.,  {Klypin} A.~A.,   {Nagai} D.,  2004,
  \mn@doi [\apj] {10.1086/424914}, \href
  {http://adsabs.harvard.edu/abs/2004ApJ...616...16G} {616, 16}

\bibitem[\protect\citeauthoryear{{Gnedin}, {Brown}, {Geller}  \&
  {Kenyon}}{{Gnedin} et~al.}{2010}]{gnedin10}
{Gnedin} O.~Y.,  {Brown} W.~R.,  {Geller} M.~J.,   {Kenyon} S.~J.,  2010,
  \mn@doi [\apjl] {10.1088/2041-8205/720/1/L108}, \href
  {http://adsabs.harvard.edu/abs/2010ApJ...720L.108G} {720, L108}

\bibitem[\protect\citeauthoryear{{G{\'o}mez}, {Minchev}, {O'Shea}, {Beers},
  {Bullock}  \& {Purcell}}{{G{\'o}mez} et~al.}{2013}]{gomez13}
{G{\'o}mez} F.~A.,  {Minchev} I.,  {O'Shea} B.~W.,  {Beers} T.~C.,  {Bullock}
  J.~S.,   {Purcell} C.~W.,  2013, \mn@doi [\mnras] {10.1093/mnras/sts327},
  \href {http://adsabs.harvard.edu/abs/2013MNRAS.429..159G} {429, 159}

\bibitem[\protect\citeauthoryear{{G{\'o}mez}, {Besla}, {Carpintero},
  {Villalobos}, {O'Shea}  \& {Bell}}{{G{\'o}mez} et~al.}{2015}]{gomez15}
{G{\'o}mez} F.~A.,  {Besla} G.,  {Carpintero} D.~D.,  {Villalobos} {\'A}.,
  {O'Shea} B.~W.,   {Bell} E.~F.,  2015, \mn@doi [\apj]
  {10.1088/0004-637X/802/2/128}, \href
  {http://adsabs.harvard.edu/abs/2015ApJ...802..128G} {802, 128}

\bibitem[\protect\citeauthoryear{{Gonz{\'a}lez}, {Kravtsov}  \&
  {Gnedin}}{{Gonz{\'a}lez} et~al.}{2013}]{gonzalez13}
{Gonz{\'a}lez} R.~E.,  {Kravtsov} A.~V.,   {Gnedin} N.~Y.,  2013, \mn@doi
  [\apj] {10.1088/0004-637X/770/2/96}, \href
  {http://adsabs.harvard.edu/abs/2013ApJ...770...96G} {770, 96}

\bibitem[\protect\citeauthoryear{{Guglielmo}, {Lewis}  \&
  {Bland-Hawthorn}}{{Guglielmo} et~al.}{2014}]{guglielmo14}
{Guglielmo} M.,  {Lewis} G.~F.,   {Bland-Hawthorn} J.,  2014, \mn@doi [\mnras]
  {10.1093/mnras/stu1549}, \href
  {http://adsabs.harvard.edu/abs/2014MNRAS.444.1759G} {444, 1759}

\bibitem[\protect\citeauthoryear{{Guo}, {White}, {Li}  \&
  {Boylan-Kolchin}}{{Guo} et~al.}{2010}]{guo10}
{Guo} Q.,  {White} S.,  {Li} C.,   {Boylan-Kolchin} M.,  2010, \mn@doi [\mnras]
  {10.1111/j.1365-2966.2010.16341.x}, \href
  {http://adsabs.harvard.edu/abs/2010MNRAS.404.1111G} {404, 1111}

\bibitem[\protect\citeauthoryear{{Guo} et~al.,}{{Guo} et~al.}{2011}]{guo11}
{Guo} Q.,  et~al., 2011, \mn@doi [\mnras] {10.1111/j.1365-2966.2010.18114.x},
  \href {http://adsabs.harvard.edu/abs/2011MNRAS.413..101G} {413, 101}

\bibitem[\protect\citeauthoryear{{Hashimoto}, {Funato}  \&
  {Makino}}{{Hashimoto} et~al.}{2003}]{hashimoto03}
{Hashimoto} Y.,  {Funato} Y.,   {Makino} J.,  2003, \mn@doi [\apj]
  {10.1086/344260}, \href {http://adsabs.harvard.edu/abs/2003ApJ...582..196H}
  {582, 196}

\bibitem[\protect\citeauthoryear{{Hernquist}}{{Hernquist}}{1990}]{hernquist90}
{Hernquist} L.,  1990, \mn@doi [\apj] {10.1086/168845}, \href
  {http://adsabs.harvard.edu/abs/1990ApJ...356..359H} {356, 359}

\bibitem[\protect\citeauthoryear{{Hinshaw} et~al.,}{{Hinshaw}
  et~al.}{2013}]{hinshaw13}
{Hinshaw} G.,  et~al., 2013, \mn@doi [\apjs] {10.1088/0067-0049/208/2/19},
  \href {http://adsabs.harvard.edu/abs/2013ApJS..208...19H} {208, 19}

\bibitem[\protect\citeauthoryear{{Ibata} \& {Lewis}}{{Ibata} \&
  {Lewis}}{1998}]{ibata98}
{Ibata} R.~A.,  {Lewis} G.~F.,  1998, \mn@doi [\apj] {10.1086/305773}, \href
  {http://adsabs.harvard.edu/abs/1998ApJ...500..575I} {500, 575}

\bibitem[\protect\citeauthoryear{{Ibata} et~al.,}{{Ibata}
  et~al.}{2013}]{ibata13}
{Ibata} R.~A.,  et~al., 2013, \mn@doi [\nat] {10.1038/nature11717}, \href
  {http://adsabs.harvard.edu/abs/2013Natur.493...62I} {493, 62}

\bibitem[\protect\citeauthoryear{{Jethwa}, {Erkal}  \& {Belokurov}}{{Jethwa}
  et~al.}{2016}]{jethwa16}
{Jethwa} P.,  {Erkal} D.,   {Belokurov} V.,  2016, preprint, \href
  {http://adsabs.harvard.edu/abs/2016arXiv160304420J} {} (\mn@eprint {arXiv}
  {1603.04420})

\bibitem[\protect\citeauthoryear{{Kallivayalil}, {van der Marel}, {Alcock},
  {Axelrod}, {Cook}, {Drake}  \& {Geha}}{{Kallivayalil} et~al.}{2006a}]{k06a}
{Kallivayalil} N.,  {van der Marel} R.~P.,  {Alcock} C.,  {Axelrod} T.,  {Cook}
  K.~H.,  {Drake} A.~J.,   {Geha} M.,  2006a, \mn@doi [\apj] {10.1086/498972},
  \href {http://adsabs.harvard.edu/abs/2006ApJ...638..772K} {638, 772}

\bibitem[\protect\citeauthoryear{{Kallivayalil}, {van der Marel}  \&
  {Alcock}}{{Kallivayalil} et~al.}{2006b}]{k06b}
{Kallivayalil} N.,  {van der Marel} R.~P.,   {Alcock} C.,  2006b, \mn@doi
  [\apj] {10.1086/508014}, \href
  {http://adsabs.harvard.edu/abs/2006ApJ...652.1213K} {652, 1213}

\bibitem[\protect\citeauthoryear{{Kallivayalil}, {Besla}, {Sanderson}  \&
  {Alcock}}{{Kallivayalil} et~al.}{2009}]{k09}
{Kallivayalil} N.,  {Besla} G.,  {Sanderson} R.,   {Alcock} C.,  2009, \mn@doi
  [\apj] {10.1088/0004-637X/700/2/924}, \href
  {http://adsabs.harvard.edu/abs/2009ApJ...700..924K} {700, 924}

\bibitem[\protect\citeauthoryear{{Kallivayalil}, {van der Marel}, {Besla},
  {Anderson}  \& {Alcock}}{{Kallivayalil} et~al.}{2013}]{k13}
{Kallivayalil} N.,  {van der Marel} R.~P.,  {Besla} G.,  {Anderson} J.,
  {Alcock} C.,  2013, \mn@doi [\apj] {10.1088/0004-637X/764/2/161}, \href
  {http://adsabs.harvard.edu/abs/2013ApJ...764..161K} {764, 161}

\bibitem[\protect\citeauthoryear{{Kim} \& {Jerjen}}{{Kim} \&
  {Jerjen}}{2015}]{kim15b}
{Kim} D.,  {Jerjen} H.,  2015, \mn@doi [\apjl] {10.1088/2041-8205/808/2/L39},
  \href {http://adsabs.harvard.edu/abs/2015ApJ...808L..39K} {808, L39}

\bibitem[\protect\citeauthoryear{{Kim}, {Staveley-Smith}, {Dopita}, {Freeman},
  {Sault}, {Kesteven}  \& {McConnell}}{{Kim} et~al.}{1998}]{kim98}
{Kim} S.,  {Staveley-Smith} L.,  {Dopita} M.~A.,  {Freeman} K.~C.,  {Sault}
  R.~J.,  {Kesteven} M.~J.,   {McConnell} D.,  1998, \mn@doi [\apj]
  {10.1086/306030}, \href {http://adsabs.harvard.edu/abs/1998ApJ...503..674K}
  {503, 674}

\bibitem[\protect\citeauthoryear{{Kim}, {Jerjen}, {Mackey}, {Da Costa}  \&
  {Milone}}{{Kim} et~al.}{2015}]{kim15a}
{Kim} D.,  {Jerjen} H.,  {Mackey} D.,  {Da Costa} G.~S.,   {Milone} A.~P.,
  2015, \mn@doi [\apjl] {10.1088/2041-8205/804/2/L44}, \href
  {http://adsabs.harvard.edu/abs/2015ApJ...804L..44K} {804, L44}

\bibitem[\protect\citeauthoryear{{Klypin}, {Trujillo-Gomez}  \&
  {Primack}}{{Klypin} et~al.}{2011}]{klypin11}
{Klypin} A.~A.,  {Trujillo-Gomez} S.,   {Primack} J.,  2011, \mn@doi [\apj]
  {10.1088/0004-637X/740/2/102}, \href
  {http://adsabs.harvard.edu/abs/2011ApJ...740..102K} {740, 102}

\bibitem[\protect\citeauthoryear{{Koposov}, {Belokurov}, {Torrealba}  \&
  {Evans}}{{Koposov} et~al.}{2015}]{koposov15}
{Koposov} S.~E.,  {Belokurov} V.,  {Torrealba} G.,   {Evans} N.~W.,  2015,
  \mn@doi [\apj] {10.1088/0004-637X/805/2/130}, \href
  {http://adsabs.harvard.edu/abs/2015ApJ...805..130K} {805, 130}

\bibitem[\protect\citeauthoryear{{Laporte}, {G{\'o}mez}, {Besla}, {Johnston}
  \& {Garavito-Camargo}}{{Laporte} et~al.}{2016}]{laporte16}
{Laporte} C.~F.~P.,  {G{\'o}mez} F.~A.,  {Besla} G.,  {Johnston} K.~V.,
  {Garavito-Camargo} N.,  2016, preprint, \href
  {http://adsabs.harvard.edu/abs/2016arXiv160804743L} {} (\mn@eprint {arXiv}
  {1608.04743})

\bibitem[\protect\citeauthoryear{{Lehner}, {Howk}  \& {Wakker}}{{Lehner}
  et~al.}{2015}]{lehner15}
{Lehner} N.,  {Howk} J.~C.,   {Wakker} B.~P.,  2015, \mn@doi [\apj]
  {10.1088/0004-637X/804/2/79}, \href
  {http://adsabs.harvard.edu/abs/2015ApJ...804...79L} {804, 79}

\bibitem[\protect\citeauthoryear{{Lewis} et~al.,}{{Lewis}
  et~al.}{2013}]{lewis13}
{Lewis} G.~F.,  et~al., 2013, \mn@doi [\apj] {10.1088/0004-637X/763/1/4}, \href
  {http://adsabs.harvard.edu/abs/2013ApJ...763....4L} {763, 4}

\bibitem[\protect\citeauthoryear{{Li} \& {White}}{{Li} \&
  {White}}{2008}]{liwhite08}
{Li} Y.-S.,  {White} S.~D.~M.,  2008, \mn@doi [\mnras]
  {10.1111/j.1365-2966.2007.12748.x}, \href
  {http://adsabs.harvard.edu/abs/2008MNRAS.384.1459L} {384, 1459}

\bibitem[\protect\citeauthoryear{{Loeb}, {Reid}, {Brunthaler}  \&
  {Falcke}}{{Loeb} et~al.}{2005}]{loeb05}
{Loeb} A.,  {Reid} M.~J.,  {Brunthaler} A.,   {Falcke} H.,  2005, \mn@doi
  [\apj] {10.1086/491644}, \href
  {http://adsabs.harvard.edu/abs/2005ApJ...633..894L} {633, 894}

\bibitem[\protect\citeauthoryear{{Lux}, {Read}  \& {Lake}}{{Lux}
  et~al.}{2010}]{lux10}
{Lux} H.,  {Read} J.~I.,   {Lake} G.,  2010, \mn@doi [\mnras]
  {10.1111/j.1365-2966.2010.16877.x}, \href
  {http://adsabs.harvard.edu/abs/2010MNRAS.406.2312L} {406, 2312}

\bibitem[\protect\citeauthoryear{{Mackey}, {Koposov}, {Erkal}, {Belokurov}, {Da
  Costa}  \& {G{\'o}mez}}{{Mackey} et~al.}{2016}]{mackey16}
{Mackey} A.~D.,  {Koposov} S.~E.,  {Erkal} D.,  {Belokurov} V.,  {Da Costa}
  G.~S.,   {G{\'o}mez} F.~A.,  2016, \mn@doi [\mnras] {10.1093/mnras/stw497},
  \href {http://adsabs.harvard.edu/abs/2016MNRAS.459..239M} {459, 239}

\bibitem[\protect\citeauthoryear{{Majewski}, {Nidever}, {Mu{\~n}oz},
  {Patterson}, {Kunkel}  \& {Carlin}}{{Majewski} et~al.}{2009}]{majewski09}
{Majewski} S.~R.,  {Nidever} D.~L.,  {Mu{\~n}oz} R.~R.,  {Patterson} R.~J.,
  {Kunkel} W.~E.,   {Carlin} J.~L.,  2009, in {Van Loon} J.~T.,  {Oliveira}
  J.~M.,  eds,  IAU Symposium Vol. 256, The Magellanic System: Stars, Gas, and
  Galaxies. pp 51--56, \mn@doi{10.1017/S1743921308028251}

\bibitem[\protect\citeauthoryear{{Martin} et~al.,}{{Martin}
  et~al.}{2009}]{martin09}
{Martin} N.~F.,  et~al., 2009, \mn@doi [\apj] {10.1088/0004-637X/705/1/758},
  \href {http://adsabs.harvard.edu/abs/2009ApJ...705..758M} {705, 758}

\bibitem[\protect\citeauthoryear{{Martin} et~al.,}{{Martin}
  et~al.}{2015}]{martin15}
{Martin} N.~F.,  et~al., 2015, \mn@doi [\apjl] {10.1088/2041-8205/804/1/L5},
  \href {http://adsabs.harvard.edu/abs/2015ApJ...804L...5M} {804, L5}

\bibitem[\protect\citeauthoryear{{Mathewson}, {Schwarz}  \&
  {Murray}}{{Mathewson} et~al.}{1977}]{mathewson77}
{Mathewson} D.~S.,  {Schwarz} M.~P.,   {Murray} J.~D.,  1977, \mn@doi [\apjl]
  {10.1086/182527}, \href {http://adsabs.harvard.edu/abs/1977ApJ...217L...5M}
  {217, L5}

\bibitem[\protect\citeauthoryear{{McConnachie} et~al.,}{{McConnachie}
  et~al.}{2009}]{mcconnachie09}
{McConnachie} A.~W.,  et~al., 2009, \mn@doi [\nat] {10.1038/nature08327}, \href
  {http://adsabs.harvard.edu/abs/2009Natur.461...66M} {461, 66}

\bibitem[\protect\citeauthoryear{{McMillan}}{{McMillan}}{2011}]{mcmillan11}
{McMillan} P.~J.,  2011, \mn@doi [\mnras] {10.1111/j.1365-2966.2011.19520.x},
  \href {http://adsabs.harvard.edu/abs/2011MNRAS.418.1565M} {418, 1565}

\bibitem[\protect\citeauthoryear{{McMonigal} et~al.,}{{McMonigal}
  et~al.}{2016}]{mcmonigal16}
{McMonigal} B.,  et~al., 2016, \mn@doi [\mnras] {10.1093/mnras/stw1657}, \href
  {http://adsabs.harvard.edu/abs/2016MNRAS.461.4374M} {461, 4374}

\bibitem[\protect\citeauthoryear{{Miyamoto} \& {Nagai}}{{Miyamoto} \&
  {Nagai}}{1975}]{mn75}
{Miyamoto} M.,  {Nagai} R.,  1975, \pasj, \href
  {http://adsabs.harvard.edu/abs/1975PASJ...27..533M} {27, 533}

\bibitem[\protect\citeauthoryear{{More}, {Diemer}  \& {Kravtsov}}{{More}
  et~al.}{2015}]{more15}
{More} S.,  {Diemer} B.,   {Kravtsov} A.~V.,  2015, \mn@doi [\apj]
  {10.1088/0004-637X/810/1/36}, \href
  {http://adsabs.harvard.edu/abs/2015ApJ...810...36M} {810, 36}

\bibitem[\protect\citeauthoryear{{Moster}, {Naab}  \& {White}}{{Moster}
  et~al.}{2013}]{moster13}
{Moster} B.~P.,  {Naab} T.,   {White} S.~D.~M.,  2013, \mn@doi [\mnras]
  {10.1093/mnras/sts261}, \href
  {http://adsabs.harvard.edu/abs/2013MNRAS.428.3121M} {428, 3121}

\bibitem[\protect\citeauthoryear{{Murai} \& {Fujimoto}}{{Murai} \&
  {Fujimoto}}{1980}]{murai80}
{Murai} T.,  {Fujimoto} M.,  1980, \pasj, \href
  {http://adsabs.harvard.edu/abs/1980PASJ...32..581M} {32, 581}

\bibitem[\protect\citeauthoryear{{Navarro}, {Frenk}  \& {White}}{{Navarro}
  et~al.}{1996}]{nfw96}
{Navarro} J.~F.,  {Frenk} C.~S.,   {White} S.~D.~M.,  1996, \mn@doi [\apj]
  {10.1086/177173}, \href {http://adsabs.harvard.edu/abs/1996ApJ...462..563N}
  {462, 563}

\bibitem[\protect\citeauthoryear{{Nelson} et~al.,}{{Nelson}
  et~al.}{2015}]{nelson15}
{Nelson} D.,  et~al., 2015, \mn@doi [Astronomy and Computing]
  {10.1016/j.ascom.2015.09.003}, \href
  {http://adsabs.harvard.edu/abs/2015A%26C....13...12N} {13, 12}

\bibitem[\protect\citeauthoryear{{Newton} \& {Emerson}}{{Newton} \&
  {Emerson}}{1977}]{newton77}
{Newton} K.,  {Emerson} D.~T.,  1977, \mn@doi [\mnras]
  {10.1093/mnras/181.3.573}, \href
  {http://adsabs.harvard.edu/abs/1977MNRAS.181..573N} {181, 573}

\bibitem[\protect\citeauthoryear{{Nidever}, {Majewski}  \& {Butler
  Burton}}{{Nidever} et~al.}{2008}]{nidever08}
{Nidever} D.~L.,  {Majewski} S.~R.,   {Butler Burton} W.,  2008, \mn@doi [\apj]
  {10.1086/587042}, \href {http://adsabs.harvard.edu/abs/2008ApJ...679..432N}
  {679, 432}

\bibitem[\protect\citeauthoryear{{Nidever}, {Majewski}, {Butler Burton}  \&
  {Nigra}}{{Nidever} et~al.}{2010}]{nidever10}
{Nidever} D.~L.,  {Majewski} S.~R.,  {Butler Burton} W.,   {Nigra} L.,  2010,
  \mn@doi [\apj] {10.1088/0004-637X/723/2/1618}, \href
  {http://adsabs.harvard.edu/abs/2010ApJ...723.1618N} {723, 1618}

\bibitem[\protect\citeauthoryear{{Oort}, {Kerr}  \& {Westerhout}}{{Oort}
  et~al.}{1958}]{oort58}
{Oort} J.~H.,  {Kerr} F.~J.,   {Westerhout} G.,  1958, \mn@doi [\mnras]
  {10.1093/mnras/118.4.379}, \href
  {http://adsabs.harvard.edu/abs/1958MNRAS.118..379O} {118, 379}

\bibitem[\protect\citeauthoryear{{Peebles}}{{Peebles}}{1996}]{peebles96}
{Peebles} P.~J.~E.,  1996, in {Lahav} O.,  {Terlevich} E.,   {Terlevich} R.~J.,
   eds, Gravitational dynamics. p.~219

\bibitem[\protect\citeauthoryear{{Putman} et~al.,}{{Putman}
  et~al.}{1998}]{putman98}
{Putman} M.~E.,  et~al., 1998, \mn@doi [\nat] {10.1038/29466}, \href
  {http://adsabs.harvard.edu/abs/1998Natur.394..752P} {394, 752}

\bibitem[\protect\citeauthoryear{{Putman}, {Staveley-Smith}, {Freeman},
  {Gibson}  \& {Barnes}}{{Putman} et~al.}{2003}]{putman03}
{Putman} M.~E.,  {Staveley-Smith} L.,  {Freeman} K.~C.,  {Gibson} B.~K.,
  {Barnes} D.~G.,  2003, \mn@doi [\apj] {10.1086/344477}, \href
  {http://adsabs.harvard.edu/abs/2003ApJ...586..170P} {586, 170}

\bibitem[\protect\citeauthoryear{{Putman} et~al.,}{{Putman}
  et~al.}{2009}]{putman09}
{Putman} M.~E.,  et~al., 2009, \mn@doi [\apj] {10.1088/0004-637X/703/2/1486},
  \href {http://adsabs.harvard.edu/abs/2009ApJ...703.1486P} {703, 1486}

\bibitem[\protect\citeauthoryear{{Roberts} \& {Whitehurst}}{{Roberts} \&
  {Whitehurst}}{1975}]{roberts75}
{Roberts} M.~S.,  {Whitehurst} R.~N.,  1975, \mn@doi [\apj] {10.1086/153889},
  \href {http://adsabs.harvard.edu/abs/1975ApJ...201..327R} {201, 327}

\bibitem[\protect\citeauthoryear{{Rodriguez-Gomez} et~al.,}{{Rodriguez-Gomez}
  et~al.}{2015}]{rg15}
{Rodriguez-Gomez} V.,  et~al., 2015, \mn@doi [\mnras] {10.1093/mnras/stv264},
  \href {http://adsabs.harvard.edu/abs/2015MNRAS.449...49R} {449, 49}

\bibitem[\protect\citeauthoryear{{R\r{u}\v{z}i\v{c}ka}, {Theis}  \&
  {Palou\v{s}}}{{R\r{u}\v{z}i\v{c}ka} et~al.}{2009}]{ruzicka09}
{R\r{u}\v{z}i\v{c}ka} A.,  {Theis} C.,   {Palou\v{s}} J.,  2009, \mn@doi [\apj]
  {10.1088/0004-637X/691/2/1807}, \href
  {http://adsabs.harvard.edu/abs/2009ApJ...691.1807R} {691, 1807}

\bibitem[\protect\citeauthoryear{{Saha} et~al.,}{{Saha} et~al.}{2010}]{saha10}
{Saha} A.,  et~al., 2010, \mn@doi [\aj] {10.1088/0004-6256/140/6/1719}, \href
  {http://adsabs.harvard.edu/abs/2010AJ....140.1719S} {140, 1719}

\bibitem[\protect\citeauthoryear{{Sakamoto}, {Chiba}  \& {Beers}}{{Sakamoto}
  et~al.}{2003}]{sakamoto03}
{Sakamoto} T.,  {Chiba} M.,   {Beers} T.~C.,  2003, \mn@doi [\aap]
  {10.1051/0004-6361:20021499}, \href
  {http://adsabs.harvard.edu/abs/2003A%26A...397..899S} {397, 899}

\bibitem[\protect\citeauthoryear{{Salem}, {Besla}, {Bryan}, {Putman}, {van der
  Marel}  \& {Tonnesen}}{{Salem} et~al.}{2015}]{salem15}
{Salem} M.,  {Besla} G.,  {Bryan} G.,  {Putman} M.,  {van der Marel} R.~P.,
  {Tonnesen} S.,  2015, \mn@doi [\apj] {10.1088/0004-637X/815/1/77}, \href
  {http://adsabs.harvard.edu/abs/2015ApJ...815...77S} {815, 77}

\bibitem[\protect\citeauthoryear{{Sales}, {Navarro}, {Cooper}, {White}, {Frenk}
   \& {Helmi}}{{Sales} et~al.}{2011}]{sales11}
{Sales} L.~V.,  {Navarro} J.~F.,  {Cooper} A.~P.,  {White} S.~D.~M.,  {Frenk}
  C.~S.,   {Helmi} A.,  2011, \mn@doi [\mnras]
  {10.1111/j.1365-2966.2011.19514.x}, \href
  {http://adsabs.harvard.edu/abs/2011MNRAS.418..648S} {418, 648}

\bibitem[\protect\citeauthoryear{{Sales}, {Wang}, {White}  \&
  {Navarro}}{{Sales} et~al.}{2013}]{sales13}
{Sales} L.~V.,  {Wang} W.,  {White} S.~D.~M.,   {Navarro} J.~F.,  2013, \mn@doi
  [\mnras] {10.1093/mnras/sts054}, \href
  {http://adsabs.harvard.edu/abs/2013MNRAS.428..573S} {428, 573}

\bibitem[\protect\citeauthoryear{{Salomon}, {Ibata}, {Famaey}, {Martin}  \&
  {Lewis}}{{Salomon} et~al.}{2016}]{salomon16}
{Salomon} J.-B.,  {Ibata} R.~A.,  {Famaey} B.,  {Martin} N.~F.,   {Lewis}
  G.~F.,  2016, \mn@doi [\mnras] {10.1093/mnras/stv2865}, \href
  {http://adsabs.harvard.edu/abs/2016MNRAS.456.4432S} {456, 4432}

\bibitem[\protect\citeauthoryear{{Sch{\"o}nrich}, {Binney}  \&
  {Dehnen}}{{Sch{\"o}nrich} et~al.}{2010}]{schonrich10}
{Sch{\"o}nrich} R.,  {Binney} J.,   {Dehnen} W.,  2010, \mn@doi [\mnras]
  {10.1111/j.1365-2966.2010.16253.x}, \href
  {http://adsabs.harvard.edu/abs/2010MNRAS.403.1829S} {403, 1829}

\bibitem[\protect\citeauthoryear{{Shaya} \& {Tully}}{{Shaya} \&
  {Tully}}{2013}]{shaya13}
{Shaya} E.~J.,  {Tully} R.~B.,  2013, \mn@doi [\mnras] {10.1093/mnras/stt1714},
  \href {http://adsabs.harvard.edu/abs/2013MNRAS.436.2096S} {436, 2096}

\bibitem[\protect\citeauthoryear{{Sohn}, {Anderson}  \& {van der Marel}}{{Sohn}
  et~al.}{2012}]{sohn12}
{Sohn} S.~T.,  {Anderson} J.,   {van der Marel} R.~P.,  2012, \mn@doi [\apj]
  {10.1088/0004-637X/753/1/7}, \href
  {http://adsabs.harvard.edu/abs/2012ApJ...753....7S} {753, 7}

\bibitem[\protect\citeauthoryear{{Springel}}{{Springel}}{2010}]{springel10}
{Springel} V.,  2010, \mn@doi [\mnras] {10.1111/j.1365-2966.2009.15715.x},
  \href {http://adsabs.harvard.edu/abs/2010MNRAS.401..791S} {401, 791}

\bibitem[\protect\citeauthoryear{{Springel}, {Yoshida}  \& {White}}{{Springel}
  et~al.}{2001a}]{gadget}
{Springel} V.,  {Yoshida} N.,   {White} S.~D.~M.,  2001a, \mn@doi [\na]
  {10.1016/S1384-1076(01)00042-2}, \href
  {http://adsabs.harvard.edu/abs/2001NewA....6...79S} {6, 79}

\bibitem[\protect\citeauthoryear{{Springel}, {White}, {Tormen}  \&
  {Kauffmann}}{{Springel} et~al.}{2001b}]{springel01}
{Springel} V.,  {White} S.~D.~M.,  {Tormen} G.,   {Kauffmann} G.,  2001b,
  \mn@doi [\mnras] {10.1046/j.1365-8711.2001.04912.x}, \href
  {http://adsabs.harvard.edu/abs/2001MNRAS.328..726S} {328, 726}

\bibitem[\protect\citeauthoryear{{Staveley-Smith}}{{Staveley-Smith}}{2002}]{staveleysmith02}
{Staveley-Smith} L.,  2002, in {Taylor} A.~R.,  {Landecker} T.~L.,   {Willis}
  A.~G.,  eds,  Astronomical Society of the Pacific Conference Series Vol. 276,
  Seeing Through the Dust: The Detection of HI and the Exploration of the ISM
  in Galaxies. p.~391

\bibitem[\protect\citeauthoryear{{Staveley-Smith}, {Kim}, {Calabretta},
  {Haynes}  \& {Kesteven}}{{Staveley-Smith} et~al.}{2003}]{staveleysmith03}
{Staveley-Smith} L.,  {Kim} S.,  {Calabretta} M.~R.,  {Haynes} R.~F.,
  {Kesteven} M.~J.,  2003, \mn@doi [\mnras] {10.1046/j.1365-8711.2003.06146.x},
  \href {http://adsabs.harvard.edu/abs/2003MNRAS.339...87S} {339, 87}

\bibitem[\protect\citeauthoryear{{Stewart}, {Bullock}, {Wechsler}, {Maller}  \&
  {Zentner}}{{Stewart} et~al.}{2008}]{stewart08}
{Stewart} K.~R.,  {Bullock} J.~S.,  {Wechsler} R.~H.,  {Maller} A.~H.,
  {Zentner} A.~R.,  2008, \mn@doi [\apj] {10.1086/588579}, \href
  {http://adsabs.harvard.edu/abs/2008ApJ...683..597S} {683, 597}

\bibitem[\protect\citeauthoryear{{Tollerud}, {Boylan-Kolchin}, {Barton},
  {Bullock}  \& {Trinh}}{{Tollerud} et~al.}{2011}]{tollerud11}
{Tollerud} E.~J.,  {Boylan-Kolchin} M.,  {Barton} E.~J.,  {Bullock} J.~S.,
  {Trinh} C.~Q.,  2011, \mn@doi [\apj] {10.1088/0004-637X/738/1/102}, \href
  {http://adsabs.harvard.edu/abs/2011ApJ...738..102T} {738, 102}

\bibitem[\protect\citeauthoryear{{Tollerud} et~al.,}{{Tollerud}
  et~al.}{2012}]{tollerud12}
{Tollerud} E.~J.,  et~al., 2012, \mn@doi [\apj] {10.1088/0004-637X/752/1/45},
  \href {http://adsabs.harvard.edu/abs/2012ApJ...752...45T} {752, 45}

\bibitem[\protect\citeauthoryear{{Tsuchiya}}{{Tsuchiya}}{2002}]{tsuchiya02}
{Tsuchiya} T.,  2002, \mn@doi [\na] {10.1016/S1384-1076(02)00138-0}, \href
  {http://adsabs.harvard.edu/abs/2002NewA....7..293T} {7, 293}

\bibitem[\protect\citeauthoryear{{U}, {Urbaneja}, {Kudritzki}, {Jacobs},
  {Bresolin}  \& {Przybilla}}{{U} et~al.}{2009}]{u09}
{U} V.,  {Urbaneja} M.~A.,  {Kudritzki} R.-P.,  {Jacobs} B.~A.,  {Bresolin} F.,
    {Przybilla} N.,  2009, \mn@doi [\apj] {10.1088/0004-637X/704/2/1120}, \href
  {http://adsabs.harvard.edu/abs/2009ApJ...704.1120U} {704, 1120}

\bibitem[\protect\citeauthoryear{{Vogelsberger} et~al.,}{{Vogelsberger}
  et~al.}{2014a}]{vogelsberger14B}
{Vogelsberger} M.,  et~al., 2014a, \mn@doi [\mnras] {10.1093/mnras/stu1536},
  \href {http://adsabs.harvard.edu/abs/2014MNRAS.444.1518V} {444, 1518}

\bibitem[\protect\citeauthoryear{{Vogelsberger} et~al.,}{{Vogelsberger}
  et~al.}{2014b}]{vogelsberger14A}
{Vogelsberger} M.,  et~al., 2014b, \mn@doi [\nat] {10.1038/nature13316}, \href
  {http://adsabs.harvard.edu/abs/2014Natur.509..177V} {509, 177}

\bibitem[\protect\citeauthoryear{{Wang}, {Li}, {Kauffmann}  \& {De
  Lucia}}{{Wang} et~al.}{2006}]{wang06}
{Wang} L.,  {Li} C.,  {Kauffmann} G.,   {De Lucia} G.,  2006, \mn@doi [\mnras]
  {10.1111/j.1365-2966.2006.10669.x}, \href
  {http://adsabs.harvard.edu/abs/2006MNRAS.371..537W} {371, 537}

\bibitem[\protect\citeauthoryear{{Watkins}, {Evans}  \& {An}}{{Watkins}
  et~al.}{2010}]{watkins10}
{Watkins} L.~L.,  {Evans} N.~W.,   {An} J.~H.,  2010, \mn@doi [\mnras]
  {10.1111/j.1365-2966.2010.16708.x}, \href
  {http://adsabs.harvard.edu/abs/2010MNRAS.406..264W} {406, 264}

\bibitem[\protect\citeauthoryear{{Weinberg}}{{Weinberg}}{1998}]{weinberg98}
{Weinberg} M.~D.,  1998, \mn@doi [\mnras] {10.1046/j.1365-8711.1998.01790.x},
  \href {http://adsabs.harvard.edu/abs/1998MNRAS.299..499W} {299, 499}

\bibitem[\protect\citeauthoryear{{Weinberg} \& {Blitz}}{{Weinberg} \&
  {Blitz}}{2006}]{weinberg06}
{Weinberg} M.~D.,  {Blitz} L.,  2006, \mn@doi [\apjl] {10.1086/503607}, \href
  {http://adsabs.harvard.edu/abs/2006ApJ...641L..33W} {641, L33}

\bibitem[\protect\citeauthoryear{{Weisz} et~al.,}{{Weisz}
  et~al.}{2011}]{weisz11}
{Weisz} D.~R.,  et~al., 2011, \mn@doi [\apj] {10.1088/0004-637X/739/1/5}, \href
  {http://adsabs.harvard.edu/abs/2011ApJ...739....5W} {739, 5}

\bibitem[\protect\citeauthoryear{{Werk} et~al.,}{{Werk} et~al.}{2014}]{werk14}
{Werk} J.~K.,  et~al., 2014, \mn@doi [\apj] {10.1088/0004-637X/792/1/8}, \href
  {http://adsabs.harvard.edu/abs/2014ApJ...792....8W} {792, 8}

\bibitem[\protect\citeauthoryear{{Wetzel}}{{Wetzel}}{2011}]{wetzel11}
{Wetzel} A.~R.,  2011, \mn@doi [\mnras] {10.1111/j.1365-2966.2010.17877.x},
  \href {http://adsabs.harvard.edu/abs/2011MNRAS.412...49W} {412, 49}

\bibitem[\protect\citeauthoryear{{Wetzel}, {Tinker}, {Conroy}  \& {van den
  Bosch}}{{Wetzel} et~al.}{2014}]{wetzel14}
{Wetzel} A.~R.,  {Tinker} J.~L.,  {Conroy} C.,   {van den Bosch} F.~C.,  2014,
  \mn@doi [\mnras] {10.1093/mnras/stu122}, \href
  {http://adsabs.harvard.edu/abs/2014MNRAS.439.2687W} {439, 2687}

\bibitem[\protect\citeauthoryear{{Wetzel}, {Tollerud}  \& {Weisz}}{{Wetzel}
  et~al.}{2015}]{wetzel15}
{Wetzel} A.~R.,  {Tollerud} E.~J.,   {Weisz} D.~R.,  2015, \mn@doi [\apjl]
  {10.1088/2041-8205/808/1/L27}, \href
  {http://adsabs.harvard.edu/abs/2015ApJ...808L..27W} {808, L27}

\bibitem[\protect\citeauthoryear{{Williams}, {Dalcanton}, {Dolphin}, {Holtzman}
   \& {Sarajedini}}{{Williams} et~al.}{2009}]{williams09}
{Williams} B.~F.,  {Dalcanton} J.~J.,  {Dolphin} A.~E.,  {Holtzman} J.,
  {Sarajedini} A.,  2009, \mn@doi [\apjl] {10.1088/0004-637X/695/1/L15}, \href
  {http://adsabs.harvard.edu/abs/2009ApJ...695L..15W} {695, L15}

\bibitem[\protect\citeauthoryear{{Zentner} \& {Bullock}}{{Zentner} \&
  {Bullock}}{2003}]{zentner03}
{Zentner} A.~R.,  {Bullock} J.~S.,  2003, \mn@doi [\apj] {10.1086/378797},
  \href {http://adsabs.harvard.edu/abs/2003ApJ...598...49Z} {598, 49}

\bibitem[\protect\citeauthoryear{{van den Bergh}}{{van den
  Bergh}}{2006}]{vdb06}
{van den Bergh} S.,  2006, \mn@doi [\aj] {10.1086/507332}, \href
  {http://adsabs.harvard.edu/abs/2006AJ....132.1571V} {132, 1571}

\bibitem[\protect\citeauthoryear{{van der Marel} \& {Guhathakurta}}{{van der
  Marel} \& {Guhathakurta}}{2008}]{vdmG08}
{van der Marel} R.~P.,  {Guhathakurta} P.,  2008, \mn@doi [\apj]
  {10.1086/533430}, \href {http://adsabs.harvard.edu/abs/2008ApJ...678..187V}
  {678, 187}

\bibitem[\protect\citeauthoryear{{van der Marel} \& {Kallivayalil}}{{van der
  Marel} \& {Kallivayalil}}{2014}]{vdmnk14}
{van der Marel} R.~P.,  {Kallivayalil} N.,  2014, \mn@doi [\apj]
  {10.1088/0004-637X/781/2/121}, \href
  {http://adsabs.harvard.edu/abs/2014ApJ...781..121V} {781, 121}

\bibitem[\protect\citeauthoryear{{van der Marel}, {Alves}, {Hardy}  \&
  {Suntzeff}}{{van der Marel} et~al.}{2002}]{vdm02}
{van der Marel} R.~P.,  {Alves} D.~R.,  {Hardy} E.,   {Suntzeff} N.~B.,  2002,
  \mn@doi [\aj] {10.1086/343775}, \href
  {http://adsabs.harvard.edu/abs/2002AJ....124.2639V} {124, 2639}

\bibitem[\protect\citeauthoryear{{van der Marel}, {Fardal}, {Besla}, {Beaton},
  {Sohn}, {Anderson}, {Brown}  \& {Guhathakurta}}{{van der Marel}
  et~al.}{2012a}]{vdm12ii}
{van der Marel} R.~P.,  {Fardal} M.,  {Besla} G.,  {Beaton} R.~L.,  {Sohn}
  S.~T.,  {Anderson} J.,  {Brown} T.,   {Guhathakurta} P.,  2012a, \mn@doi
  [\apj] {10.1088/0004-637X/753/1/8}, \href
  {http://adsabs.harvard.edu/abs/2012ApJ...753....8V} {753, 8}

\bibitem[\protect\citeauthoryear{{van der Marel}, {Besla}, {Cox}, {Sohn}  \&
  {Anderson}}{{van der Marel} et~al.}{2012b}]{vdm12iii}
{van der Marel} R.~P.,  {Besla} G.,  {Cox} T.~J.,  {Sohn} S.~T.,   {Anderson}
  J.,  2012b, \mn@doi [\apj] {10.1088/0004-637X/753/1/9}, \href
  {http://adsabs.harvard.edu/abs/2012ApJ...753....9V} {753, 9}

\makeatother
\end{thebibliography}
\appendix

\section{Forward Orbit Integrations}
\label{sec:A}
For the massive satellite analogs which do not contain a pericentre and/or apocentre between their crossing time and $z=0$ in the Illustris-Dark merger tree data, we integrate their orbits forward in time for 6 Gyr. The host haloes are modeled as NFW dark matter haloes (Equation~\ref{eq:nfw}) and the satellites are modeled by Plummer spheres (Equation~\ref{eq:plummer}). We aim to match the orbits from the Illustris-Dark simulation, which contains only dark matter, so we do not model the baryons in the host galaxies unlike the analytic constructions in Section~\ref{sec:analyticmethods}. Consequently, the NFW host haloes are not adiabatically contracted. We do, however, implement dynamical friction (Equation~\ref{eq:df}) and allow the host haloes to move as a result of the gravitational force exerted by the satellites. 

The satellite gravitational softening lengths, $\rm k_{sat}$, for the massive satellite analogs are computed by fitting the following equation:
\begin{equation} \rm M(r_{half}) = \frac{M_{tot}r_{half}^3}{(r_{half}^2 + k_{sat}^2)^{3/2}}. \label{eq:ksat} \end{equation}
$\rm M_{tot}$ is the subhalo mass at $z=0$, M(r$\rm_{half}$) is half of the subhalo mass, and $\rm r_{half}$ is the radius at which half of the subhalo mass is enclosed. $\rm r_{half}$ is provided in the Illustris-Dark halo catalogs. With these quantities, $\rm k_{sat}$ is calculated to match the mass enclosed within $\rm r_{half}$, given $\rm M_{tot}$.

All forward orbits are calculated for 6 Gyr, except in Section~\ref{subsec:M33closeencounter} where we integrate forward for 3 Gyr to recover only recently accreted satellites. The future orbital trajectories are then analysed to find a true pericentre/and or apocentre as described in Section~\ref{subsubsec:forward}. In the event that a pericentre exists in the merger tree data, it is used in combination with the next apocentre from the forward orbit. Therefore, the merger trees and forward orbits are used in unison and act as a complete past and future orbital history. 

For 19 (4 per cent) of the massive satellite analogs, the forward orbit integration fails to find a pericentre and apocentre. We suspect these are fly by satellites that only remain in the vicinity of their host halo's \Rvir for a short period of time. Upon inspecting these 19 analogs further, we find that all satellites which are energetically unbound in Figs~\ref{fig:m33energy} and~\ref{fig:lmcenergy} are members of this population, providing a nice consistency check within our data sample. 

\begin{figure}
\centering
\label{fig:backwardorbit}
\includegraphics[scale=0.57, trim=10mm 10mm 0mm 0mm]{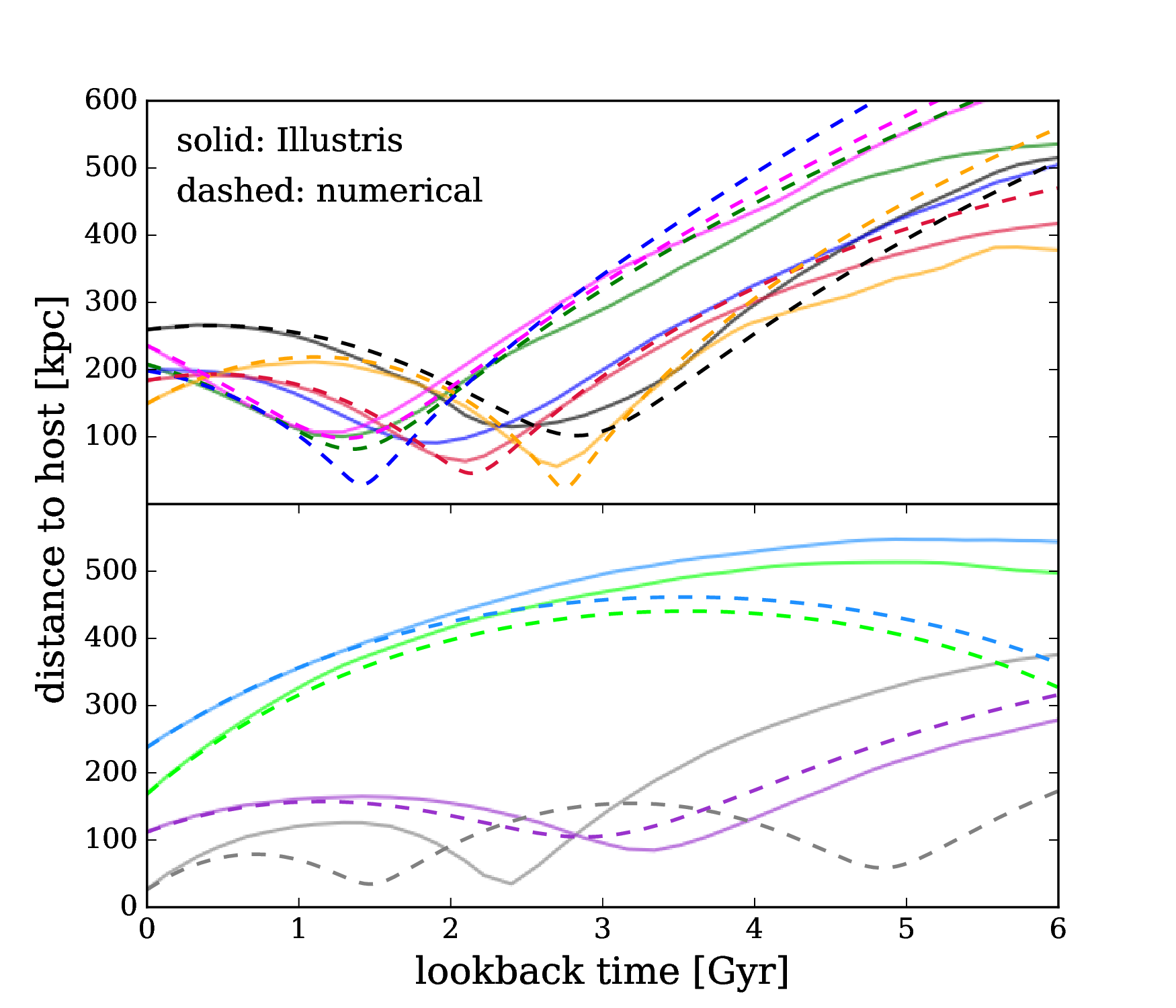}
\caption{Orbital histories for ten randomly chosen massive satellite analogs. The solid lines indicate the true distance of each analog relative to its host as a function of time from the Illustris-Dark merger tree data. The dashed lines indicate the corresponding numerical orbit integration using the $z=0$ properties of each host-satellite system. The hosts are modeled as NFW dark matter haloes and the satellites are approximated by Plummer spheres. The top panel shows the subset of orbits on first infall which reach a pericentric distance $\lesssim$100 kpc from their hosts in the last 3 Gyr, while the bottom panel shows the remaining orbits. There is good agreement for the first infall scenarios with a close pericentric passage (top) and first infall with no pericentric passage (bottom--light blue and green), which are the orbits relevant to our analysis. Therefore, numerical orbit integrations are a reasonable approximation to the future orbits of massive satellite analogs from Illustris-Dark.}
\end{figure}

While analytic orbits are often questioned as suitable matches to the true orbits in a cosmological volume, we show here that our backwards orbit integration methods result in acceptable solutions, especially for recent, first infall scenarios. Fig.~\ref{fig:backwardorbit} shows the merger tree data (solid lines) and corresponding numerical orbits (dashed lines) for ten randomly chosen massive satellite analogs in our Illustris sample. The top panel separates the six massive satellite analogs whose orbits exhibit a recent, first infall scenario. For these satellites, the orbital trajectories are recovered very well by the numerical integrations. 

The bottom panel of Fig.~\ref{fig:backwardorbit} indicates the remaining four orbits which are either accreted early and make multiple pericentric passages (purple and gray) or are on first infall with no recent pericentre (light blue and green). The latter are the analogs whose forward orbit integrations are important to our analysis, and they show good agreement between the merger tree data and backwards orbit integration. Therefore, integrating their orbits forward in time using their $z=0$ properties to recover their first pericentre is justified.

The numerical integrations are least effective at reproducing the orbits of massive satellite analogs that were accreted early and make several pericentric passages in the last 6 Gyr \citep{lux10}, but these are scarce in our massive satellite analogs population. Furthermore, we can recover orbital histories for them in the merger tree data, so there is no need to integrate their orbits forward in time. There is little concern with regards to cosmology in the future orbits since large scale structure is changing minimally at $z=0$ and therefore in the future. Furthermore, the sample is chosen such that there is only one massive host in the vicinity of each satellite analog. These numerical orbit integrations also confirm that the new implementation of dynamical friction used in this work (which allows for varying softening lengths based on mass) is efficient at predicting the orbital decay of massive satellites accurately. Finally, we relax our assumption that host haloes are truncated at their \Rvir and allow dynamical friction to continue to larger radii. By doing so, we find that there is little to no difference in our ability to match analytic orbits of massive satellites to their cosmological counterparts within $\sim$500 kpc of their hosts.

\bsp	
\label{lastpage}
\end{document}